\documentclass[twocolumn,showpacs,aps,prl,superscriptaddress]{revtex4}

\usepackage{graphicx} 
\usepackage{dcolumn} 
\usepackage{amsmath} 
\usepackage{epsfig} 
\usepackage{hhline} 
\usepackage{soul} 
\usepackage{tabularx,pifont,multirow} 
\usepackage{colordvi} 
 
\RequirePackage{xspace} 
 
\input babarsym 
\newcommand{\de}{\ensuremath{\mbox{$\Delta E$}}\xspace} 
\newcommand{\fis}{\ensuremath{\mbox{$\mathcal{F}$}}\xspace} 
\newcommand{\pvec}{{\bf p}}

\def\beq{\begin{equation}} 
\def\eeq{\end{equation}} 
\def\bea{\begin{eqnarray}} 
\def\eea{\end{eqnarray}} 
\def\bq{\begin{quote}} 
\def\eq{\end{quote}} 
\def\ben{\begin{enumerate}} 
\def\een{\end{enumerate}} 
\def\nn{\nonumber}

\newcommand{\phm}{\ensuremath{\phantom{-}}} 
 
\newcommand{\phz}{\ensuremath{\phantom{0}}} 
\newcommand{\phzz}{\ensuremath{\phantom{00}}} 

\def\kspipi{\ensuremath{\KS \pip \pim}\xspace} 
 
\def\kskk{\ensuremath{\KS \Kp \Km}\xspace} 
 
\def\kshh{\ensuremath{\KS h^+ h^-}\xspace} 

\def\D       {\ensuremath{D}\xspace} 
\def\K  {\ensuremath{K}\xspace} 

\def\DDstar   {\ensuremath{D^{(\ast)}}\xspace} 

\def\Dztokshh{\ensuremath{\Dz \to \kshh}\xspace} 
 
\def\Dtokspipi{\ensuremath{\D \to \kspipi}\xspace} 
\def\Dtokskk{\ensuremath{\D \to \kskk}\xspace} 

\def\splus{\ensuremath{s_{\rm +}}\xspace} 
\def\sminus{\ensuremath{s_{\rm -}}\xspace} 
 
\def\splusminus{\ensuremath{s_{\rm \pm}}\xspace} 
\def\sminusplus{\ensuremath{s_{\rm \mp}}\xspace} 
\def\hp{\ensuremath{h^{+}}\xspace} 
\def\hm{\ensuremath{h^{-}}\xspace} 
\def\h{\ensuremath{h}\xspace} 
\def \rb {\ensuremath {r_\B}\xspace} 
\def \rbmp {\ensuremath {r_{\Bmp}}\xspace} 
\def \rbst {\ensuremath {r^\ast_\B}\xspace} 
\def \rbrbst {\ensuremath {r^{(\ast)}_\B}\xspace} 
 
\def \rbrbstmp {\ensuremath {r^{(\ast)}_{\Bmp}}\xspace} 
\def \rs {\ensuremath {r_s}\xspace} 
 
\def \rsmp {\ensuremath {r_{s\mp}}\xspace} 
 
\def \krs {\ensuremath {\kappa \rs}\xspace} 
\def \deltab {\ensuremath {\delta_\B}\xspace} 
\def \deltabst {\ensuremath {\delta^\ast_\B}\xspace} 
\def \deltabdeltabst {\ensuremath {\delta^{(\ast)}_\B}\xspace} 
\def \deltas {\ensuremath {\delta_s}\xspace} 
\def \xbp {\ensuremath {x_+}\xspace} 
\def \xbm {\ensuremath {x_-}\xspace}

\def \ybp {\ensuremath {y_+}\xspace} 
\def \ybm {\ensuremath {y_-}\xspace}

\def \zbp {\ensuremath {{\mathsf z}_+}\xspace} 
\def \zbm {\ensuremath {{\mathsf z}_-}\xspace} 
 
\def \zbmp {\ensuremath {{\mathsf z}_\mp}\xspace} 
    \def \zpibmp {\ensuremath {{\mathsf z}_{\mp,\pi}}\xspace} 

\def \xbstp {\ensuremath {x_+^\ast}\xspace} 
\def \xbstm {\ensuremath {x_-^\ast}\xspace}

\def \ybstp {\ensuremath {y_+^\ast}\xspace} 
\def \ybstm {\ensuremath {y_-^\ast}\xspace}

\def \zbstp {\ensuremath {{\mathsf z}_+^\ast}\xspace} 
\def \zbstm {\ensuremath {{\mathsf z}_-^\ast}\xspace} 
 
\def \zbstmp {\ensuremath {{\mathsf z}_\mp^\ast}\xspace} 
     \def \zpibstmp {\ensuremath {{\mathsf z}_{\mp,\pi}^\ast}\xspace}

\def \xbxbstmp {\ensuremath {x_\mp^{(\ast)}}\xspace}

\def \ybybstmp {\ensuremath {y_\mp^{(\ast)}}\xspace}

\def \zbzbstpm {\ensuremath {{\mathsf z}_\pm^{(\ast)}}\xspace} 
\def \zbzbstmp {\ensuremath {{\mathsf z}_\mp^{(\ast)}}\xspace} 

\def \xsp {\ensuremath {x_{s+}}\xspace} 
\def \xsm {\ensuremath {x_{s-}}\xspace} 
 
\def \xsmp {\ensuremath {x_{s\mp}}\xspace} 
 
\def \ysp {\ensuremath {y_{s+}}\xspace} 
\def \ysm {\ensuremath {y_{s-}}\xspace} 
 
\def \ysmp {\ensuremath {y_{s\mp}}\xspace} 
 
\def \zsp {\ensuremath {{\mathsf z}_{s+}}\xspace} 
\def \zsm {\ensuremath {{\mathsf z}_{s-}}\xspace} 
\def \zspm {\ensuremath {{\mathsf z}_{s\pm}}\xspace} 
\def \zsmp {\ensuremath {{\mathsf z}_{s\mp}}\xspace} 

\def \zvec {\ensuremath{{\bf{z}}}\xspace} 
\def \zvecbest {\ensuremath {{\bf{z_{\rm best}}}}\xspace} 
\def \pvec {\ensuremath{{\bf{p}}}\xspace}

\def\fisher    {\ensuremath{{\cal F}}\xspace}

\newcommand{\kevc}{\ensuremath{{\mathrm{\,ke\kern -0.1em V\!/}c}}\xspace} 
\newcommand{\kevcc}{\ensuremath{{\mathrm{\,ke\kern -0.1em V\!/}c^2}}\xspace} 

\newcommand{\npa}       [1]  {\npBase\ A~{\bf #1}}

\newcommand{\BABARPubYear}    {10} 
\newcommand{\BABARPubNumber}  {005} 
 
\newcommand{\SLACPubNumber} {14090}

\def\figurebox#1#2#3{% 
    \def\arg{#3}% 
    \ifx\arg\empty 
    {\hfill\vbox{\hsize#2\hrule\hbox to #2{\vrule\hfill\vbox to #1{\hsize#2\vfill}\vrule}\hrule}\hfill}% 
    \else 
    {\hfill\epsfbox{#3}\hfill}% 
    \fi}

\long\def\inst#1{\par\nobreak\kern 4pt\nobreak 
  {\it #1}\par\vskip 10pt plus 3pt minus 3pt}

\begin{document}

\begin{flushleft} 
\babar-PUB-\BABARPubYear/\BABARPubNumber ~~~~~~~ SLAC-PUB-\SLACPubNumber   \\ 
\end{flushleft}

\title{ 
{ 
\large \bf \boldmath  
Evidence for direct \CP violation in the measurement of the CKM angle $\gamma$ with $\Bmp \to D^{(*)} K^{(*)\mp}$ decays 
} 
} 

\author{P.~del~Amo~Sanchez} 
\author{J.~P.~Lees} 
\author{V.~Poireau} 
\author{E.~Prencipe} 
\author{V.~Tisserand} 
\affiliation{Laboratoire d'Annecy-le-Vieux de Physique des Particules (LAPP), Universit\'e de Savoie, CNRS/IN2P3,  F-74941 Annecy-Le-Vieux, France} 
\author{J.~Garra~Tico} 
\author{E.~Grauges} 
\affiliation{Universitat de Barcelona, Facultat de Fisica, Departament ECM, E-08028 Barcelona, Spain } 
\author{M.~Martinelli$^{ab}$} 
\author{A.~Palano$^{ab}$ } 
\author{M.~Pappagallo$^{ab}$ } 
\affiliation{INFN Sezione di Bari$^{a}$; Dipartimento di Fisica, Universit\`a di Bari$^{b}$, I-70126 Bari, Italy } 
% ...
\collaboration{The \babar\ Collaboration} 
\noaffiliation

\date{\today}% It is always \today, today, but you may specify any date with \date. 

\begin{abstract}  
\noindent 
We report the measurement of the Cabibbo-Kobayashi-Maskawa \CP-violating angle $\gamma$ through a Dalitz plot analysis 
of neutral \D meson decays to \kspipi and \kskk produced in the processes $\Bmp \to \D \Kmp$, 
$\Bmp \to \D^{*} \Kmp$ with $\Dstar \to D\piz,D\g$, and $\Bmp \to \D\Kstarmp$ with $\Kstarmp \to \KS \pimp$, 
using  
468 million \BB pairs collected by the \babar\ detector at the  
\pep2 asymmetric-energy \epem\ collider at SLAC.  
We measure 
$\gamma=(68 \pm 14 \pm 4 \pm 3)^\circ$ (\mbox{modulo $180^\circ$}), 
where the first error is statistical, the second is the experimental 
systematic uncertainty and the third reflects the uncertainty in  
the description of the neutral \D decay amplitudes. 
This result is inconsistent with $\gamma = 0$ (no direct \CP violation)  
with a significance of $3.5$ standard deviations. 
\end{abstract} 
 
\pacs{13.25.Hw, 11.30.Er, 12.15.Hh, 13.25.Ft} 
\maketitle 
The breaking of the \CP symmetry 
in the quark sector of the electroweak interactions arises  
in the standard model  
(SM) from a single irreducible phase in the Cabibbo-Kobayashi-Maskawa (CKM) quark-mixing matrix~\cite{ref:CKM}. 
This phase can be  
measured 
using a variety of methods involving \B-meson decays mediated by either only tree-level or both tree- and loop-level amplitudes. 
The comparison of these two classes of measurements tests the CKM mechanism,  
thus offering a strategy to search for new physics~\cite{ref:globalCKMfits}. 
The angle $\gamma$ of the unitarity triangle, defined as $\arg{\left[-V_{ud}^{}V_{ub}^{*}/V_{cd}^{}V_{cb}^{*}\,\right]}$,  
where $V_{ij}$ are elements of the CKM matrix,  
is particularly relevant 
since it is the  
only \CP-violating  
parameter that can be cleanly determined using  
solely 
tree-level \B-meson decays. Its precise determination constitutes an important goal 
of present and future  
experiments in flavor physics. 
 
In  
$\Bmp\to\D \Kmp$ decays~\cite{ref:carter-sanda-bigi,ref:DDstar} 
the color-favored $\Bm \to \Dz K^-$ ($\b \to \c\ubar\s$) and the color-suppressed $\Bm\to \Dzb K^-$ ($\b \to \u \cbar \s$)  
transitions~\cite{ref:chargeconj} interfere when the \Dz and \Dzb decay to a common final state~\cite{ref:gronau-soni-ggsz_ads}. 
The two interfering amplitudes differ by a factor $\rb e^{i(\deltab \mp \gamma)}$, where \rb is the magnitude of the ratio of the amplitudes  
${\cal A}(\Bm \to \Dzb \Km)$ and ${\cal A}(\Bm \to \Dz \Km)$, and \deltab is their relative strong phase.  
An amplitude analysis of the Dalitz plot (DP) of \Dz and \Dzb mesons decaying into the \kspipi and \kskk self-conjugate 
final states  
from 
$\Bmp\to\D \Kmp$ decays 
offers a unique way to access the complex amplitude ratios and thus the weak and strong phases, and \rb. 
The experimental sensitivity to $\gamma$ arises mostly from regions in the DP where 
Cabibbo-favored (CF) and doubly-Cabibbo-suppressed (DCS) amplitudes interfere, and from regions 
populated by \CP eigenstates, 
thus the uncertainty on $\gamma$ depends on $1/\rb$ ($\rb\sim0.1-0.2$). 

In this Letter we study the interference between color-favored and color-suppressed transitions 
as a function of the position in the DP of squared invariant masses $\sminus = m^2(\KS \hm)$, $\splus = m^2(\KS \hp)$, where 
\h represents $\pi$ or \K, for three related \B decays, 
$\Bmp \to \D \Kmp$, $\Bmp \to \Dstar \Kmp$, and $\Bmp \to \D \Kstarmp$~\cite{ref:DDstar,ref:KstarDstar}, 
and report the most precise single measurement of the complex amplitude ratios 
and evidence for direct \CP violation. 
We use the complete data sample of 425~\invfb of integrated luminosity at the \FourS, 
corresponding to $468\times10^6$ \BB pairs, and 45~\invfb at a center-of-mass (c.m.) energy 40~\mev below the \FourS, 
recorded by the \babar\ experiment~\cite{ref:detector} at the \pep2 asymmetric-energy \epem\ collider at SLAC 
from 1999 to 2008. 
This measurement updates our previous results based on a partial sample of $383\times10^6$ \BB pairs, 
from which we reported a significance of direct \CP violation ($\gamma \ne 0$) of $3.0$ standard deviations, 
while most of the analysis details remain unchanged~\cite{ref:babar_dalitzpub2008}. 
The Belle Collaboration using $\Bmp\to\DDstar\Kmp$, \Dtokspipi alone~\cite{ref:belle_dalitzpub}  
has also reported $\gamma \ne 0$ with a significance of $3.5$ standard deviations.

We reconstruct a total of eight signal samples,  
$\Bmp \to \DDstar \Kmp$ and $\Bmp \to \D \Kstarmp$, 
with $\Dstar \rightarrow \D\piz,\D\gamma$, $\Kstarmp \to \KS\pimp$,  
with selection criteria nearly identical to our previous analysis. 
The $\D \Kstarmp$ final state, for \Dtokskk, has been considered for the first time. 
For $\KS\to\pip\pim$ candidates, we further require the decay length  
(defined by the \KS production and decay vertices) projected along the \KS momentum  
to be greater than 10 times its  
error. 
This additional 
requirement 
helps to reduce to a negligible level background events 
from $\D\to\pip\pim\hp\hm$ decays, and from $a_1(1260)^\mp$ misreconstructed as \Kstarmp. 
After all the selection criteria the background is completely dominated by random combinations of tracks arising from  
continuum events, $e^+e^- \to \qqbar$ ($\q = \u, \d, \s, \ {\rm or} \ \c$). 
Background contributions from $\D\to\KS\KS$ decays are found to be negligible. 
The \Bmp candidates are characterized 
using the beam-energy substituted \B mass \mes,  
the difference between the reconstructed energy of the \Bmp candidate and the beam energy in the $e^+e^-$ c.m. frame \de, 
and a Fisher discriminant \fis that combines four topological variables 
optimized to separate continuum events~\cite{ref:babar_dalitzpub2008}.  
We retain candidates  
with the loose requirements 
\mbox{$\mes>5.2$}~\gevcc, \mbox{$-80<\de<120$}~\mev, and \mbox{$| \fis |<1.4$}, 
which provide signal and sideband regions while removing poorly reconstructed candidates~\cite{ref:epaps}. 
The reconstruction efficiencies 
in a signal region with \mbox{$\mes>5.272$}~\gevcc and \mbox{$|\de|<30$}~\mev 
are $26\%$, $12\%$, $15\%$, and $14\%$, for the   % these values are with cut on mES>5.272 and |de|<30 MeV, but no cut on Fisher 
$\D \Kmp$,  
$\Dstar[\D\piz]\Kmp$,  
$\Dstar[\D\gamma]\Kmp$,  
and $\D \Kstarmp$ final states, respectively, 
for $\D\to\kspipi$ (and slightly lower for $\D\to\kskk$). 
These values 
are about 30\%, 40\%, 30\%, and 20\% larger than in our previous analysis,  
with similar  
background levels, 
reflecting improvements in tracking and particle identification. 
The \mes, \de, \fis, and $(\sminus,\splus)$ distributions for events in the signal region can be found in~\cite{ref:epaps}.

The \Dztokshh decay amplitudes ${\cal A}(\sminus,\splus)$ are determined using the same data sample 
through 
DP analyses of \Dz mesons from $\Dstarp \to \Dz \pip$ decays produced  
in $e^+ e^- \to \c\cbar$ events~\cite{ref:babar_dalitzpub2008,ref:dmixing-kshh}.  
The charge of the low momentum $\pip$ from  
the \Dstarp decay identifies  
the flavor of the \D meson.  
The signal purities of the samples are $98.5\%$ and $99.2\%$,  
with about $541\hspace{2pt}000$ and $80\hspace{2pt}000$ candidates, for \kspipi and \kskk, respectively. 
The dynamical properties of the P- and D-wave amplitudes 
are parameterized through intermediate resonances with mass-dependent relativistic Breit-Wigner (BW) or  
Gounaris-Sakurai 
propagators, 
Blatt-Weisskopf centrifugal barrier factors, and 
Zemach tensors for the angular distributions~\cite{ref:RevDalitzPlotFormalism}. 
The $\pi\pi$ S-wave dynamics is described 
through a K-matrix formalism with the P-vector approximation and 5 poles~\cite{ref:AS,ref:babar_dalitzpub2008}.  
For the $\K\pi$ S-wave we include a BW for the $K^{*}_0(1430)^\mp$ state with a coherent  
non-resonant contribution 
parameterized  
by a scattering length and effective range similar to those used to describe $\K\pi$ scattering data~\cite{ref:LASS}. 
For the $\K\Kbar$ S-wave, a coupled-channel BW is used for the $a_0(980)$ with single BWs for $f_0(1370)$ and $a_0(1450)$ states. 
Overall, the amplitude models reproduce well the DP distributions~\cite{ref:dmixing-kshh}. 
MC studies show that a significant contribution to the  
discrepancies arise from 
imperfections modeling the efficiency variations at the boundaries of the DP and the invariant mass resolution. 
We account for these and other imperfections in the modeling of the \Dz decay amplitudes through our model 
systematic uncertainties. 

We perform a simultaneous, unbinned, and extended maximum-likelihood fit (referred to as \CP fit)  
to the $\Bmp \to \DDstar \Kmp$ and $\Bmp \to \D \Kstarmp$ 
decay rates $\Gamma^{(*)}_{\mp}$ and $\Gamma_{s\mp}$ 
as a function of \mes, \de, \fis, and $(\sminus,\splus)$~\cite{ref:babar_dalitzpub2008,ref:epaps}. 
We extract the signal and background yields,  
along with the  
\CP-violating parameters  
$\zbzbstmp \equiv \xbxbstmp + i\ybybstmp$ and $\zsmp \equiv \xsmp + i\ysmp$, 
defined as the \Bmp complex amplitude ratios $\zbzbstmp = \rbrbstmp e^{i(\deltabdeltabst \mp \g)}$ and  
$\zsmp = \kappa \rsmp e^{i(\deltas \mp \g)}$,  
respectively. 
Here, $\rbrbstmp$ and $\rsmp$ are the corresponding magnitude ratios 
between the $\b\to\u$ and $\b\to\c$ amplitudes for \Bmp decays, 
$\deltabdeltabst$ and $\deltas$ the relative strong phases, and $\kappa$ an effective hadronic parameter 
that accounts for the interference between $\Bmp \to\D \Kstarmp$ and other $\Bmp \to\D \KS\pi^\mp$ decays, 
as a consequence of the \Kstarmp natural width~\cite{ref:gronau2003,ref:BR-BtoDKst,ref:babar_dalitzpub2008}. 
Assuming no \CP violation and neglecting $\Dz-\Dzb$ mixing in $\Dztokshh$  
decays~\cite{ref:dmixing-kshh,ref:dmixing-kspipi-belle,ref:mixingeffects}, 
the relation  
$\overline{{\cal A}}(\sminus,\splus) = {\cal A}(\splus,\sminus)$ holds,  
where $\overline{{\cal A}}$ is the \Dzb decay amplitude.  
The $\Bmp \to \DDstar \Kmp$  
(and similarly for $\Bmp \to \D \Kstarmp$ replacing \zbzbstmp and \rbrbstmp by \zsmp and \rsmp, respectively)  
signal decay rates  
are then 
\bea 
\Gamma^{(*)}_{\mp}(\sminus,\splus) \propto  |{\cal A}_{\mp}|^2+{\rbrbstmp}^2 |{\cal A}_{\pm}|^2 + 2 \lambda \zbzbstmp {\cal A}_{\mp} {\cal A}^*_{\pm},~\nn 
\label{eq:ampgen} 
\eea 
with ${\cal A}_{\mp} \equiv {\cal A}(\sminusplus,\splusminus)$, and $\lambda=+1$ except  
for $\Bmp \to \Dstar[\D\gamma] \Kmp$ where $\lambda=-1$~\cite{ref:bondar_gershon}. 
We apply corrections for efficiency variations and neglect the invariant mass resolution across the DP~\cite{ref:babar_dalitzpub2008}. 
For each signal sample,  
the following 
background components are considered: 
continuum events, $\Bmp \to \DDstar \pimp$ decays where the pion is misidentified as a kaon 
(only for $\Bmp \to \DDstar \Kmp$ samples), and $\FourS \to \BB$  
(other than $\Bmp\to\DDstar\pimp$) decays.  
The reference \CP fit requires events to satisfy \mbox{$|\de|<30$}~\mev, 
but alternative fits are performed varying the requirements on the \mes, \de, and \fis variables 
(e.g. \mbox{$\mes>5.272$}~\gevcc or \mbox{$\fis>-0.1$}) to study the stability of the results. 
The probability density functions (PDFs) introduced to describe  
the signal, continuum, and $\K/\pi$ misidentification components, 
along with the $\K/\pi$ misidentification yields, 
are determined using events  
from signal and $\Bmp \to \DDstar \pimp, \D a_1(1260)^\mp$ control  
samples. 
The PDFs for \BB background events are obtained from  
large Monte Carlo (MC) samples with full detector simulations~\cite{ref:babar_dalitzpub2008}.

\begin{figure*}[htb!] 
\begin{tabular}{ccc} 
\includegraphics[width=0.28\textwidth]{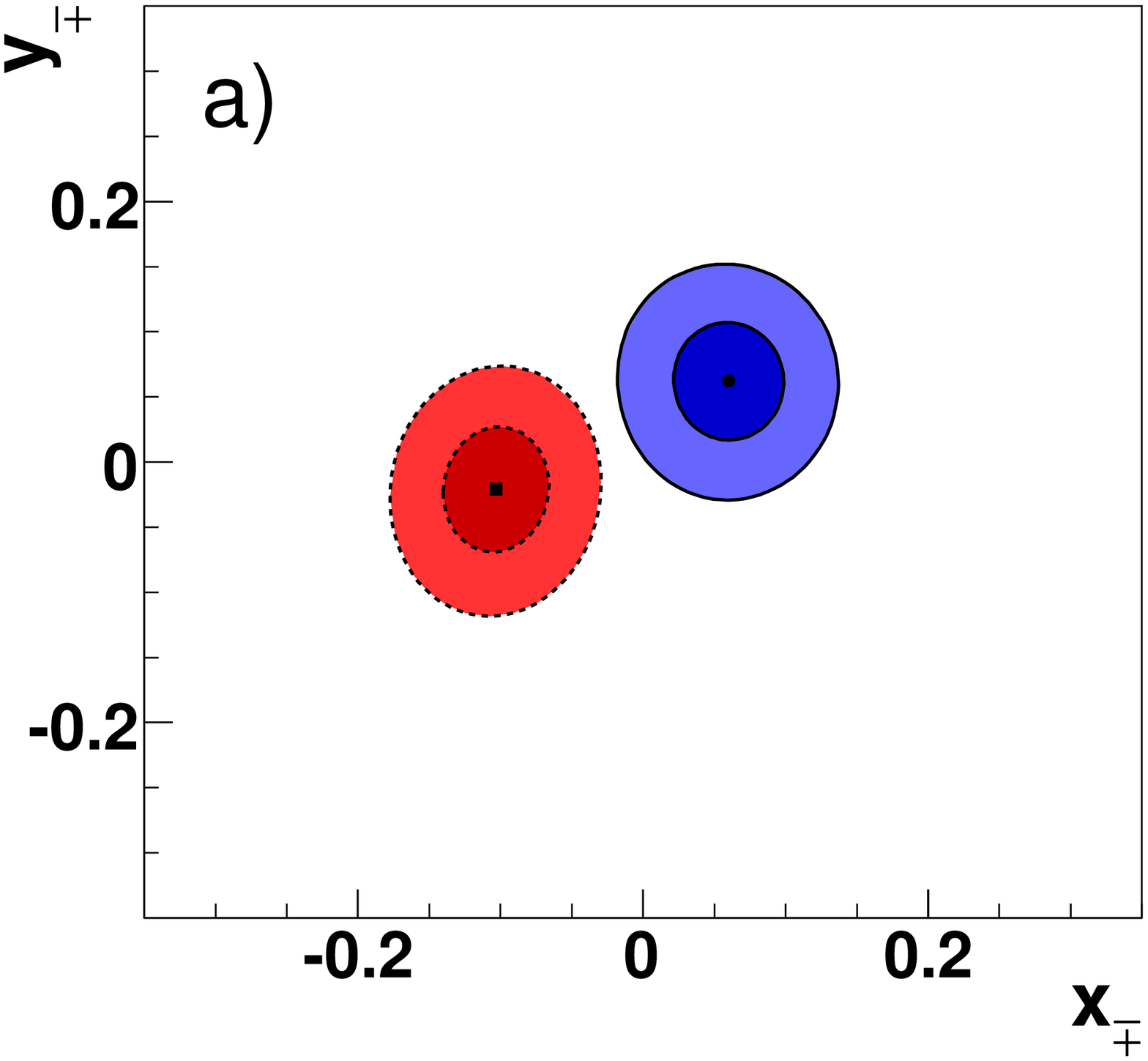}& 
\includegraphics[width=0.28\textwidth]{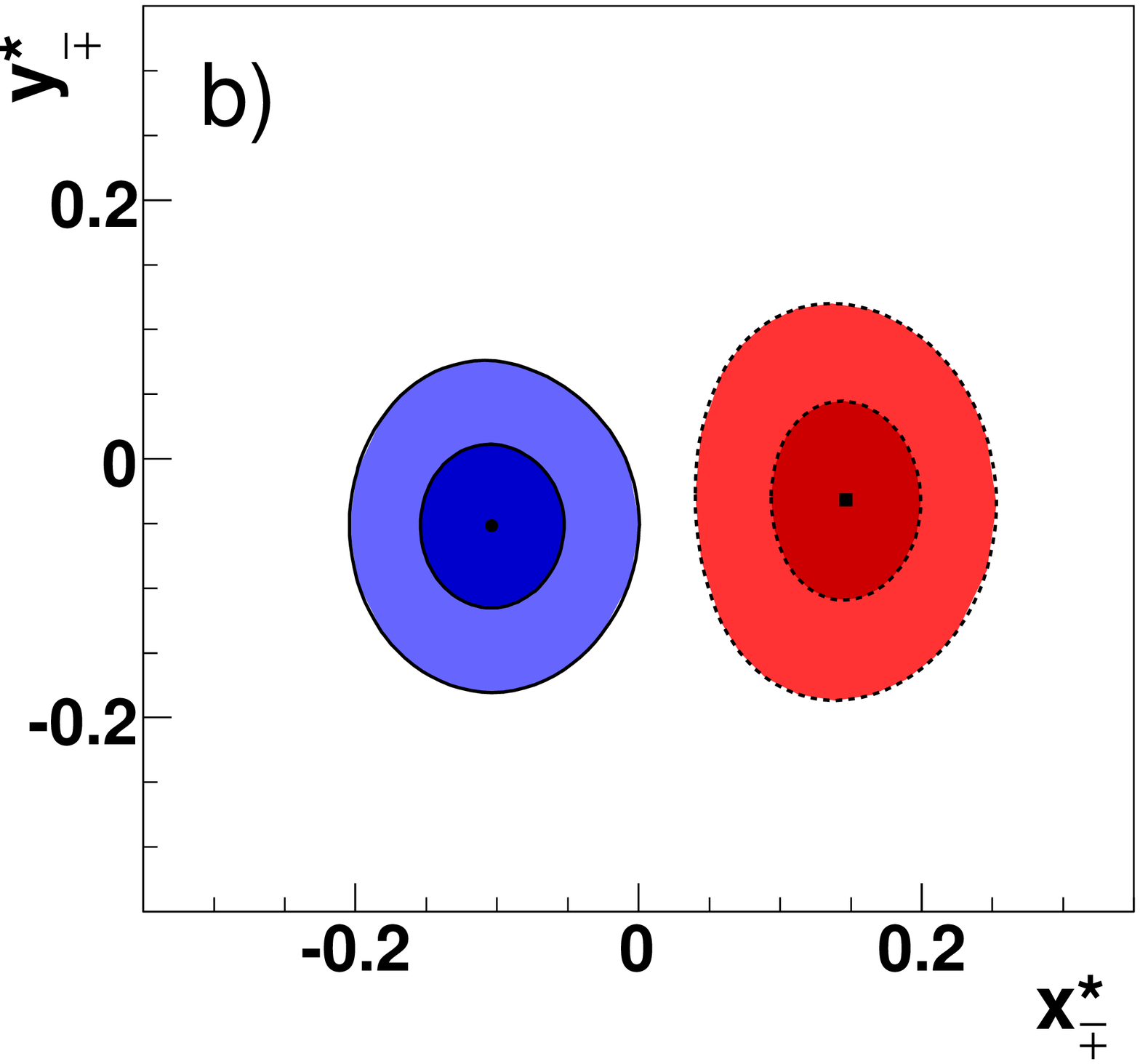} & 
\includegraphics[width=0.28\textwidth]{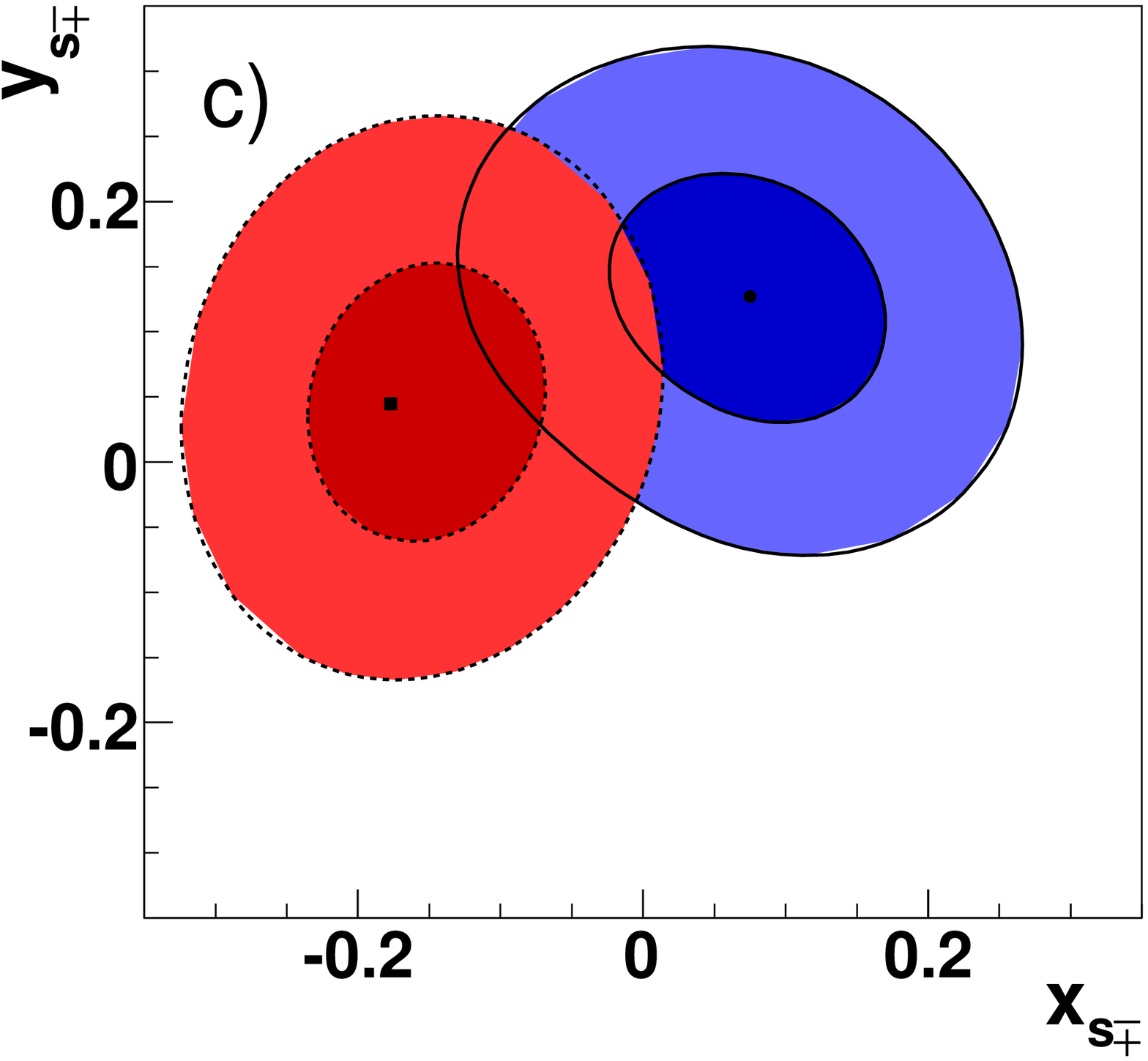} \\ 
\end{tabular} 
\caption{\label{fig:contours} (color online). Contours at $39.3\%$ (dark) and $86.5\%$ (light) 2-dimensional CL in the (a) \zbmp,   
(b) \zbstmp, and (c) \zsmp planes, corresponding to one- and two-standard deviation regions (statistical only),  
for \Bm (solid lines) and \Bp (dotted lines) decays. 
} 
\end{figure*}

The \CP fit yields $896\pm35$ ($154\pm14$), $255\pm21$ ($56\pm11$), $193\pm19$ ($30\pm7$), and $163\pm18$ ($28\pm6$) signal  
$\D\Kmp$, $\Dstar[\D\piz]\Kmp$, $\Dstar[\D\g]\Kmp$, and $\D\Kstarmp$ events, respectively, for the \kspipi (\kskk) final state. 
The results for the \CP-violating parameters  
$\zbzbstpm$ and $\zspm$ are summarized in Table~\ref{tab:xyresults}. 
Figure~\ref{fig:contours} 
shows 
the $39.3\%$ and $86.5\%$ 2-dimensional confidence-level (CL) contours 
in the \zbmp, \zbstmp, and \zsmp planes, corresponding to one- and two-standard deviation regions, 
including statistical errors only. 
The distance between the \zbm and \zbp central values (and similarly for \zbstmp and \zsmp)  
is equal to $2 \rbmp|\sin\gamma|$, and 
the angle defined by the lines connecting  
the central values 
with the origin is $2\gamma$,  
and thus is a measurement of direct \CP violation.  
Fitting separately the data for \kspipi and \kskk final states we find consistent results  
for all the \CP-violating parameters~\cite{ref:epaps}. 
 
\begin{table}[!htb] 
\caption{\label{tab:xyresults} 
\CP-violating complex parameters  
$\zbzbstmp = \xbxbstmp + i \ybybstmp$ 
and  
$\zsmp = \xsmp + i \ysmp$ 
as obtained from the \CP fit.  
The first error is statistical, the second is the experimental systematic uncertainty and the third is  
the systematic uncertainty associated with the \Dz decay amplitude models.} 
\begin{center} 
\begin{ruledtabular} 
\begin{tabular}{lrr} 
           & Real part (\%)\phzz\phzz\phz &  Imaginary part (\%)\phzz \\ [0.025in] \hline  
 $\zbm$    & $\phz\phm6.0\pm3.9\pm0.7\pm0.6\phzz$    & $\phz\phm6.2\pm\phz4.5\pm0.4\pm0.6\phz$ \\ 
 $\zbp$    & $-10.3\pm3.7\pm0.6\pm0.7\phzz$          & $\phz-2.1\pm\phz4.8\pm0.4\pm0.9\phz$ \\ 
 $\zbstm$  & $-10.4\pm5.1\pm1.9\pm0.2\phzz$          & $\phz-5.2\pm\phz6.3\pm0.9\pm0.7\phz$    \\ 
 $\zbstp$  & $\phm14.7\pm5.3\pm1.7\pm0.3\phzz$       & $\phz-3.2\pm\phz7.7\pm0.8\pm0.6\phz$ \\ 
 $\zsm$    & $\phz\phm7.5\pm9.6\pm2.9\pm0.7\phzz$    & $\phm12.7\pm\phz9.5\pm2.7\pm0.6\phz$    \\ 
 $\zsp$    & $-15.1\pm8.3\pm2.9\pm0.6\phzz$          & $\phz\phm4.5\pm10.6\pm3.6\pm0.8\phz$    \\ 
\end{tabular} 
\end{ruledtabular} 
\end{center} 
\end{table}

Experimental systematic errors~\cite{ref:babar_dalitzpub2008,ref:epaps} originate from uncertainties in the description of the efficiency variations across the DP,  
the modeling of the DP distributions for background events containing misreconstructed \D mesons,  
the fractions of continuum and \BB background events containing a real \D meson with 
either a negatively- or positively-charged kaon (or \Kstar), 
and from residual direct \CP violation in the $\Bmp \to \DDstar \pimp$ and \BB background components. 
We also account for  
statistical and systematic uncertainties 
in the \mes, \de, and \fis\ PDF shapes for signal and background components, and the $\K/\pi$ misidentification yields. 
These uncertainties account for effects that arise from the dependence of the \mes and \fis PDF shapes on the chosen \de signal region, 
the differences in \BB background for real and misreconstructed \D mesons, 
and our limited knowledge of the \mes endpoint, the peaking contributions to the small \BB background, and the $e^+e^-$ c.m. frame. 
Smaller 
systematic uncertainties originate from the  
DP resolution,  
wrongly reconstructed  
signal events with a real \D and a kaon (or \Kstar) from the other \B meson decay,  
the selection of \B candidates sharing tracks with other candidates, 
and numerical precision in the evaluation of the PDF integrals. 
We also account for residual cross-feed of $\Bmp\to\Dstar[\D\piz]\Kmp$ events into the  
$\Bmp\to\Dstar[\D\gamma]\Kmp$ sample  
(about 5\%),  
and the estimated uncertainty on the hadronic parameter $\kappa = 0.9\pm0.1$ in the 
$\Bmp\to\D\Kstarmp$ sample~\cite{ref:babar_dalitzpub2008,ref:glwadsDzKstarm}. 

Assumptions in the \Dz decay amplitude models are also a source of systematic uncertainty~\cite{ref:babar_dalitzpub2008,ref:epaps,ref:dmixing-kshh}. 
We use alternative ${\cal A}(\sminus,\splus)$ models where the BW parameters  
are varied according to their uncertainties or  
within the ranges allowed by measurements from other experiments, 
the reference K-matrix solution~\cite{ref:babar_dalitzpub2008} is replaced by other solutions~\cite{ref:AS},  
and the standard parameterizations are substituted  
by 
other related choices. 
These include replacing the  
Gounaris-Sakurai  
and $K\pi$ S-wave parameterizations by BW lineshapes,  
removing the mass dependence in the P-vector~\cite{ref:Kmatrix-Pvector}, 
changes in form factors such as changes in the Blatt-Weisskopf radius, 
and adopting a helicity formalism~\cite{ref:RevDalitzPlotFormalism} to describe the angular dependence. 
Other models are built  
by removing or adding resonances with small or negligible fractions.  
We find that the overall amplitude model uncertainty on the \CP parameters 
are dominated by alternative models built to account for experimental systematic 
effects in the determination of ${\cal A}(\sminus,\splus)$ using tagged \D mesons~\cite{ref:dmixing-kshh}. 
The statistical errors and variations in the ${\cal A}(\sminus,\splus)$ model  
parameters with and without \Dz-\Dzb mixing are also propagated to \zbzbstmp and \zsmp. 

Experimental and amplitude model systematic uncertainties~\cite{ref:epaps} have been reduced with respect to our  
previous measurement~\cite{ref:babar_dalitzpub2008} as consequence of the use of larger data and Monte Carlo samples,  
and the smaller experimental systematic contributions to the model uncertainty resulting from the improvements in the  
analysis of tagged \D mesons~\cite{ref:dmixing-kshh}. 

\begin{figure}[htb!] 
\includegraphics[width=0.35\textwidth]{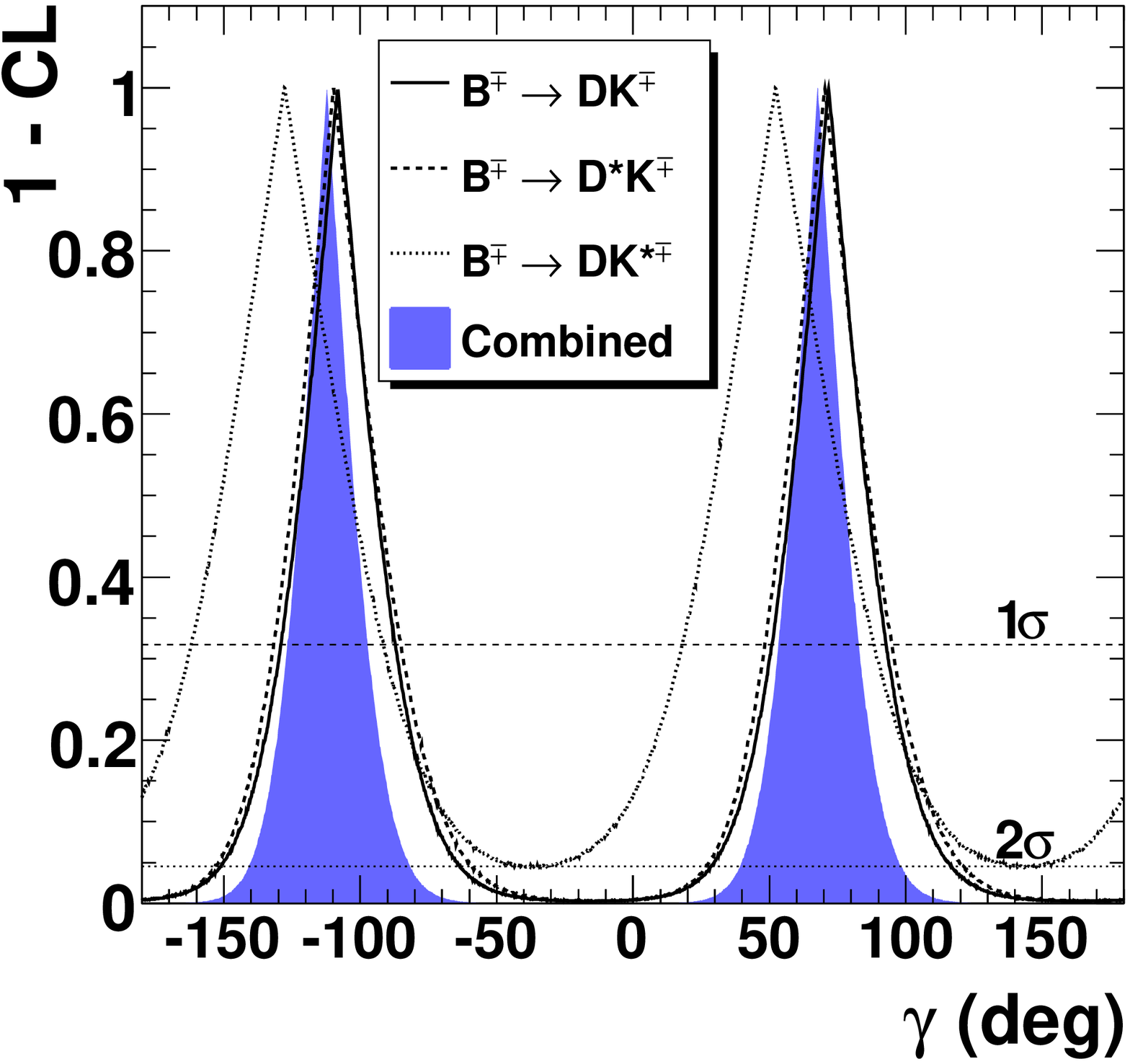} 
\caption{\label{fig:CL-gamma} (color online). $1 - {\rm CL}$ as a  
function of $\gamma$ for $\Bmp \to \D\Kmp$, $\Bmp \to \Dstar\Kmp$, and  
$\Bmp \to \D\Kstarmp$ decays separately, and their combination, including statistical and 
systematic uncertainties. The dashed (upper) and dotted (lower) horizontal lines correspond to the one- and two-standard  
deviation intervals, respectively. 
} 
\end{figure}

A frequentist construction of 1-dimensional confidence intervals of the physically relevant parameters  
$\pvec \equiv (\g, \rb, \rbst, \krs, \deltab, \deltabst, \deltas)$ based on the vector of measurements  
$\zvec =(\zbm, \zbp, \zbstm, \zbstp, \zsm, \zsp)$  
and their  
correlations~\cite{ref:epaps}  
has been adopted~\cite{ref:babar_dalitzpub2008}.  
The  
procedure takes into account unphysical regions  
which may arise since we allow \Bm and \Bp events to have different \rbrbstmp, \rsmp in the \zvec measurements. 
Figure~\ref{fig:CL-gamma} shows 
$1 - {\rm CL}$  
as a function of $\gamma$ for each of the three \B decay channels separately and their combination. 
Similar scans for \rbrbst, \krs, \deltabdeltabst, and \deltas can be found in~\cite{ref:epaps}. 
The method has a  
single 
ambiguity in the weak and strong phases. 
The results for all the \pvec parameters are listed in Table~\ref{tab:polarresults}. 
The significances of direct \CP violation ($\gamma \ne 0$) 
are $1-$CL=$6.8\times10^{-3}$, $5.4\times10^{-3}$, $6.3\times10^{-2}$, and $4.6\times10^{-4}$, 
which correspond to $2.7$, $2.8$, $1.9$, and $3.5$ standard deviations,  
for $\Bmp\to\D\Kmp$, $\Bmp\to\Dstar\Kmp$, $\Bmp\to\D\Kstarmp$, and their combination, respectively. 
 
\begin{table}[!htb] 
\caption{\label{tab:polarresults} The $68.3\%$ and $95.4\%$ 1-dimensional CL regions, 
equivalent to one- and two-standard deviation intervals,  
for \g, \deltabdeltabst, \deltas, \rbrbst, and \krs, 
including %statistical, experimental systematic and amplitude model uncertainties.  
all sources of uncertainty. 
The values inside $\{\}$ brackets indicate the symmetric error contributions to the total error coming from 
experimental and amplitude model systematic uncertainties. 
} 
\begin{center} 
\begin{ruledtabular} 
\begin{tabular}{lcc} 
Parameter            & $68.3\%$ CL  & $95.4\%$ CL \\ [0.035in] \hline 
\g $(^\circ)$        & $68^{+15}_{-14}$ $\{4,3\}$                  & $[39,98]$ \\  
\rb $(\%)$           & $9.6 \pm 2.9$ $\{0.5,0.4\}$                 & $[3.7,15.5]$ \\ 
\rbst $(\%)$         & $13.3^{+4.2}_{-3.9}$ $\{1.3,0.3\}$          & $[4.9,21.5]$ \\ 
\krs $(\%)$          & $14.9^{+6.6}_{-6.2}$ $\{2.6,0.6\}$          & $< 28.0$ \\ 
\deltab   $(^\circ)$ & $119^{+19}_{-20}$ $\{3,3\}$                 & $[75,157]$ \\ 
\deltabst $(^\circ)$ & $-82\pm21$ $\{5,3\}$                        & $[-124,-38]$ \\ 
\deltas   $(^\circ)$ & $111\pm32$ $\{11,3\}$                       & $[42,178]$ \\ 
\end{tabular} 
\end{ruledtabular} 
\end{center} 
\end{table}

We have presented a measurement of the  
$\b \to \u$ to $\b \to \c$ complex amplitude ratios 
in the processes $\Bmp\to\DDstar\Kmp$ and $\Bmp\to\D\Kstarmp$, 
using a combined DP analysis of \Dtokspipi and \Dtokskk decays.  
The results have improved precision and are consistent with our previous measured values~\cite{ref:babar_dalitzpub2008} 
and with those reported by the Belle Collaboration with  
\Dtokspipi alone~\cite{ref:belle_dalitzpub}, 
and with determinations based on other \D meson final states~\cite{ref:glwadsDzDstarzKm,ref:glwadsDzKstarm,ref:otherexp_glwads}. 
From  
our measurement 
we determine $\g=(68 \pm 14 \pm 4 \pm 3)^\circ$ (\mbox{modulo $180^\circ$}), 
exclude the no direct \CP-violation hypothesis (i.e., $\gamma = 0$) with a CL 
equivalent to $3.5$ standard deviations, and derive the most precise single 
determinations of the magnitude ratios \rbrbst and \krs.

\indent 
We are grateful for the excellent luminosity and machine conditions 
provided by our \pep2\ colleagues,  
and for the substantial dedicated effort from 
the computing organizations that support \babar. 
The collaborating institutions wish to thank  
SLAC for its support and kind hospitality.  
This work is supported by 
DOE 
and NSF (USA), 
NSERC (Canada), 
CEA and 
CNRS-IN2P3 
(France), 
BMBF and DFG 
(Germany), 
INFN (Italy), 
FOM (The Netherlands), 
NFR (Norway), 
MES (Russia), 
MICIIN (Spain), 
STFC (United Kingdom).  
Individuals have received support from the 
Marie Curie EIF (European Union), 
the A.~P.~Sloan Foundation (USA) 
and the Binational Science Foundation (USA-Israel).

%                     M.~Gronau and D.~Wyler, \plb{265}, 172 (1991). 

%                              M.~Staric {\em et al.} (Belle Collaboration), \jprl{98}, 211803 (2007). %evidence for mixing in D0->Kpi 
%                              % these also include a test for direct CP violation 

%                     E.~Farhi, \jprl{39}, 1587 (1977). %Phys.\ Rev.\ Lett.\ {\bf 39}, 1587 (1977). 

%                         S.~U.~Chung {\it et al.}, \jprd{48}, 1225 (1993).  

%                      S.~U.~Chung {\it et al.}, \annp{4}, 404 (1995).   

%                        J.~M.~Link {\em et al.} (FOCUS Collaboration), \plb{653}, 1 (2007); 
%                        G.~Bonvicini {\it et al.} (CLEO Collaboration), arXiv:0802.4214 [hep-ex], submitted to Phys.~Rev.~D. 

%                 Belle Collaboration, K.~Abe {\it et al.}, \jprd{66}, 071102 (2002). 
%                    Belle Collaboration, Y.~Chao {\it et al.}, \jprl{93}, 191802 (2004). 
%                       M.~Gronau and D.~Wyler, \plb{265}, 172 (1991); 

\onecolumngrid 
\newpage 
 
\setcounter{page}{1} 
\setcounter{table}{0} 
\setcounter{figure}{0} 
 
\begin{center} 
{\large \bfseries \boldmath 
Evidence for direct \CP violation in the measurement of the CKM angle $\gamma$ with $\Bmp \to D^{(*)} K^{(*)\mp}$ decays}\\ 
 
The \babar\ Collaboration 
\end{center}

\begin{center} 
The following includes supplementary material for the Electronic 
Physics Auxiliary Publication Service.  
\end{center}

\begin{figure}[hbt!] 
\begin{tabular}{ccc} 
\includegraphics[width=0.24\textwidth]{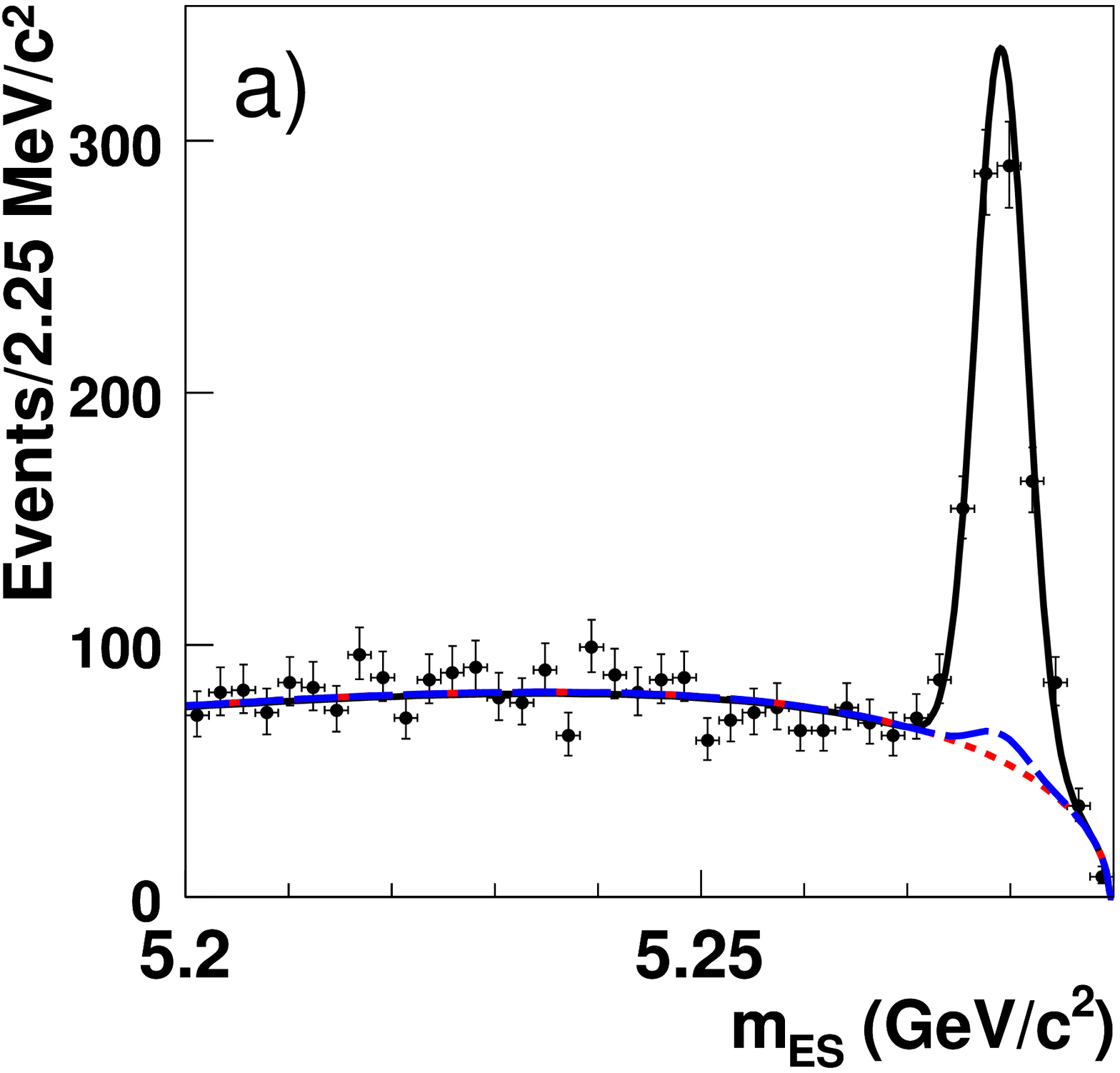} & 
\includegraphics[width=0.24\textwidth]{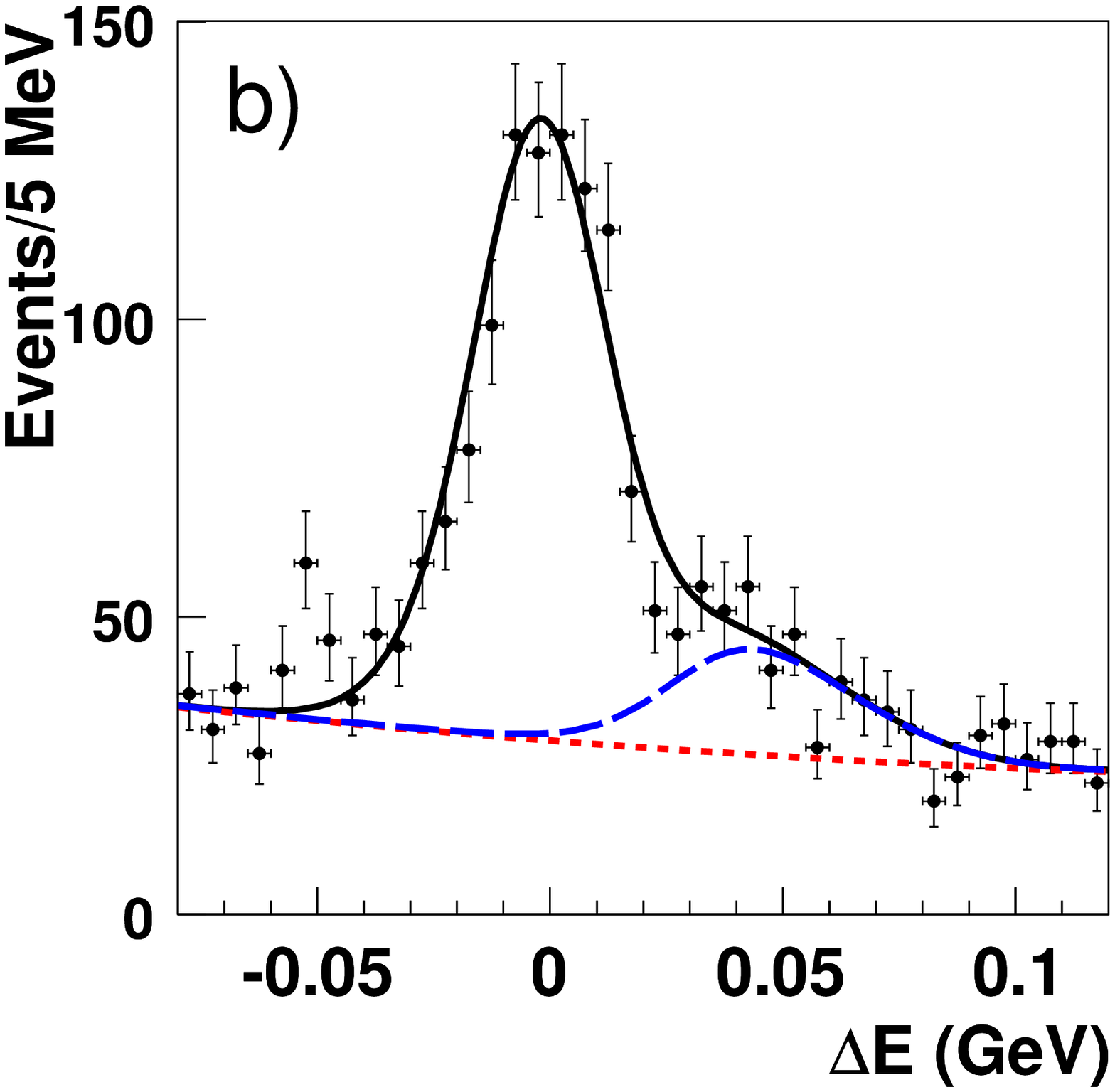} & 
\includegraphics[width=0.24\textwidth]{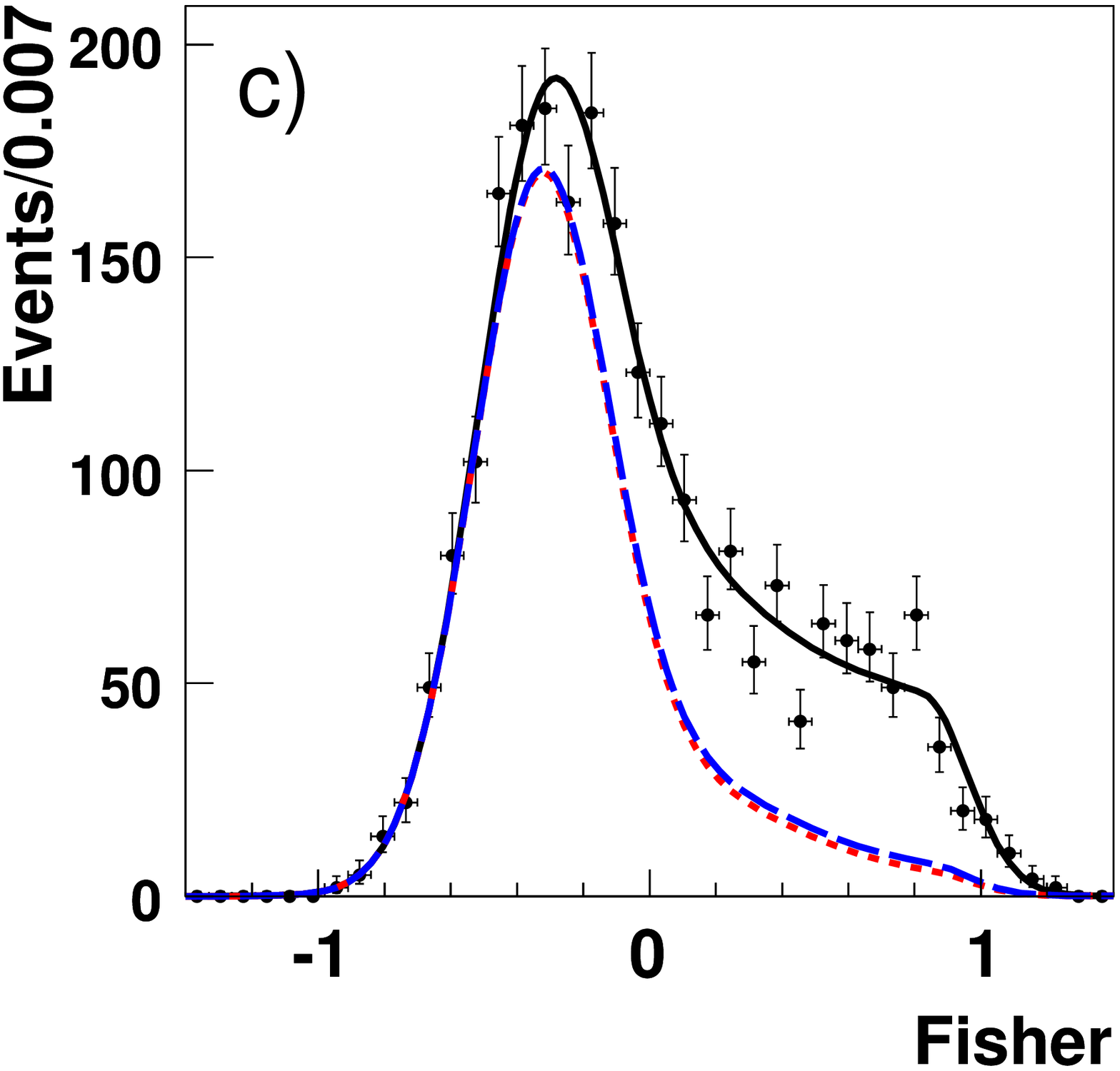} \\ 
\includegraphics[width=0.24\textwidth]{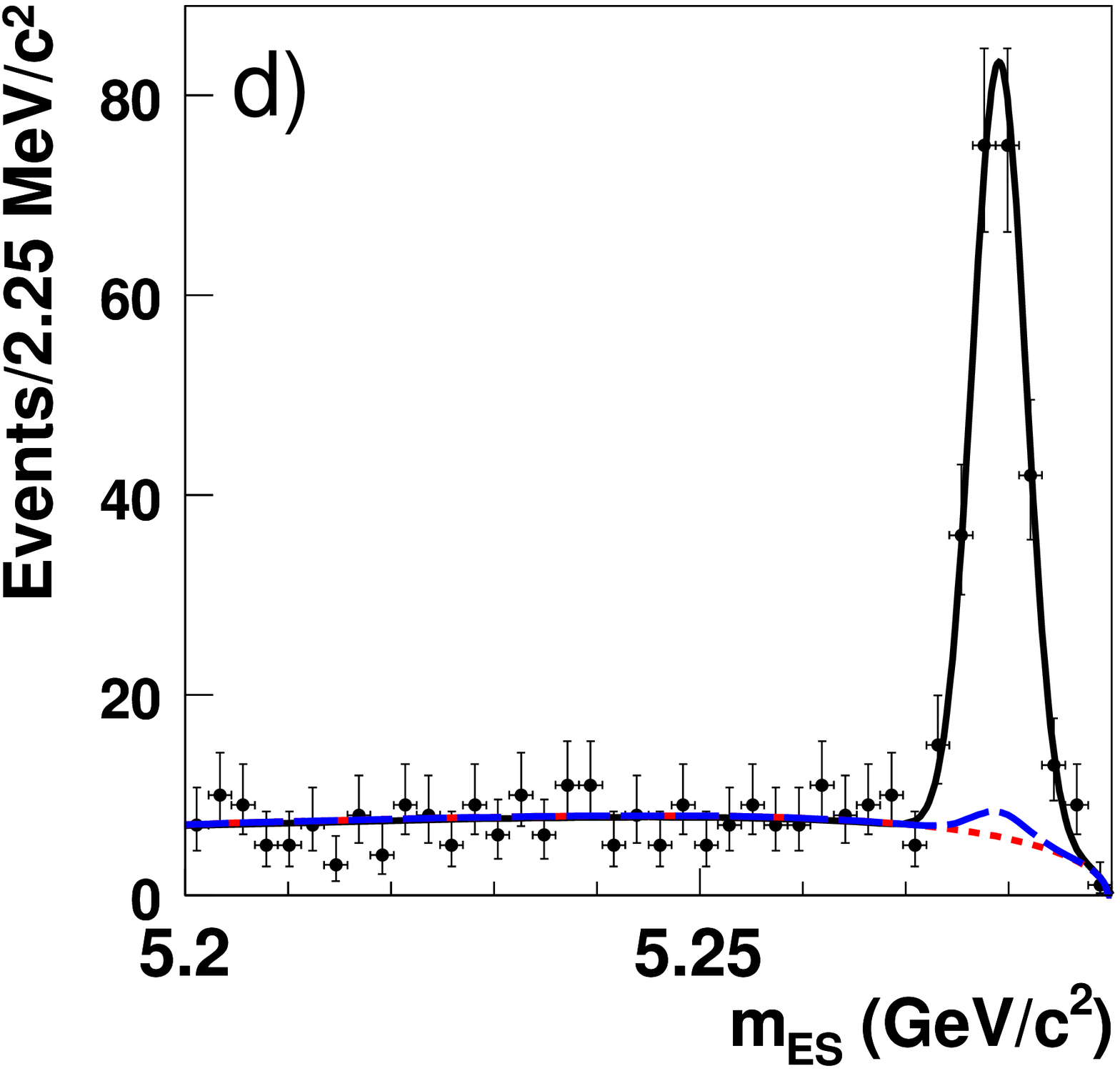} & 
\includegraphics[width=0.24\textwidth]{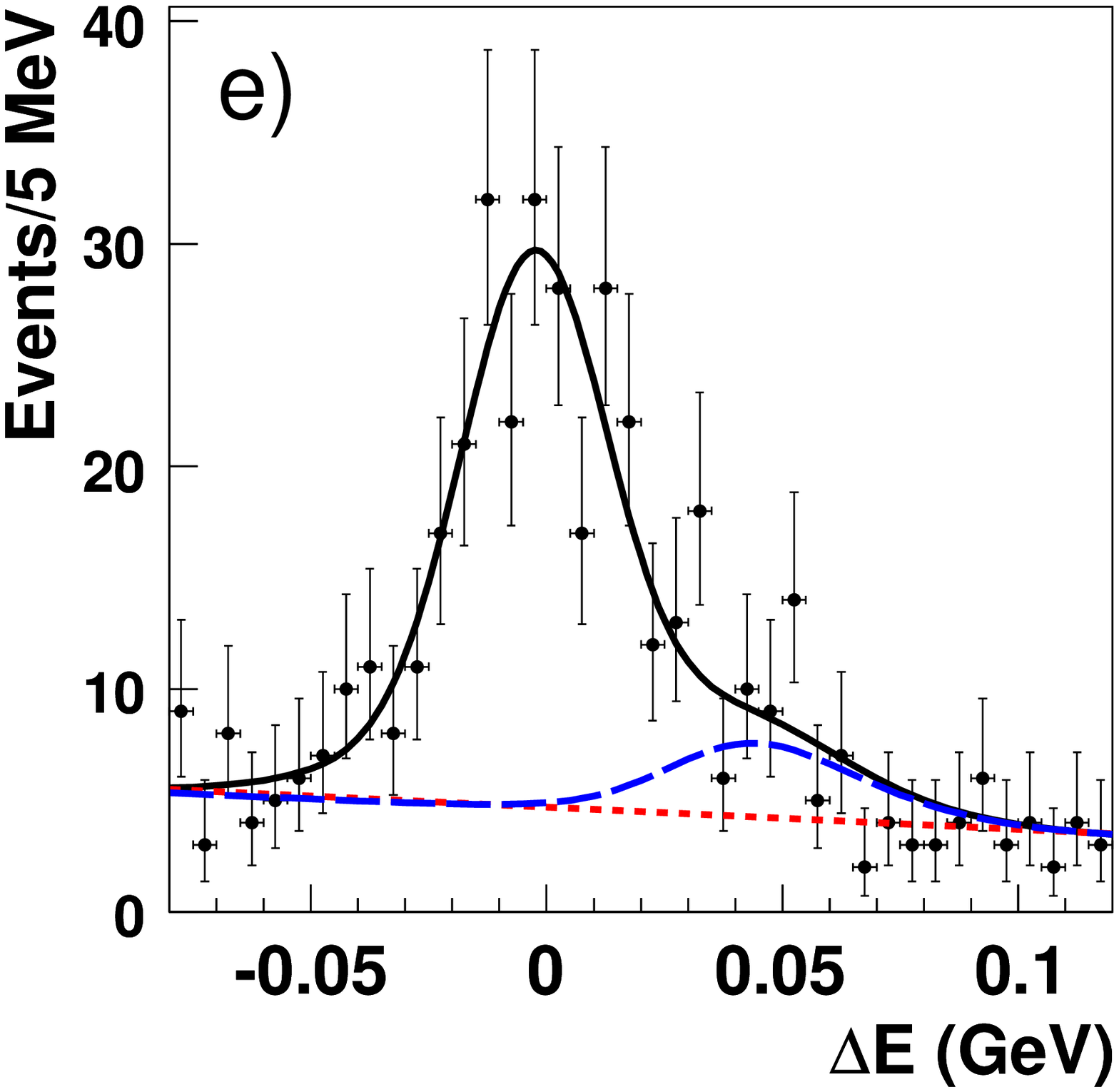} & 
\includegraphics[width=0.24\textwidth]{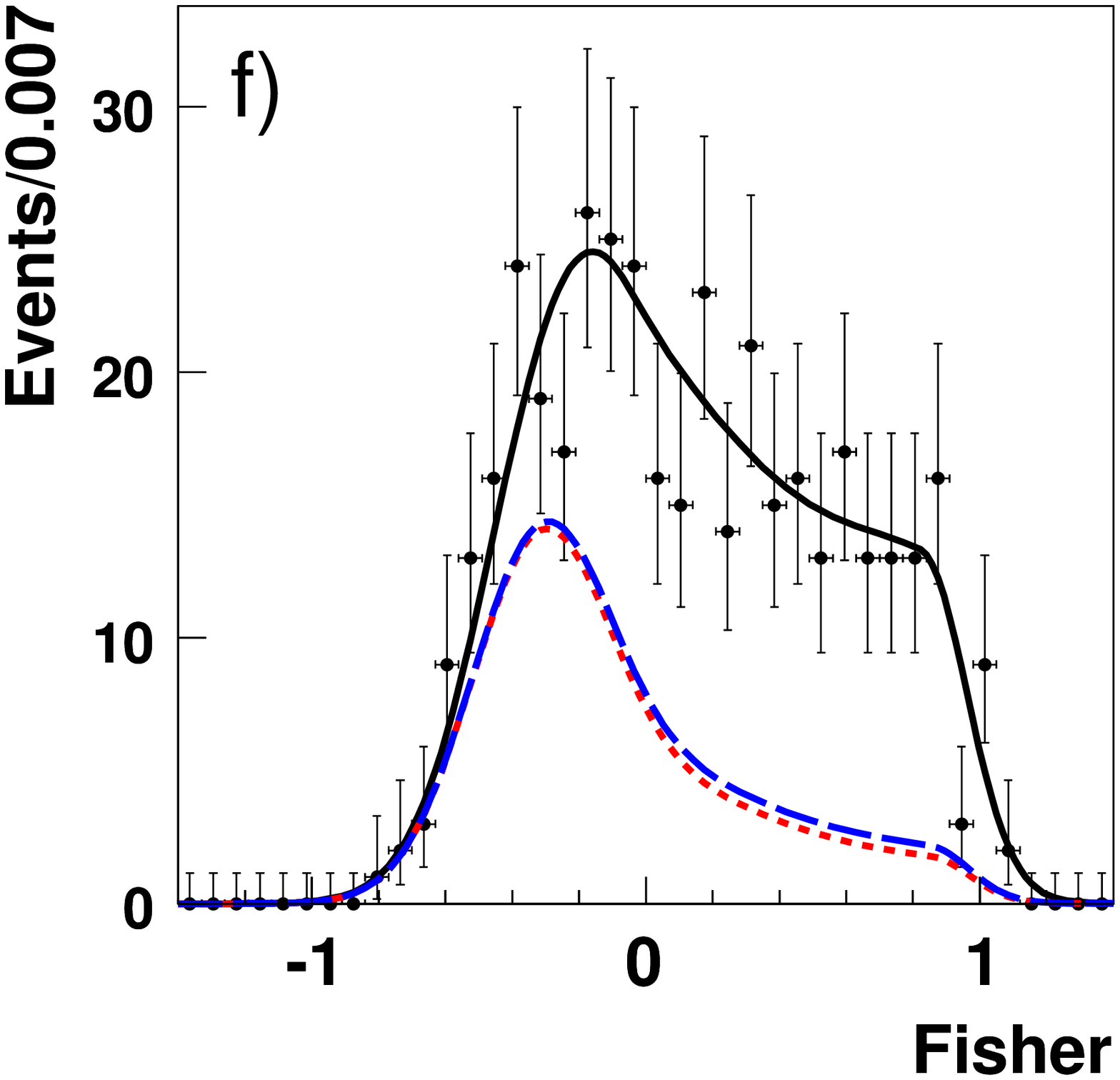} \\ 
\includegraphics[width=0.24\textwidth]{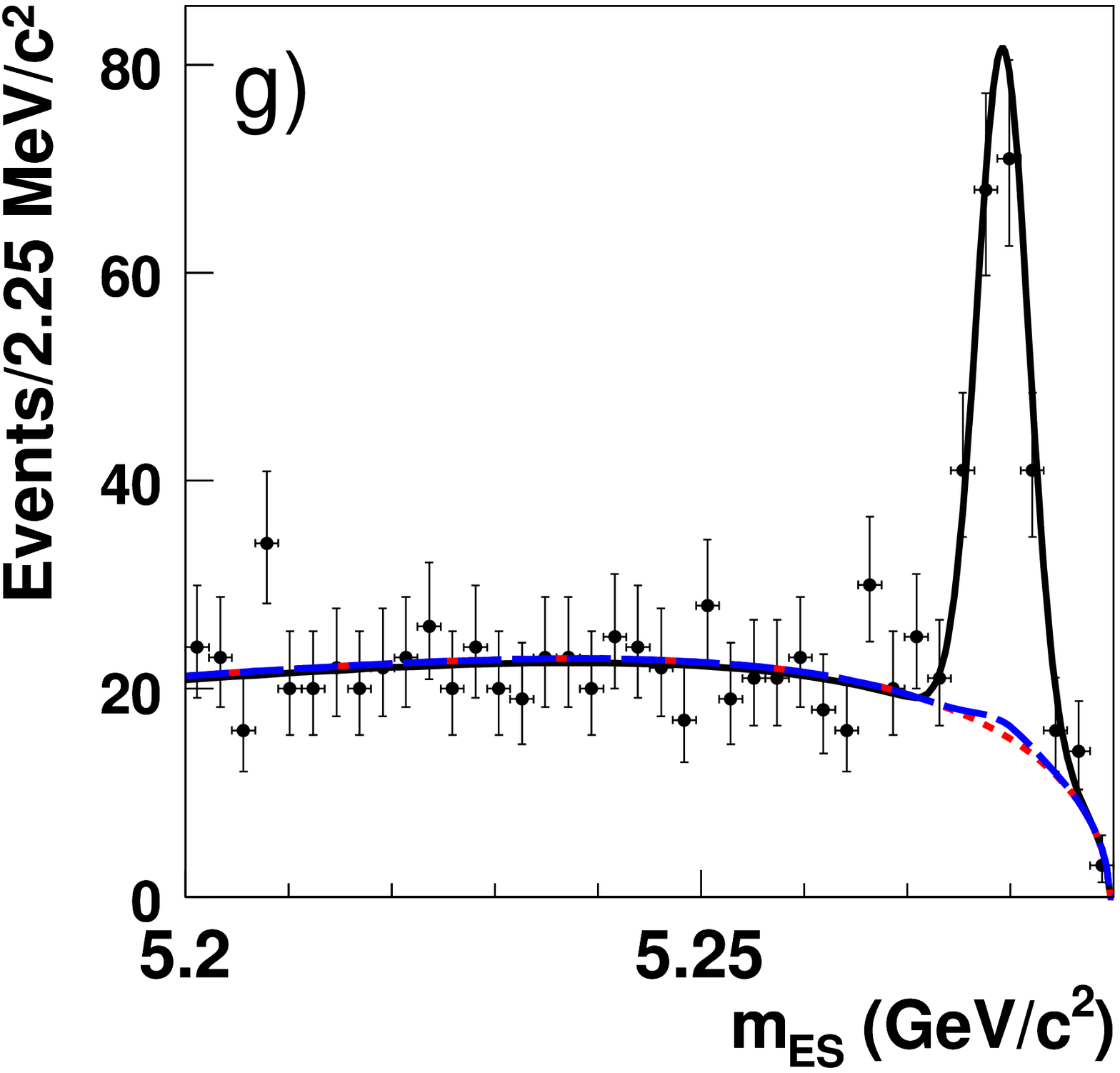} & 
\includegraphics[width=0.24\textwidth]{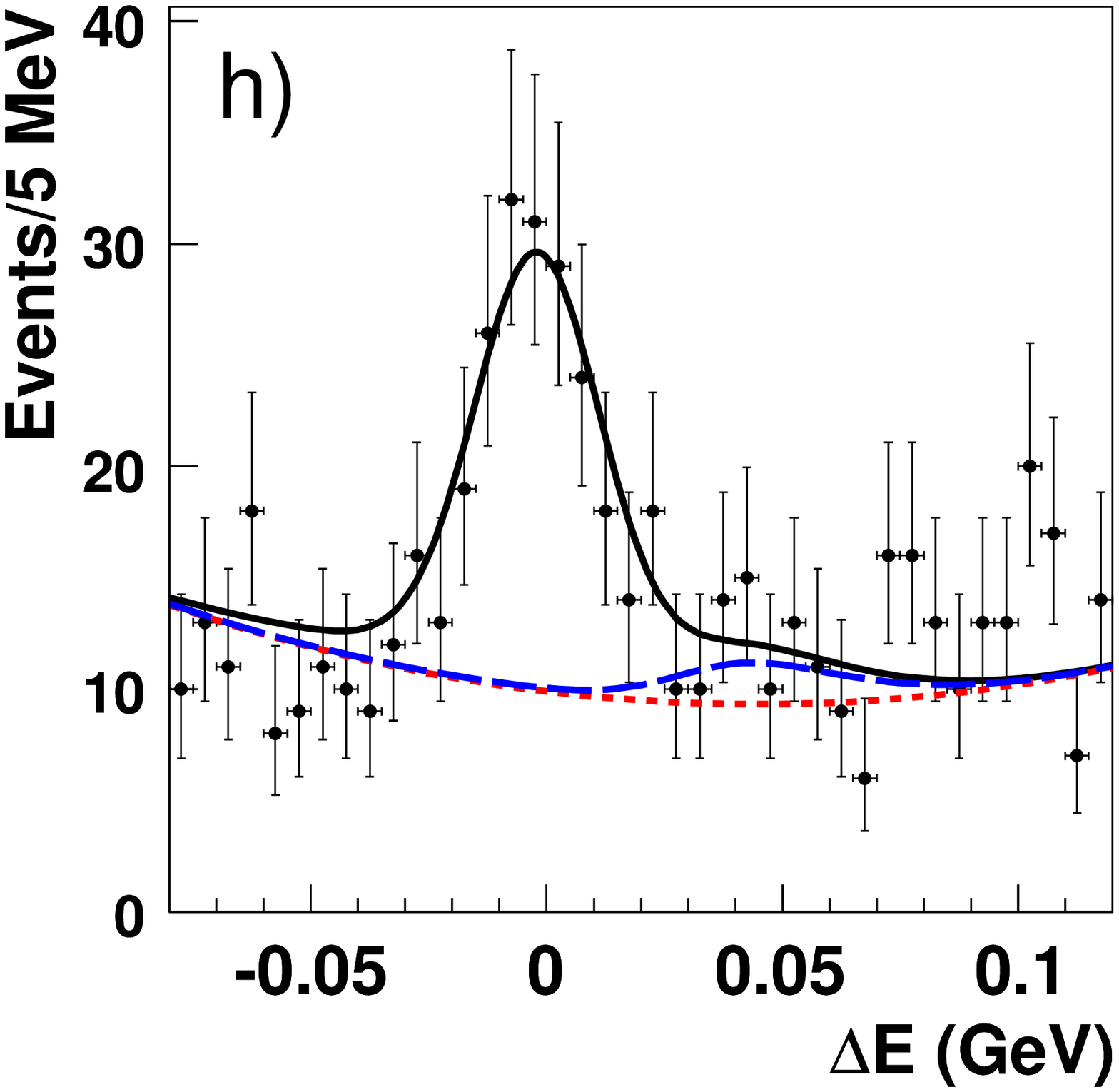} & 
\includegraphics[width=0.24\textwidth]{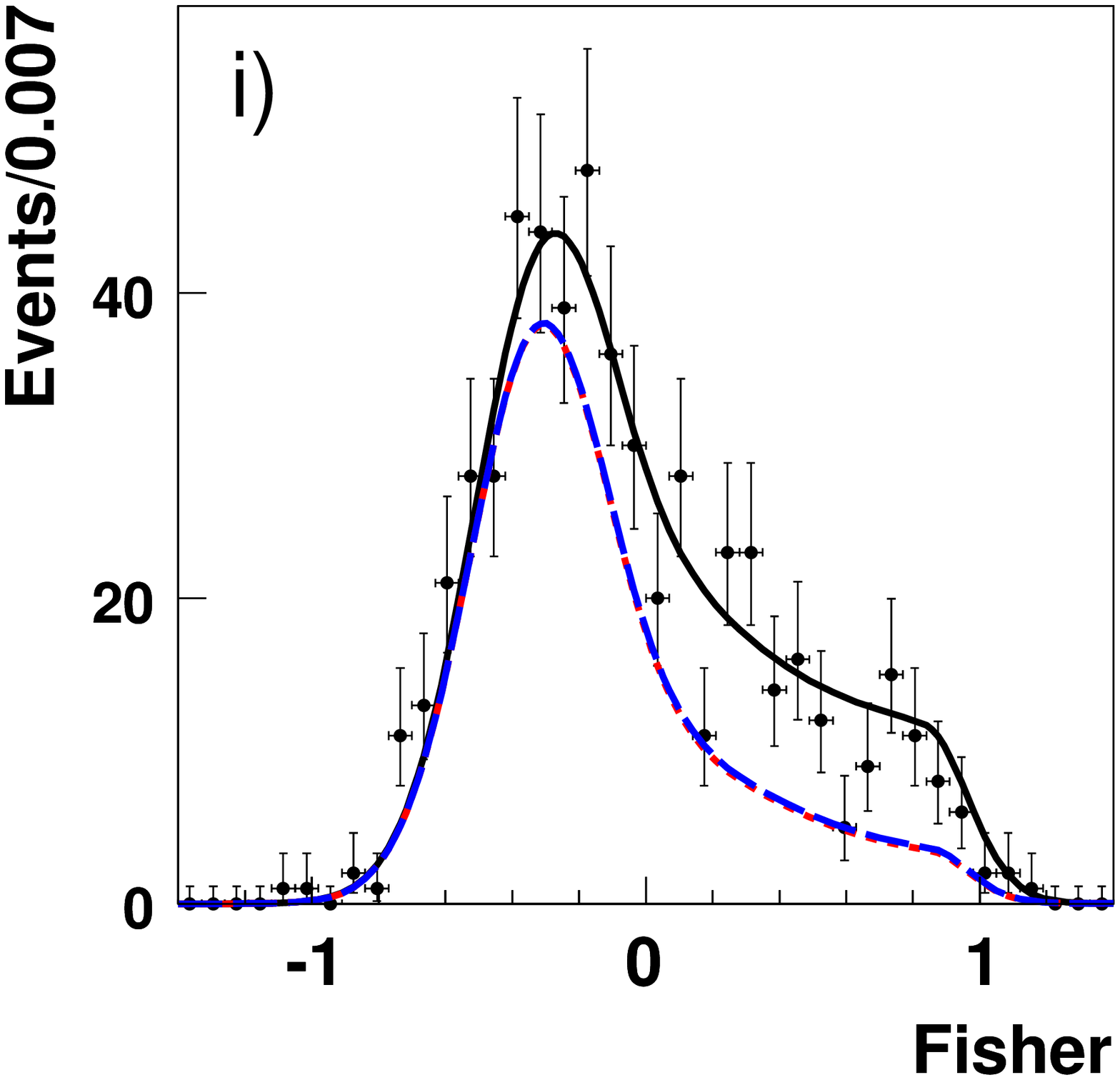} \\ 
\includegraphics[width=0.24\textwidth]{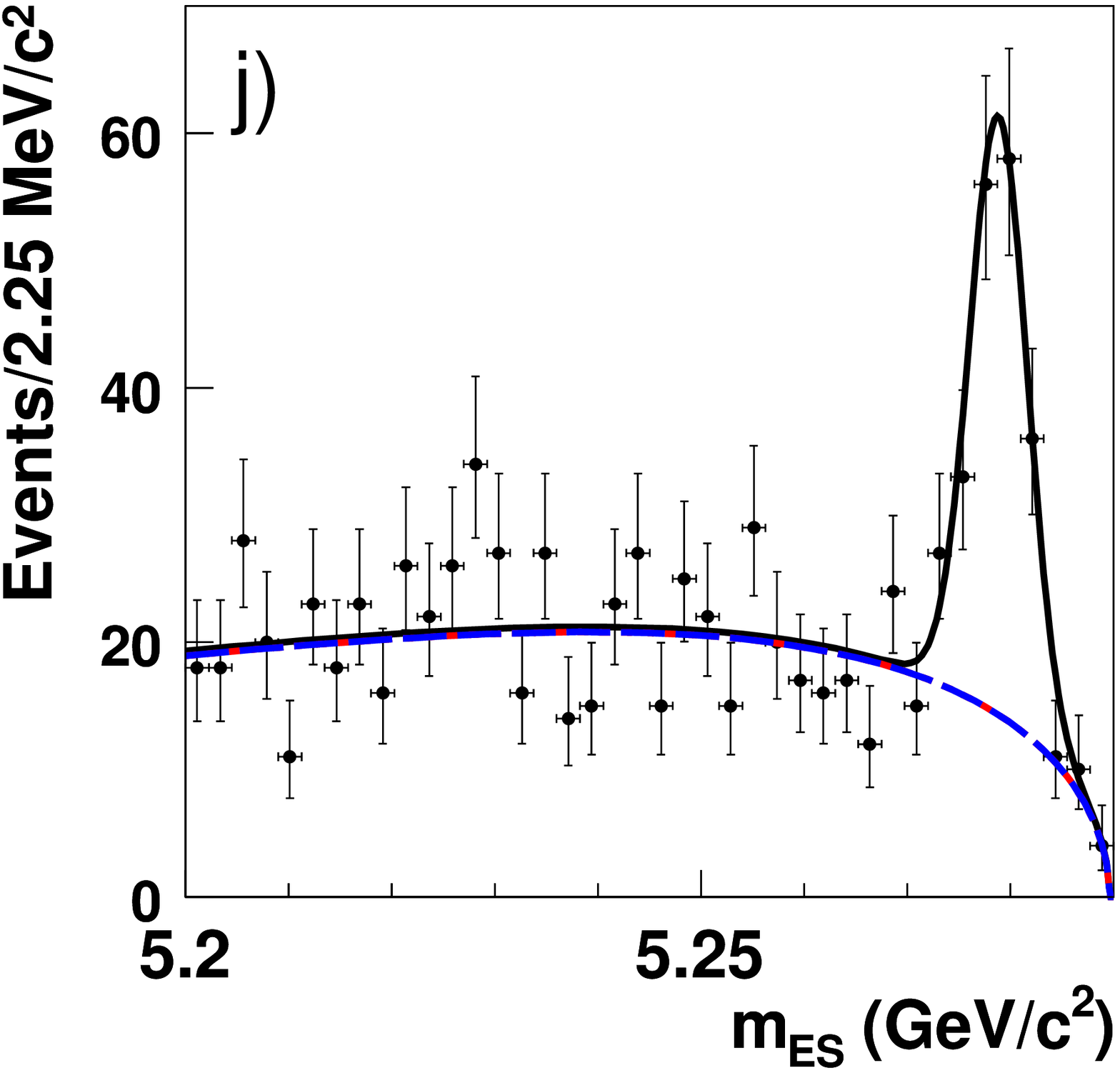} & 
\includegraphics[width=0.24\textwidth]{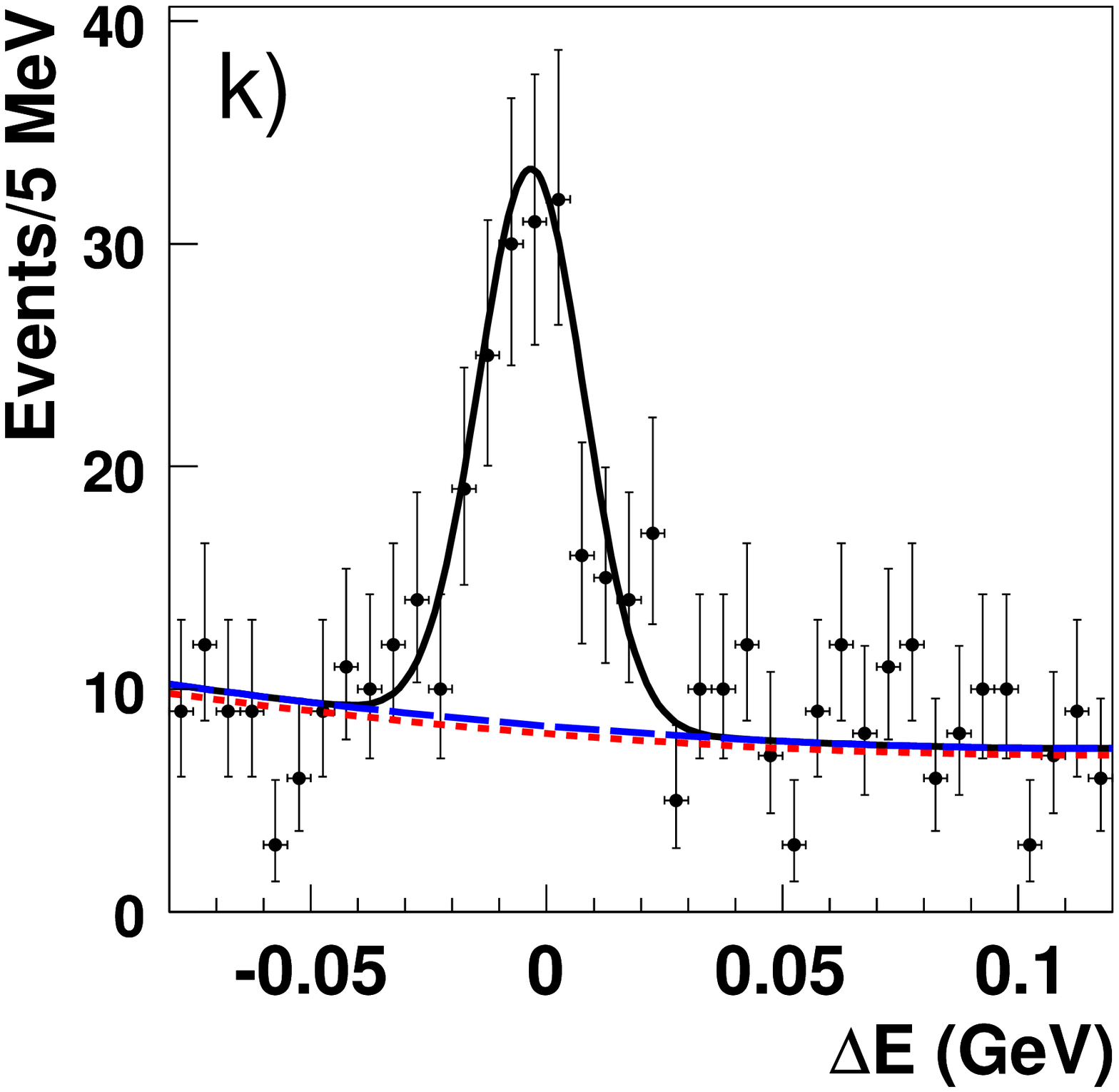} & 
\includegraphics[width=0.24\textwidth]{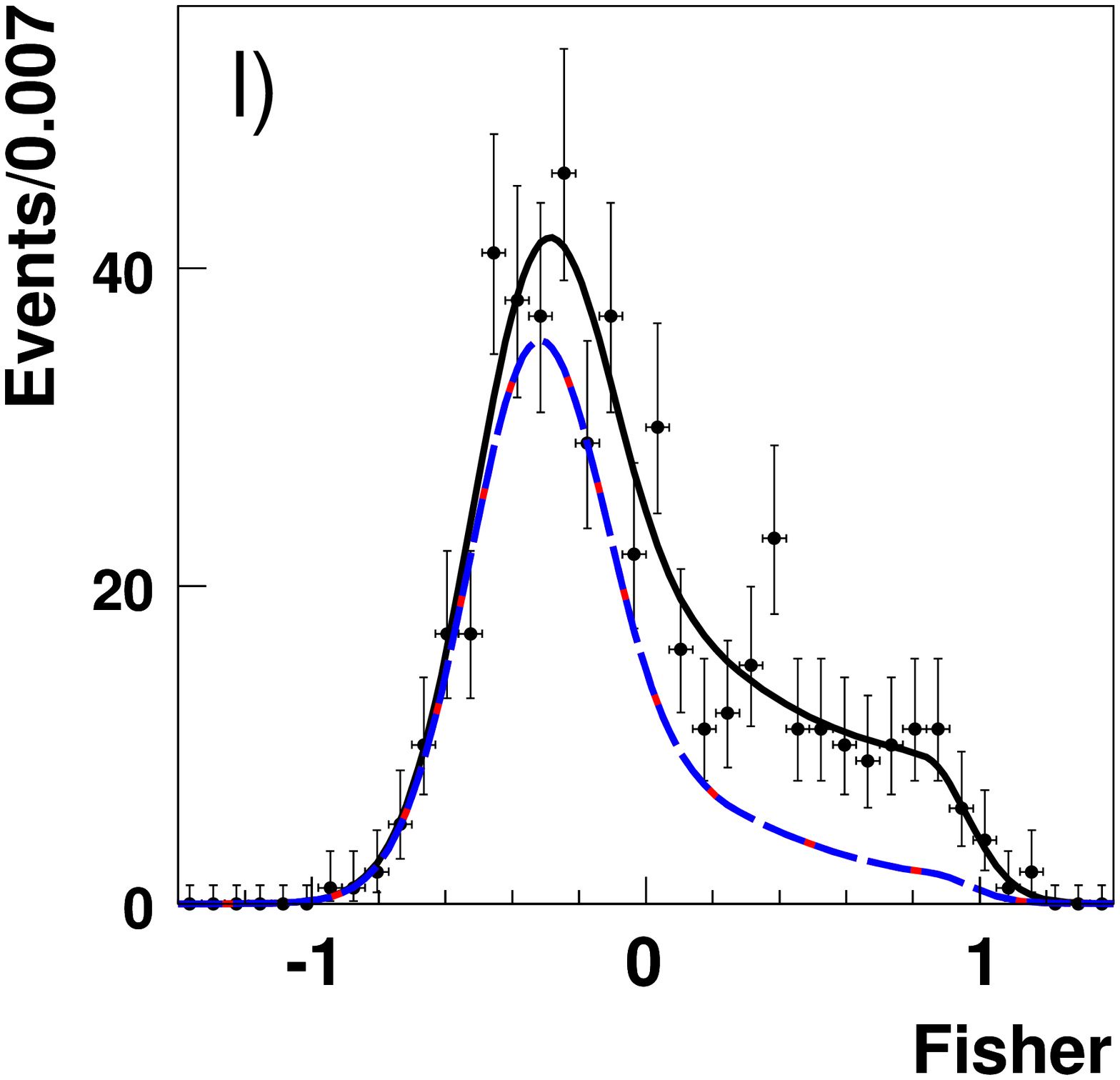} \\ 
\end{tabular} 
\caption{\label{fig:btodk_mes_kspipi} (color online).  
The \mes (first column), \de (second column), and \fis (third column) distributions for  
(a)-(c) $\Bmp\to \D \Kmp$, 
(d)-(f) $\Bmp\to \Dstar[\D\piz] \Kmp$, 
(g)-(i) $\Bmp\to \Dstar[\D\gamma] \Kmp$, and 
(j)-(l) $\Bmp\to \D \Kstarmp$ decays, with \Dtokspipi. 
The distributions are for events in the signal region defined through the requirements 
$\mes>5.272$~\gevcc, $|\de|<30$~\mev, and $\fisher>-0.1$, except the one on the plotted variable,  
after all the selection criteria are applied. 
The curves superimposed represent the projections of the \CP fit:  
signal plus background (solid black lines), the continuum plus \BB background contributions (dotted red lines),  
and the sum of the continuum, \BB, and $K/\pi$ misidentification background components (dashed blue lines). 
The reconstruction efficiencies (purities) in the signal region, based on simulation studies, are  
$22\%$ ($68\%$), $10\%$ ($81\%$), $12\%$ ($55\%$), and $12\%$ ($58\%$), respectively. 
} 
\end{figure}

\begin{figure}[hbt!] 
\begin{tabular}{ccc} 
\includegraphics[width=0.24\textwidth]{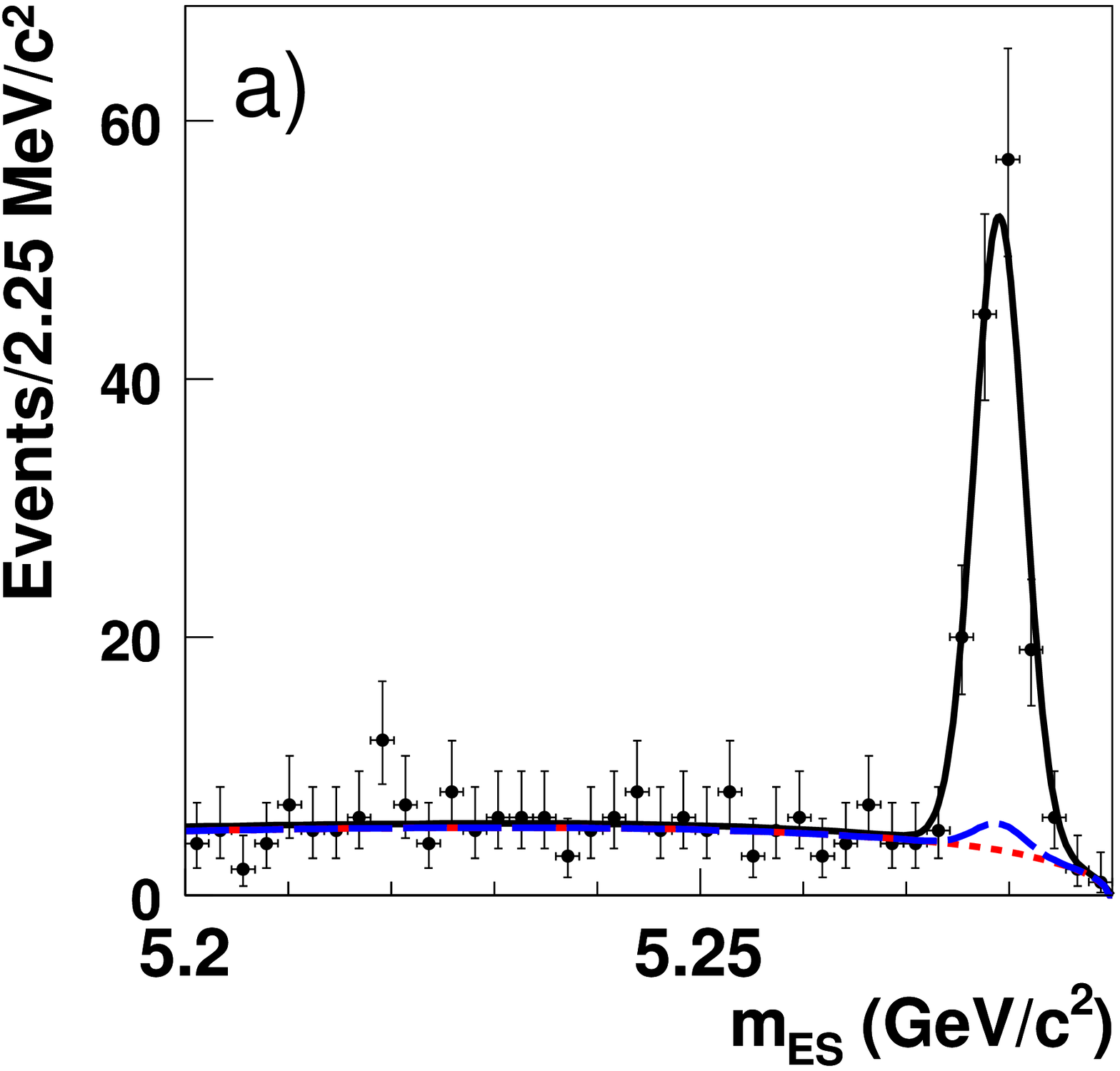} & 
\includegraphics[width=0.24\textwidth]{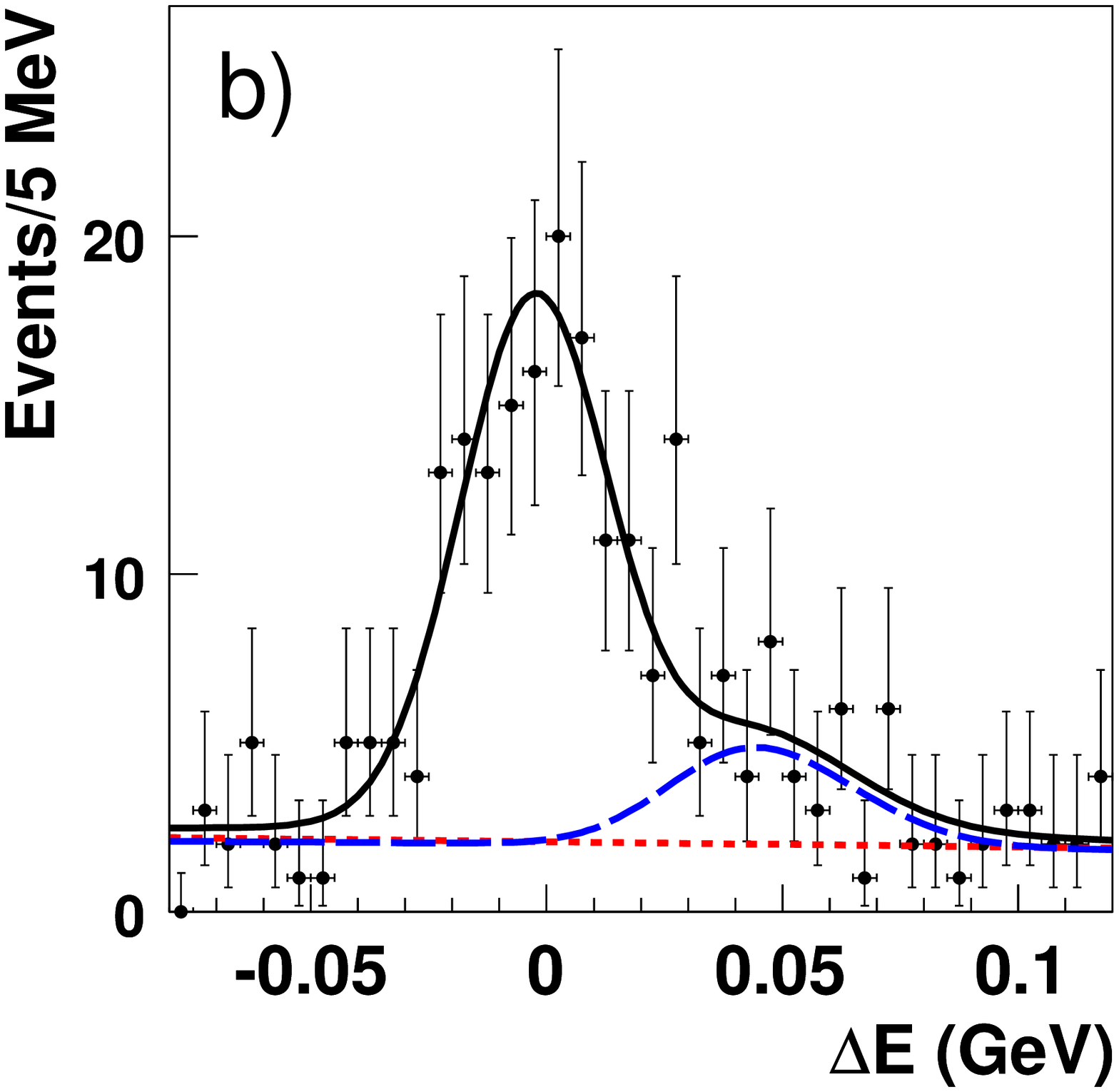} & 
\includegraphics[width=0.24\textwidth]{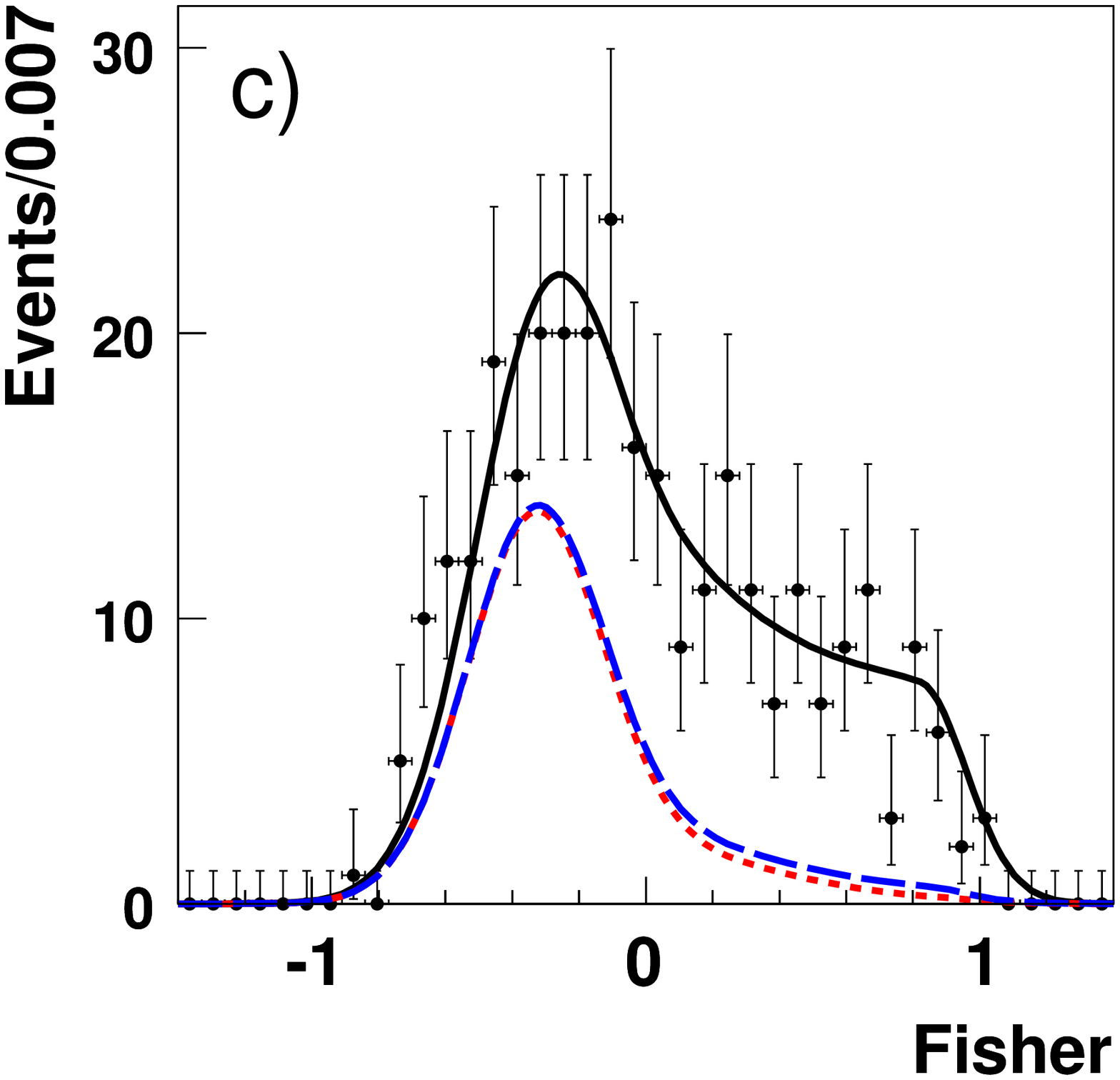} \\ 
\includegraphics[width=0.24\textwidth]{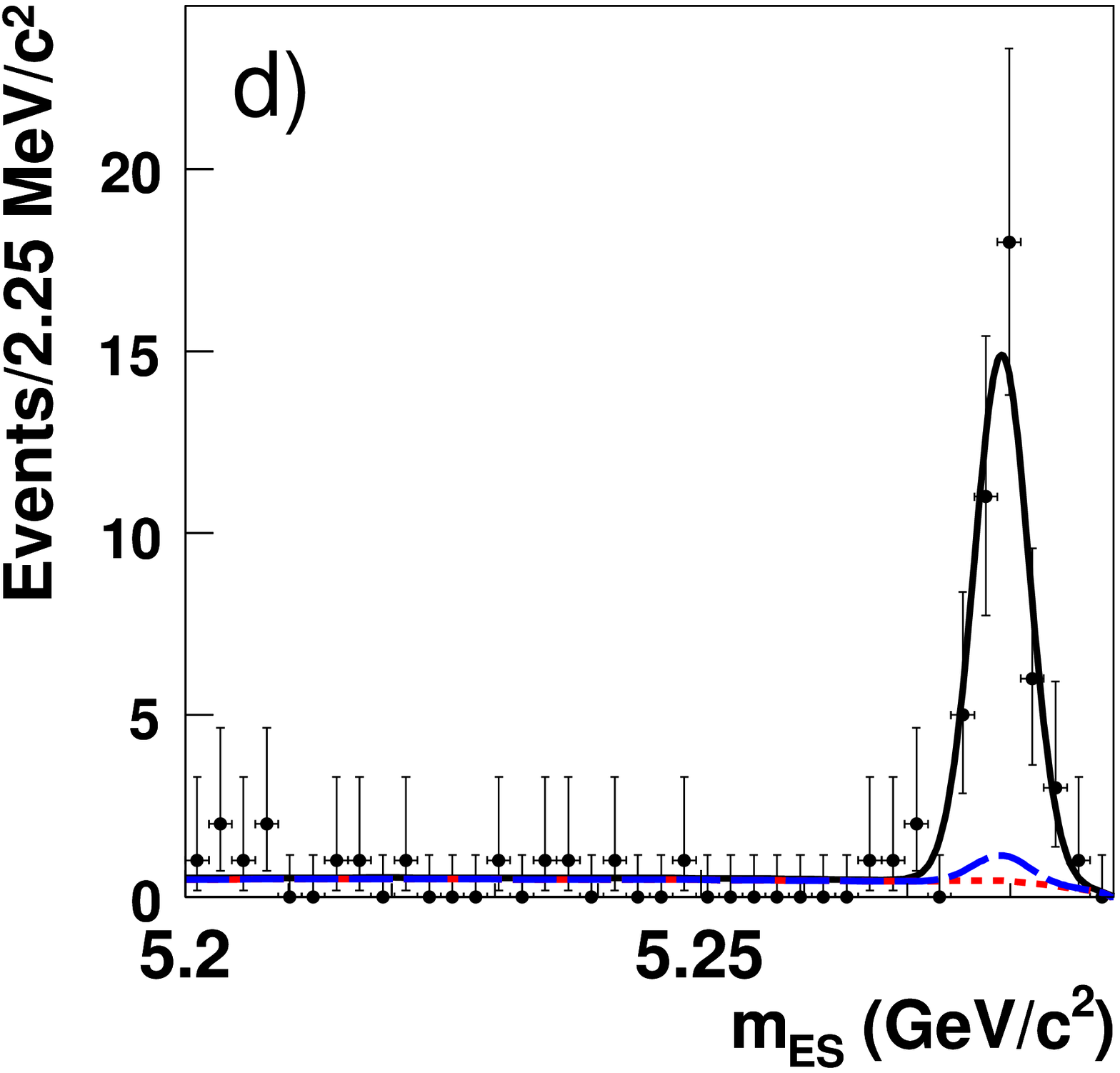} & 
\includegraphics[width=0.24\textwidth]{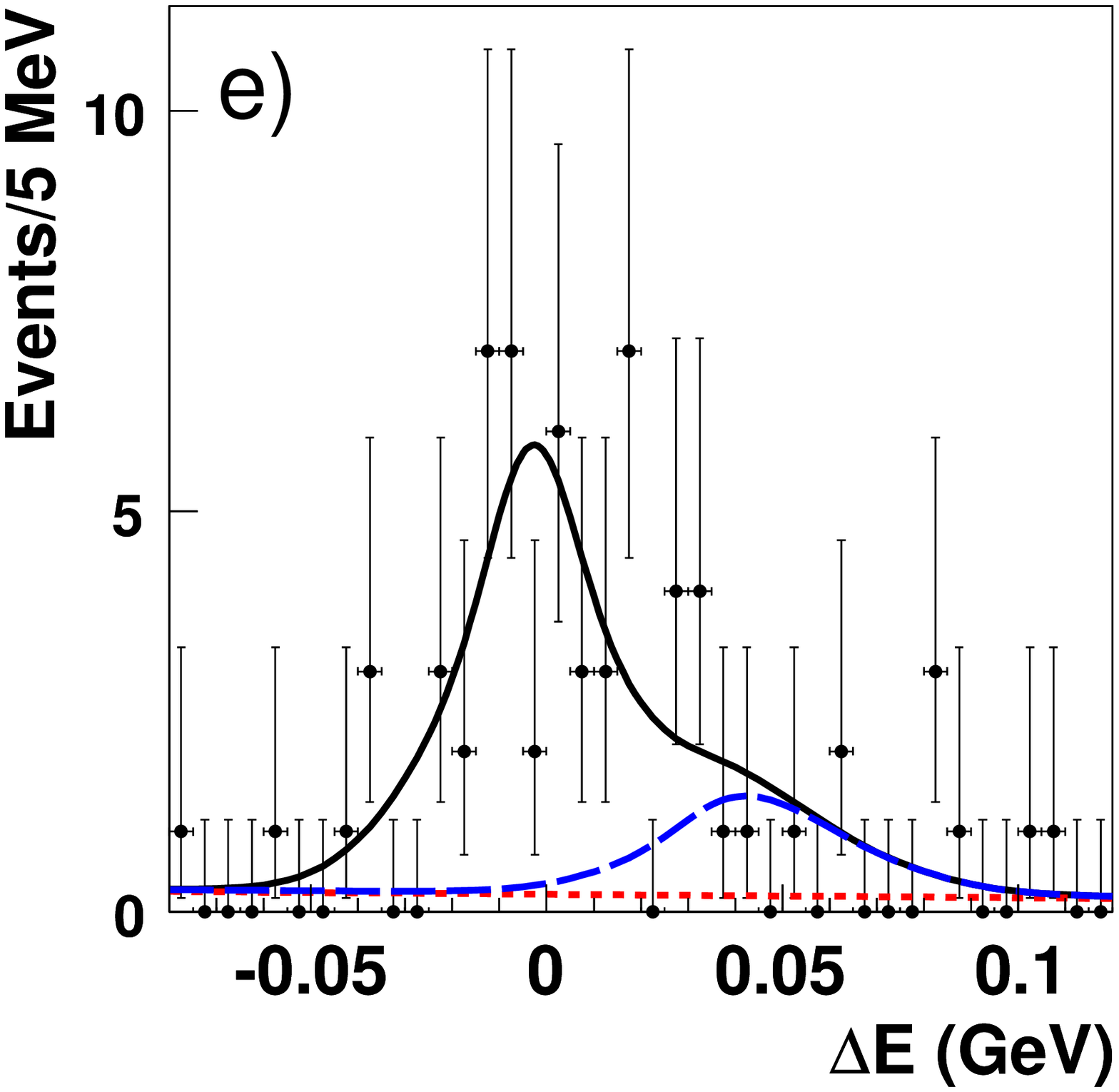} & 
\includegraphics[width=0.24\textwidth]{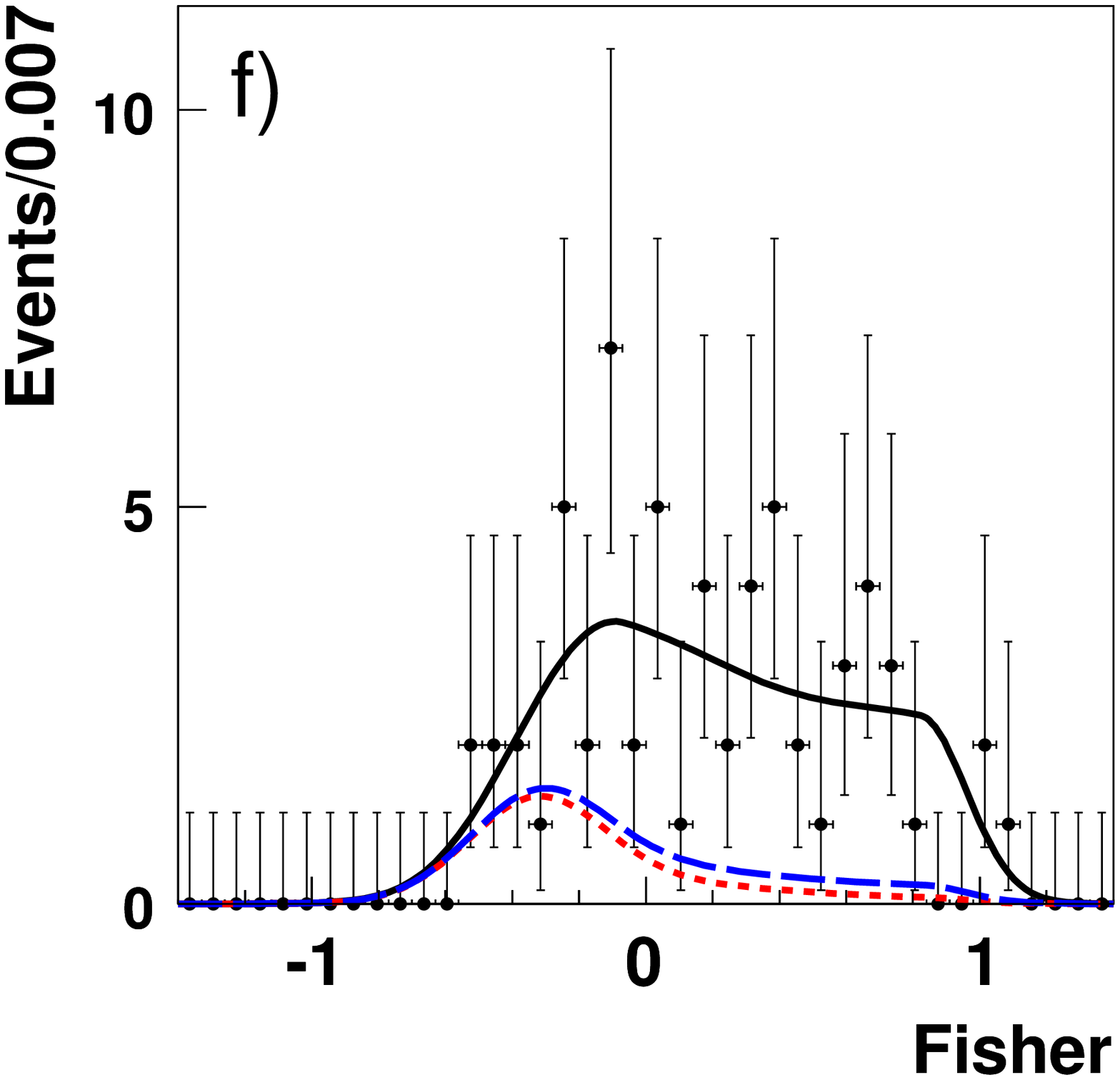} \\ 
\includegraphics[width=0.24\textwidth]{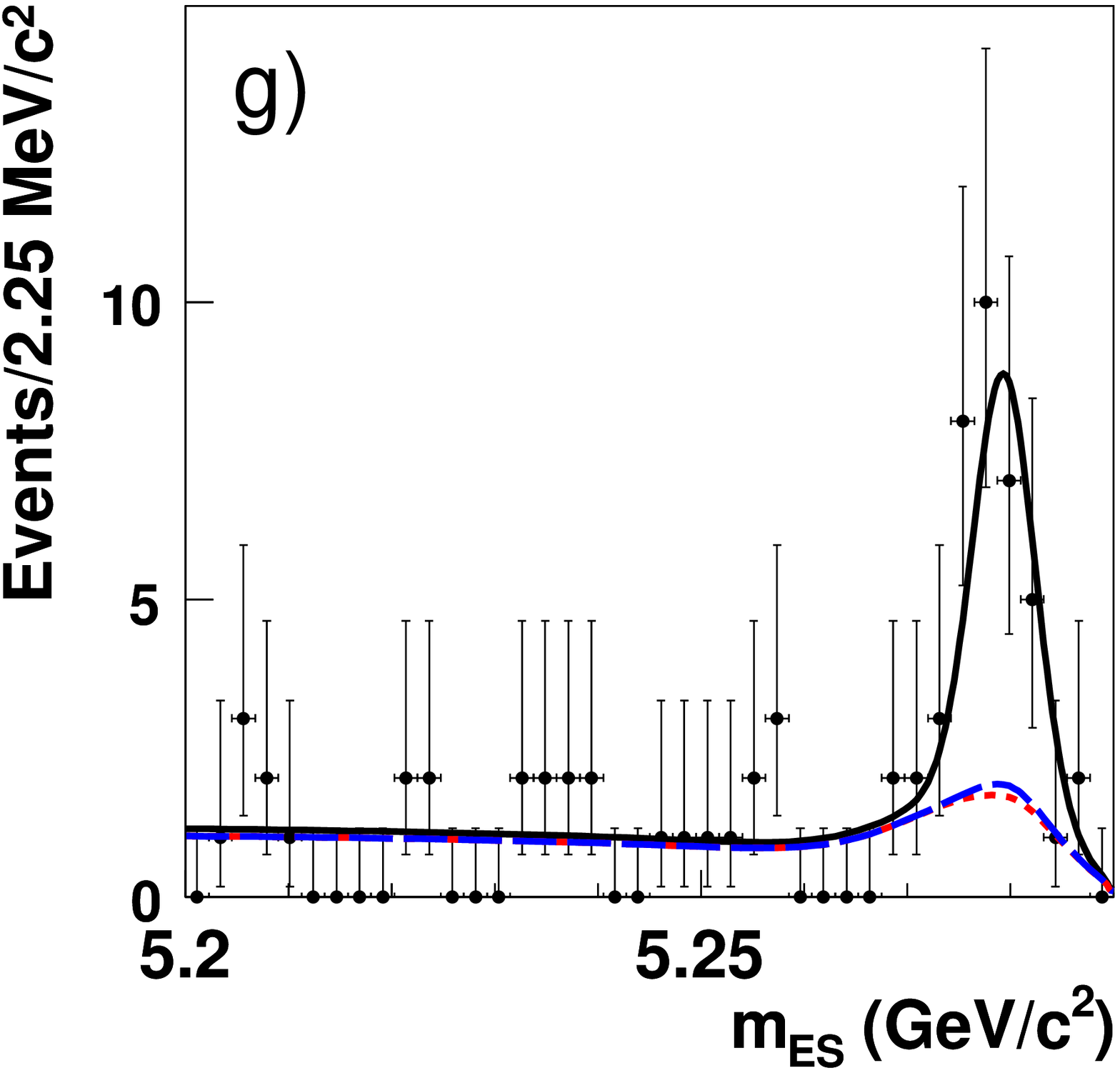} & 
\includegraphics[width=0.24\textwidth]{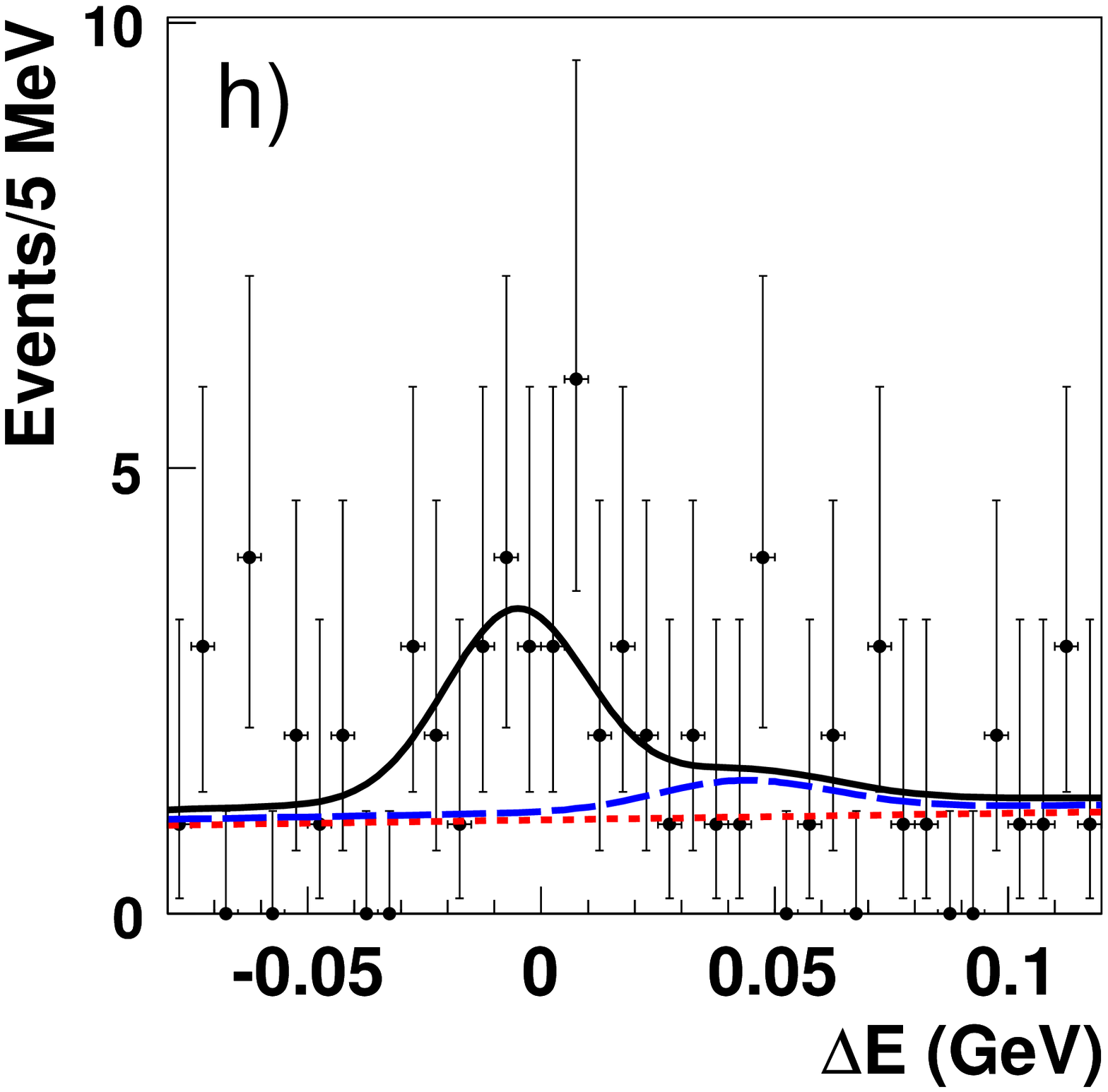} & 
\includegraphics[width=0.24\textwidth]{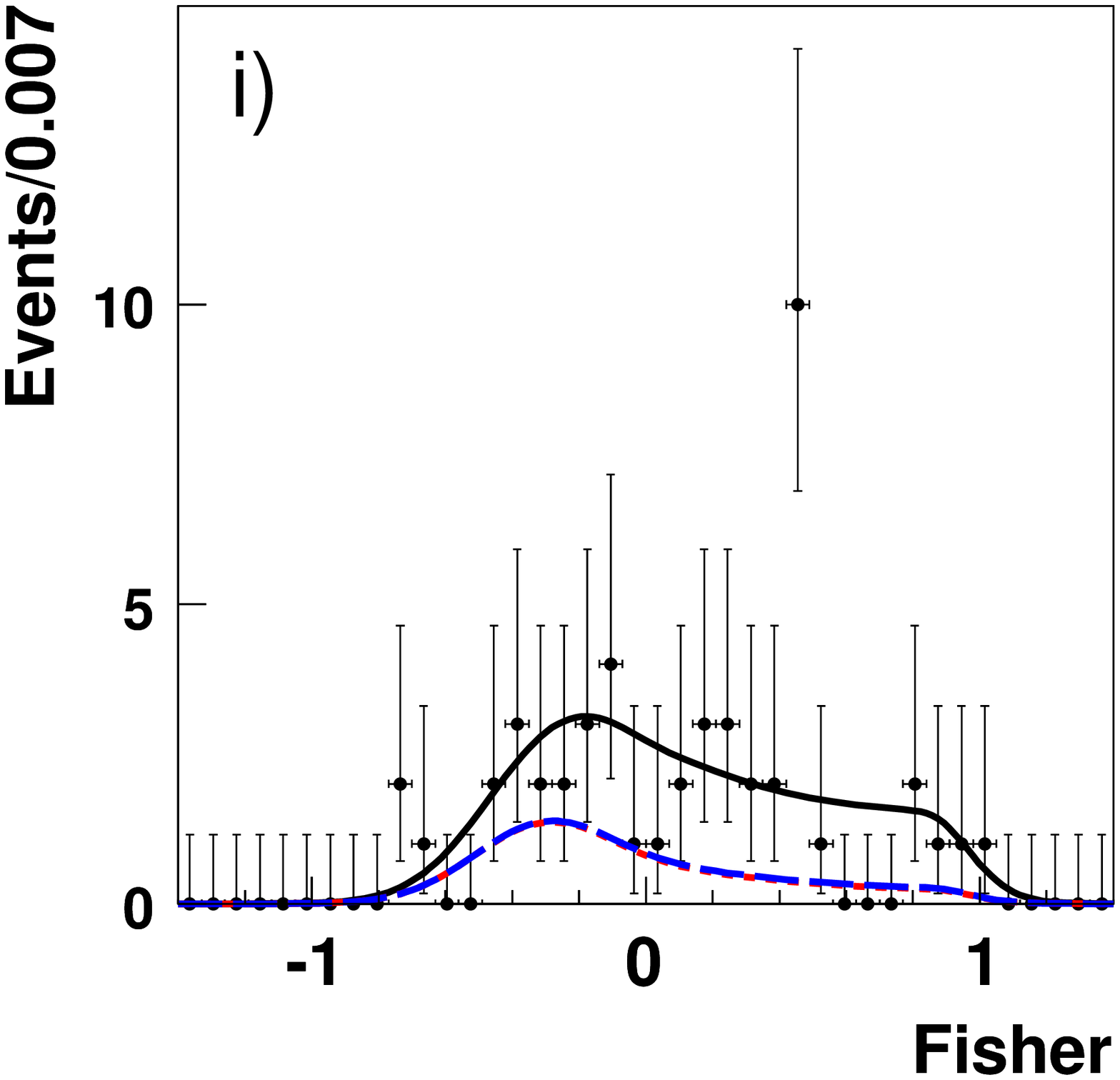} \\ 
\includegraphics[width=0.24\textwidth]{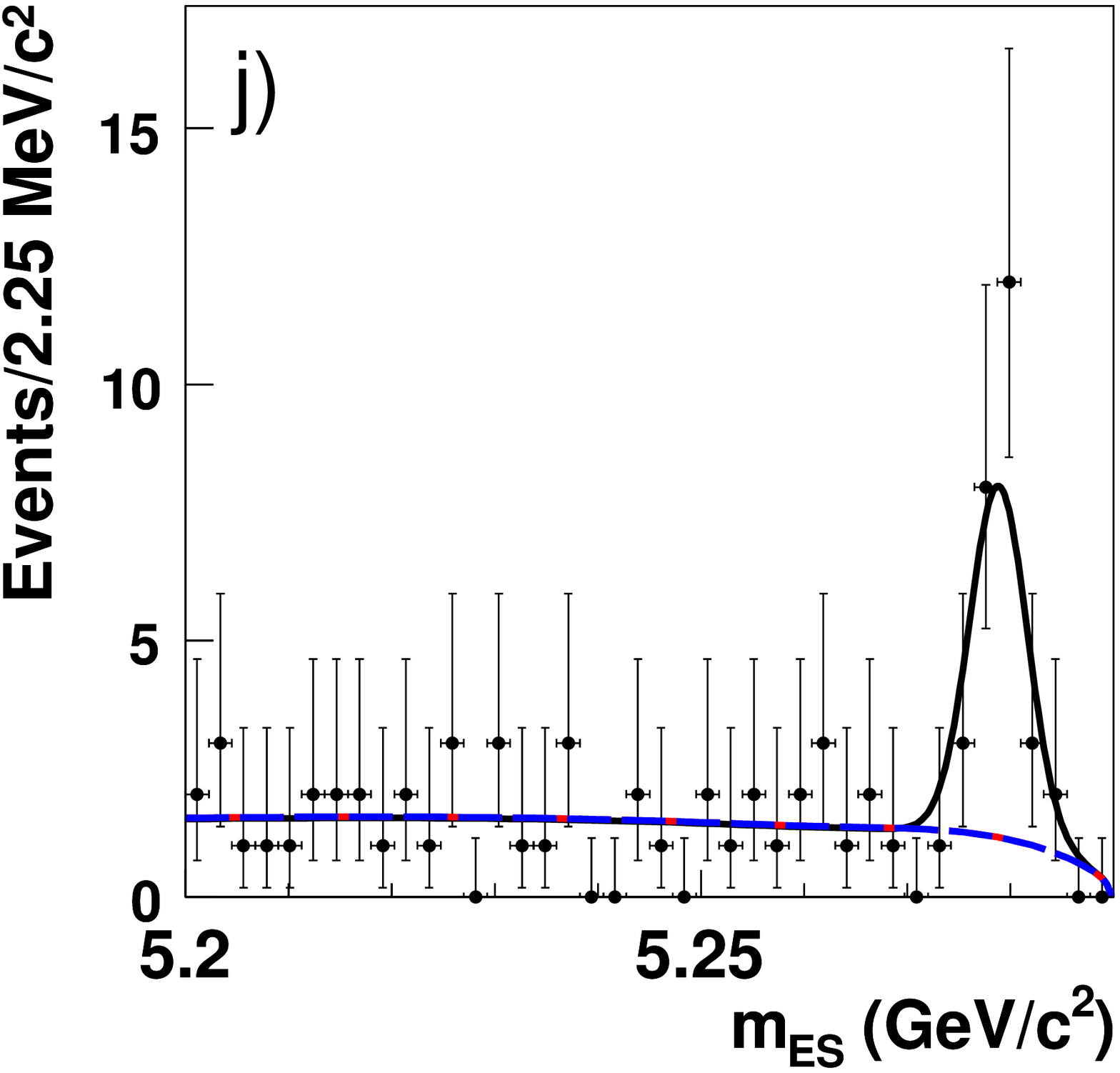} & 
\includegraphics[width=0.24\textwidth]{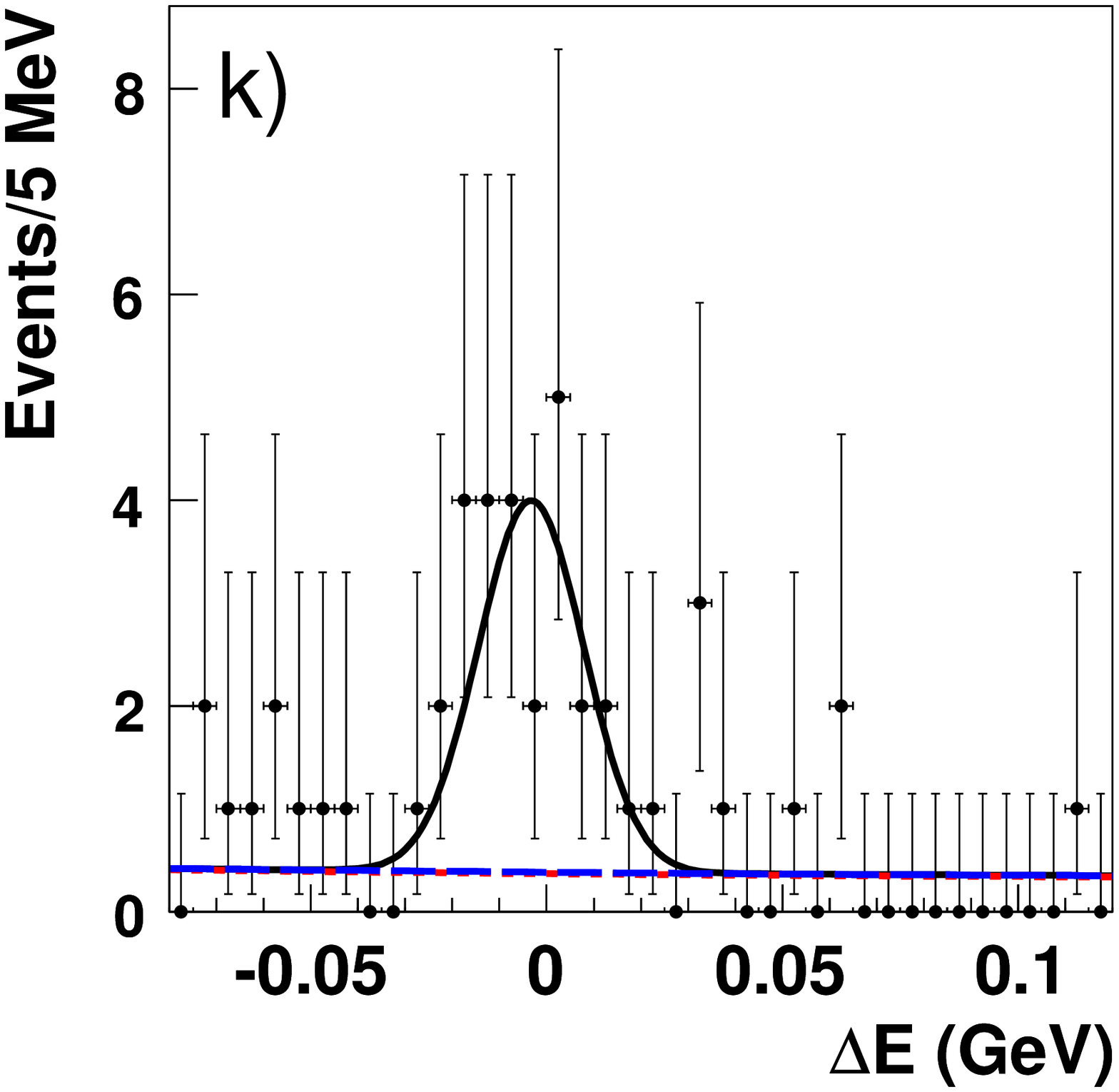} & 
\includegraphics[width=0.24\textwidth]{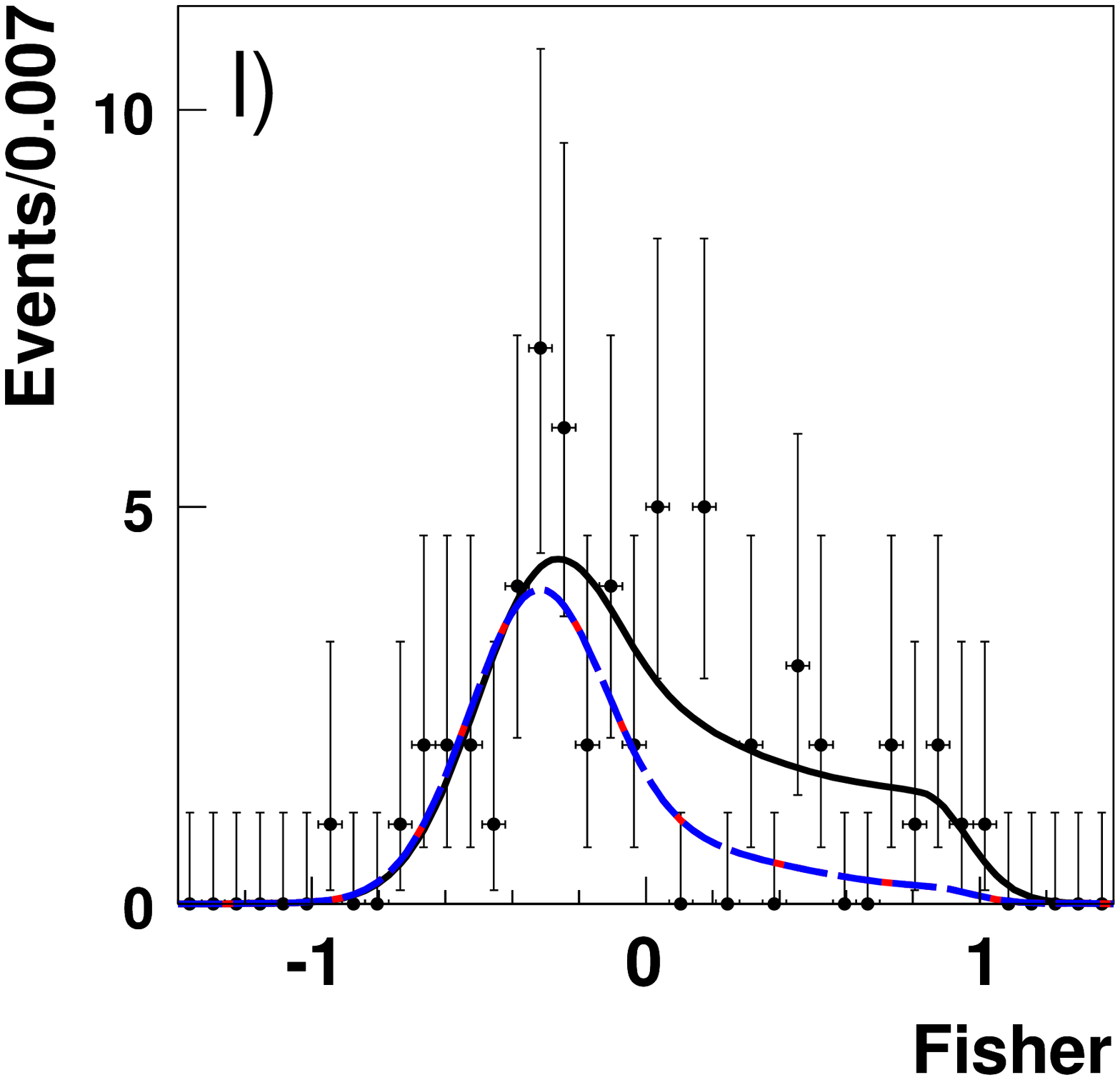} \\ 
\end{tabular} 
\caption{\label{fig:btodk_mes_kskk} (color online).  
Same as in Fig.~\ref{fig:btodk_mes_kspipi} but for 
(a)-(c) $\Bmp\to \D \Kmp$, 
(d)-(f) $\Bmp\to \Dstar[\D\piz] \Kmp$, 
(g)-(i) $\Bmp\to \Dstar[\D\gamma] \Kmp$, and 
(j)-(l) $\Bmp\to \D \Kstarmp$ decays, with \Dtokskk. 
The reconstruction efficiencies (purities) in the signal region, based on simulation studies, are in this case  
$20\%$ ($82\%$), $9\%$ ($87\%$), $12\%$ ($78\%$), and $11\%$ ($81\%$), respectively. 
} 
\end{figure}

% DP for Kspipi and KsKK 

\begin{figure}[hbt!] 
\begin{tabular}{cc|cc} 
\includegraphics[width=0.24\textwidth]{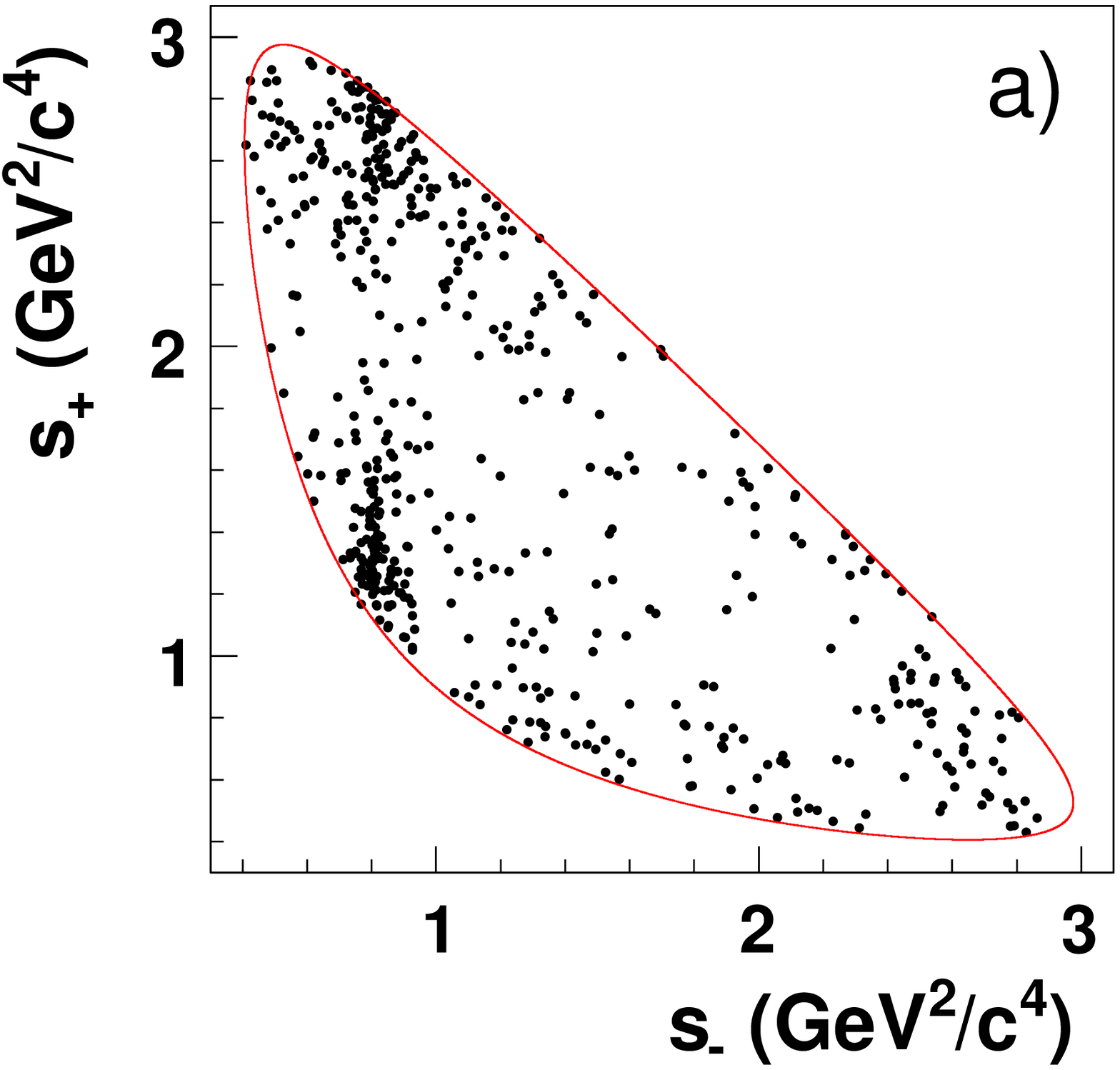}& 
\includegraphics[width=0.24\textwidth]{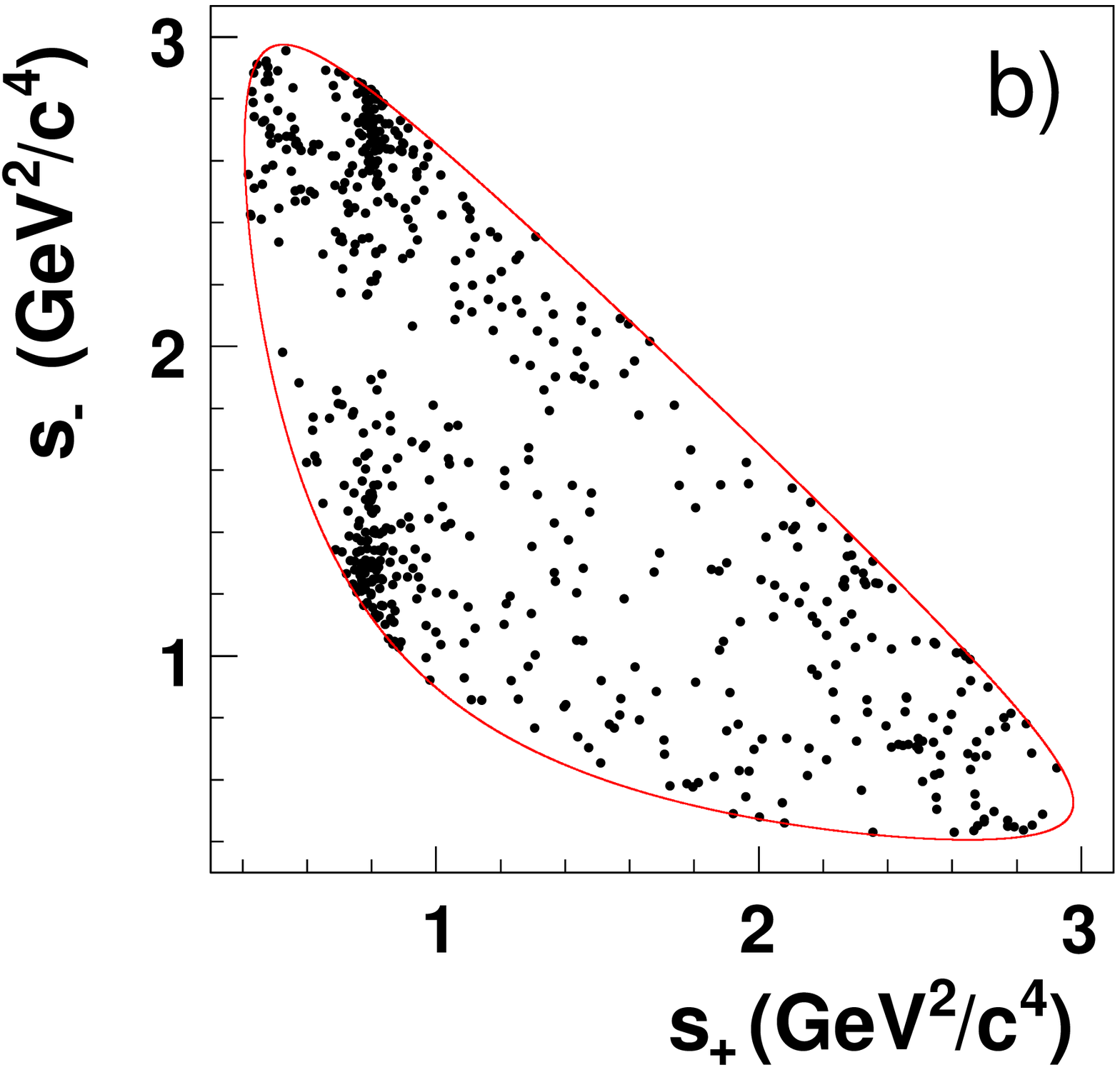}& 
\includegraphics[width=0.24\textwidth]{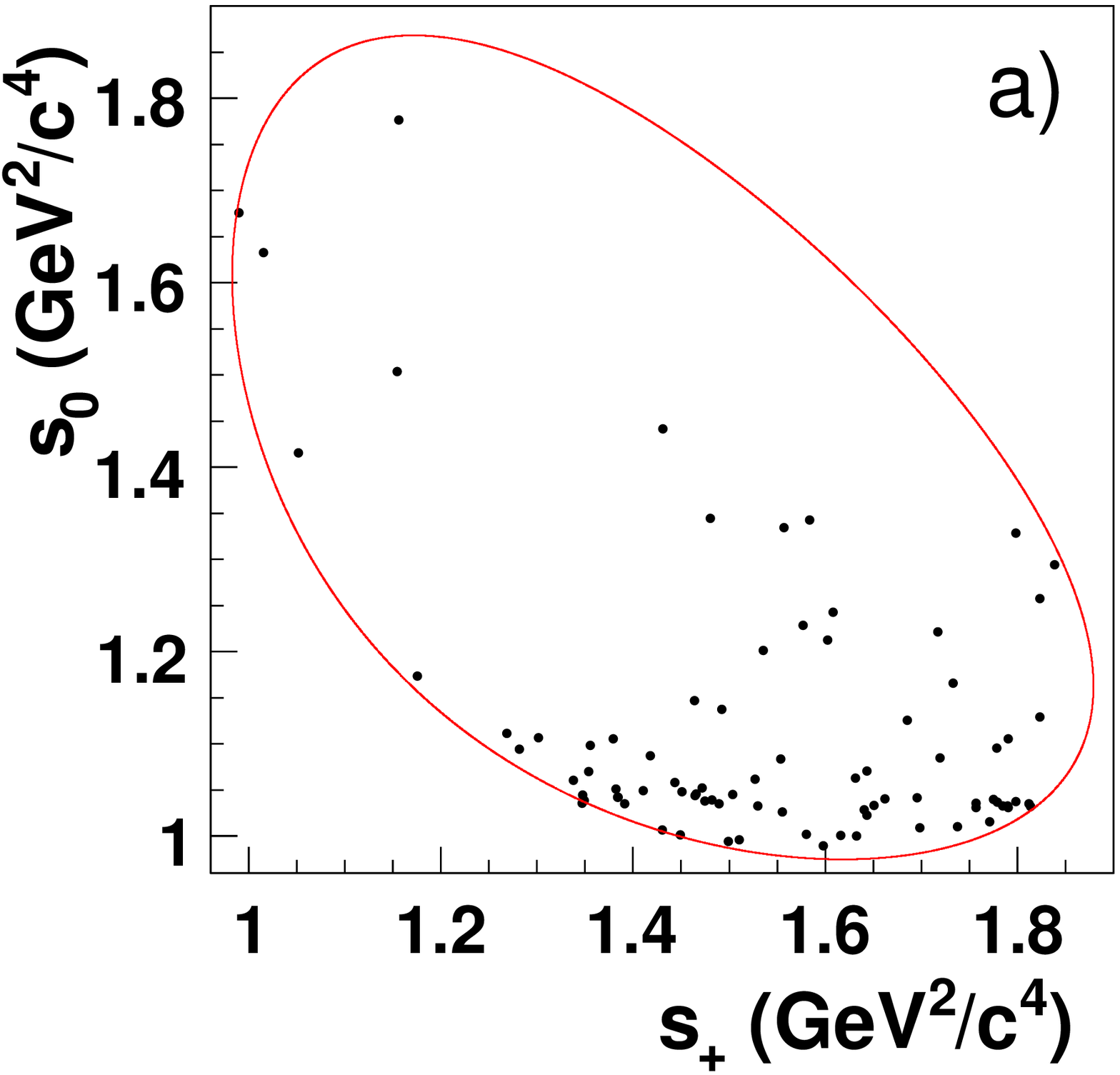}& 
\includegraphics[width=0.24\textwidth]{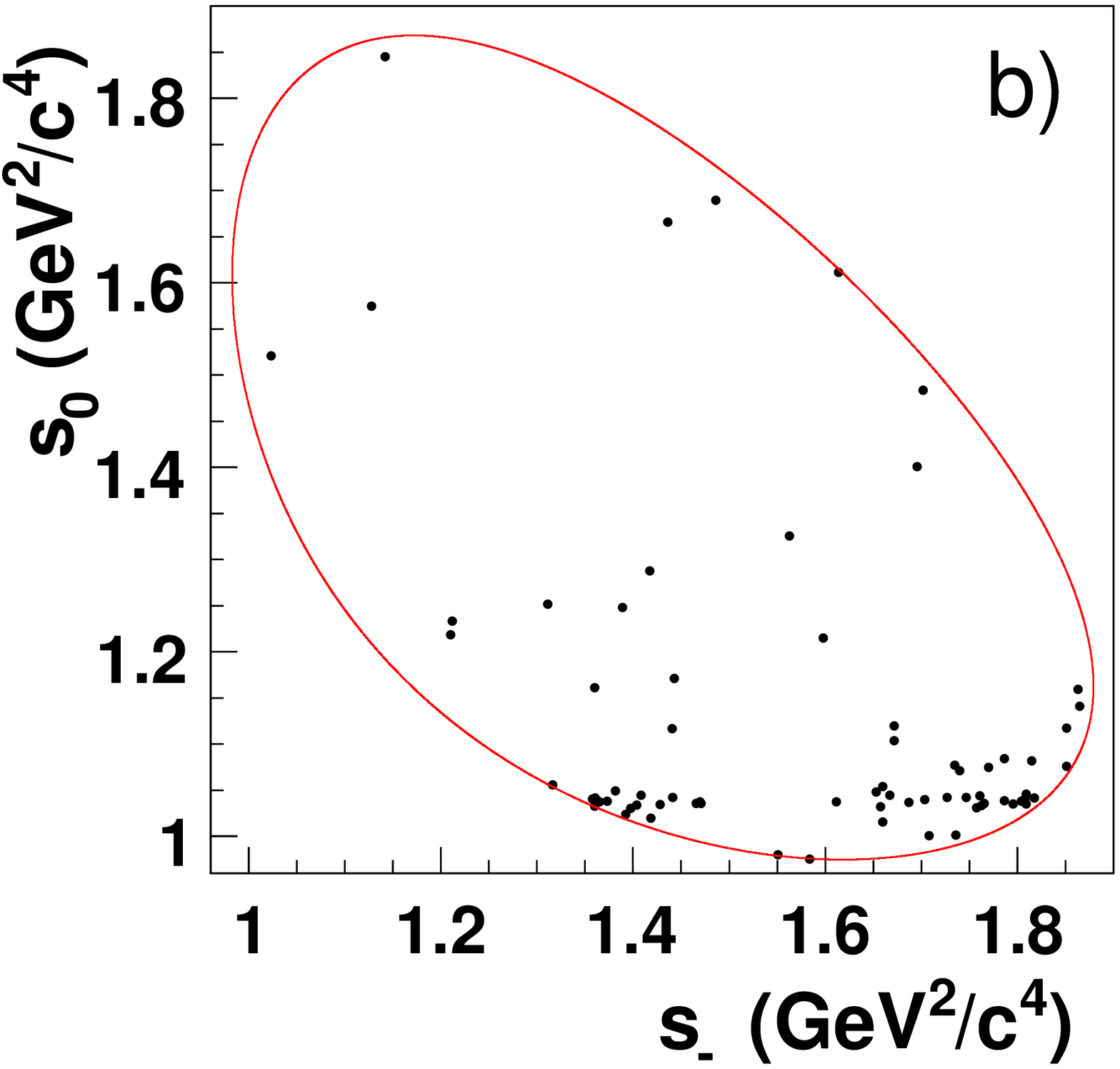} \\ 
\includegraphics[width=0.24\textwidth]{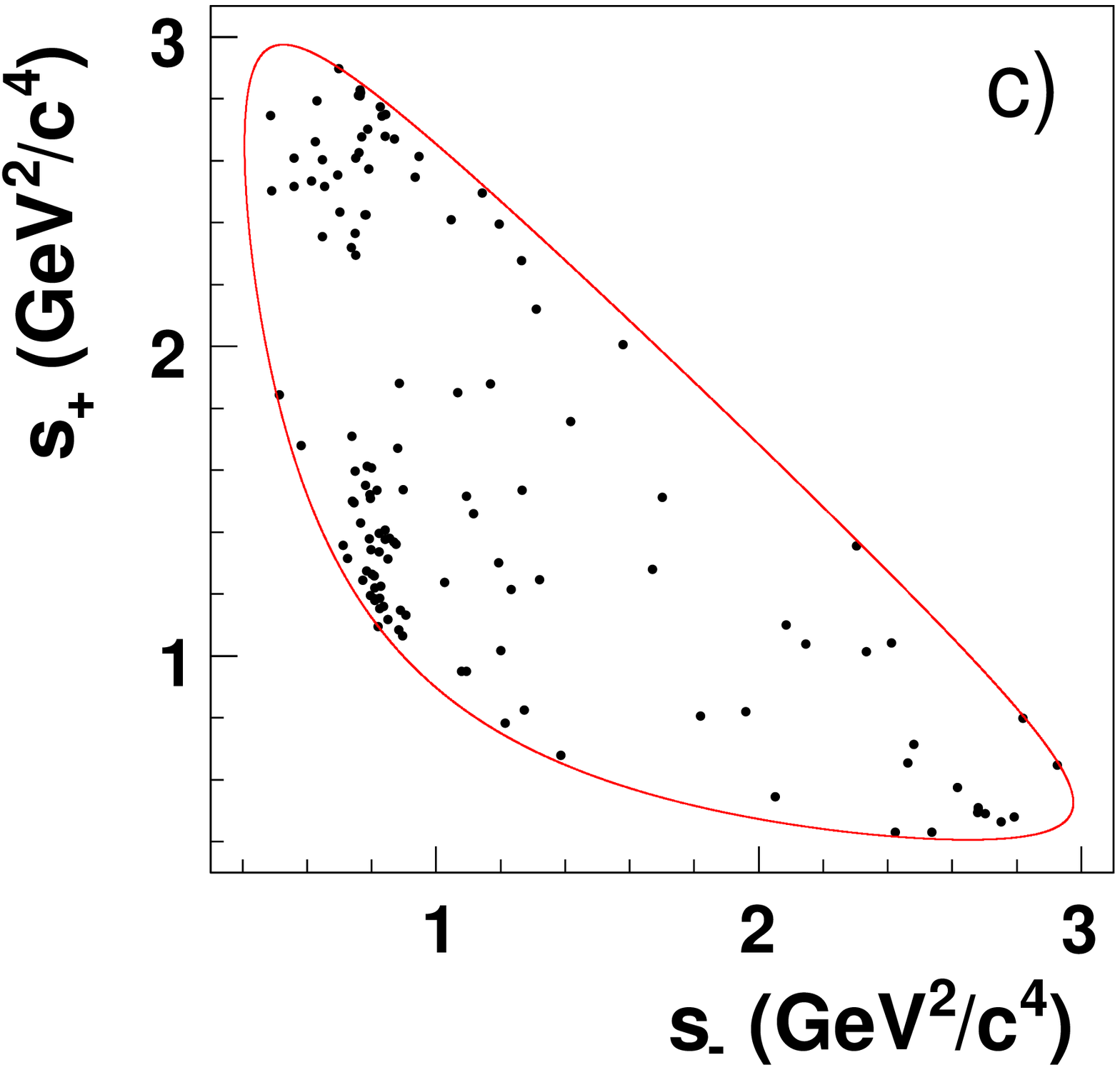}& 
\includegraphics[width=0.24\textwidth]{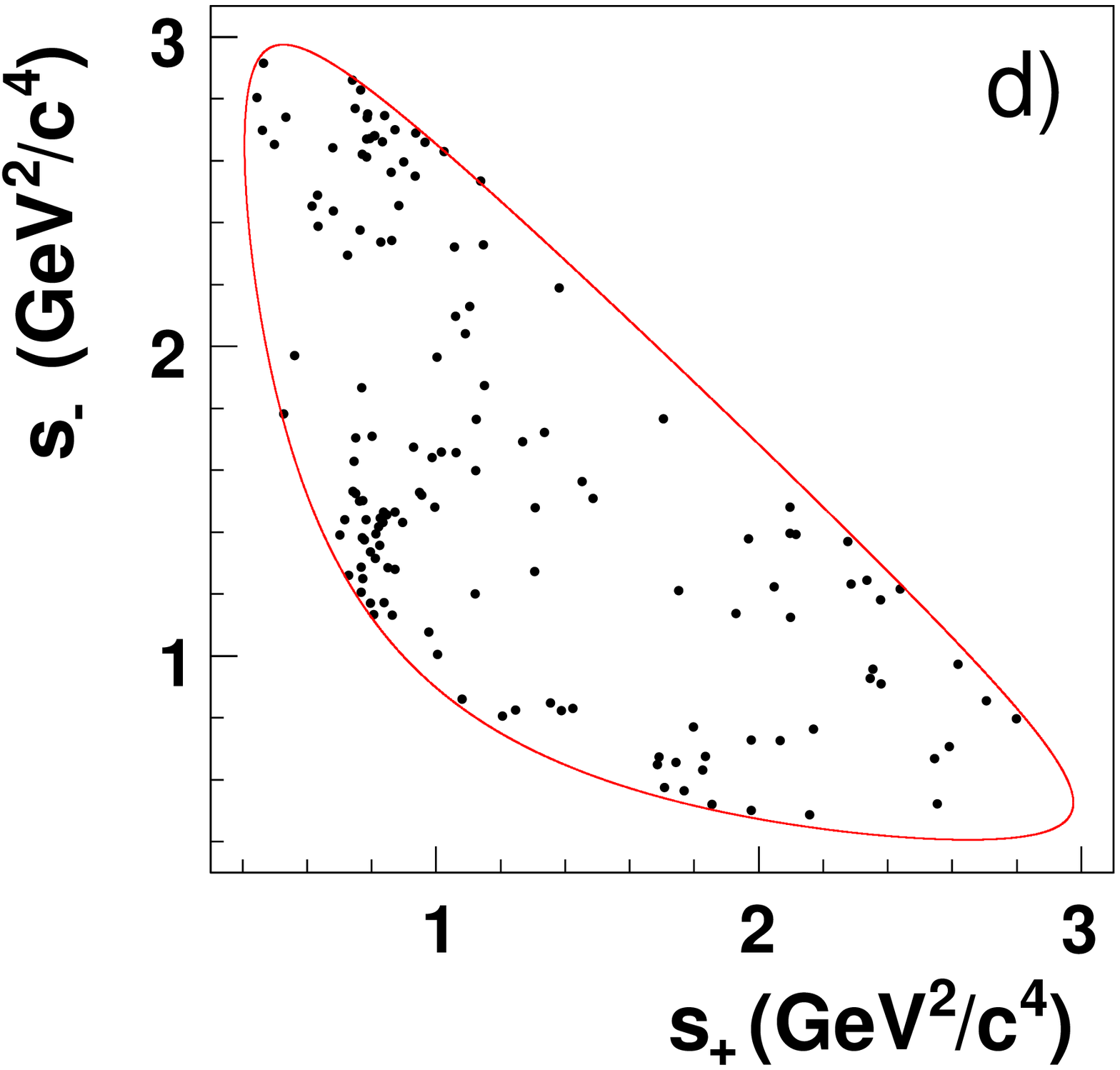} & 
\includegraphics[width=0.24\textwidth]{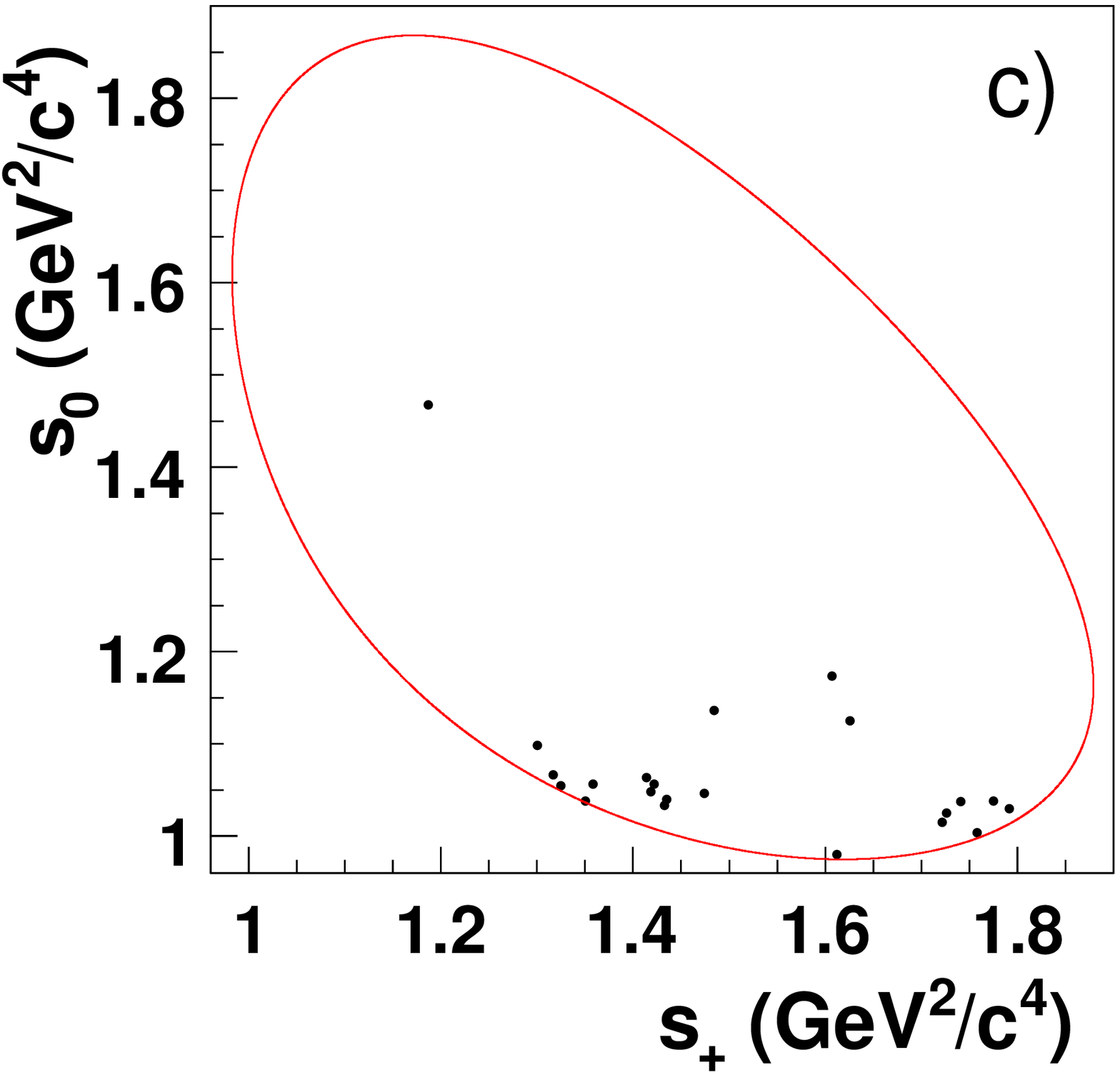}& 
\includegraphics[width=0.24\textwidth]{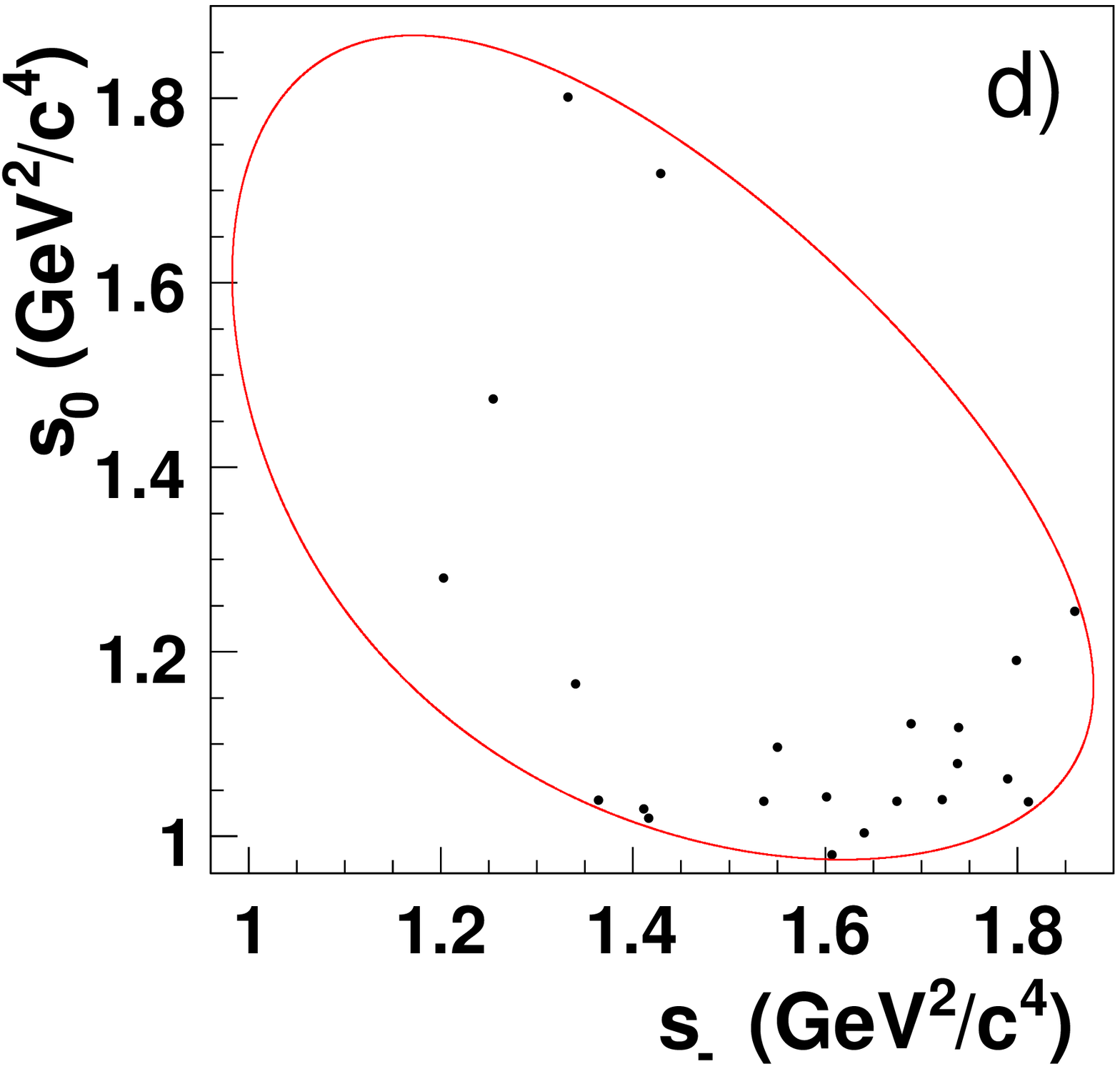} \\ 
\includegraphics[width=0.24\textwidth]{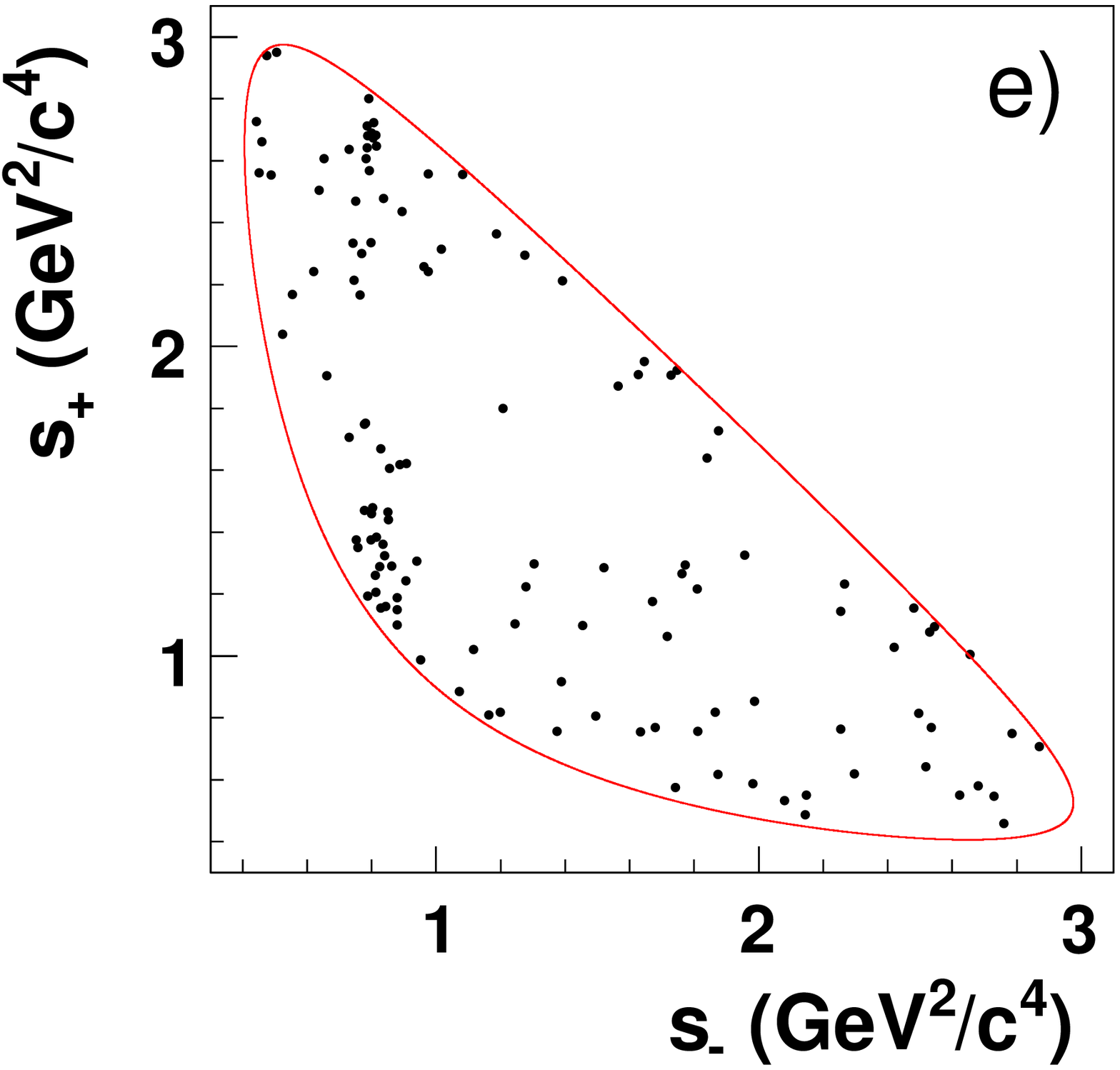}& 
\includegraphics[width=0.24\textwidth]{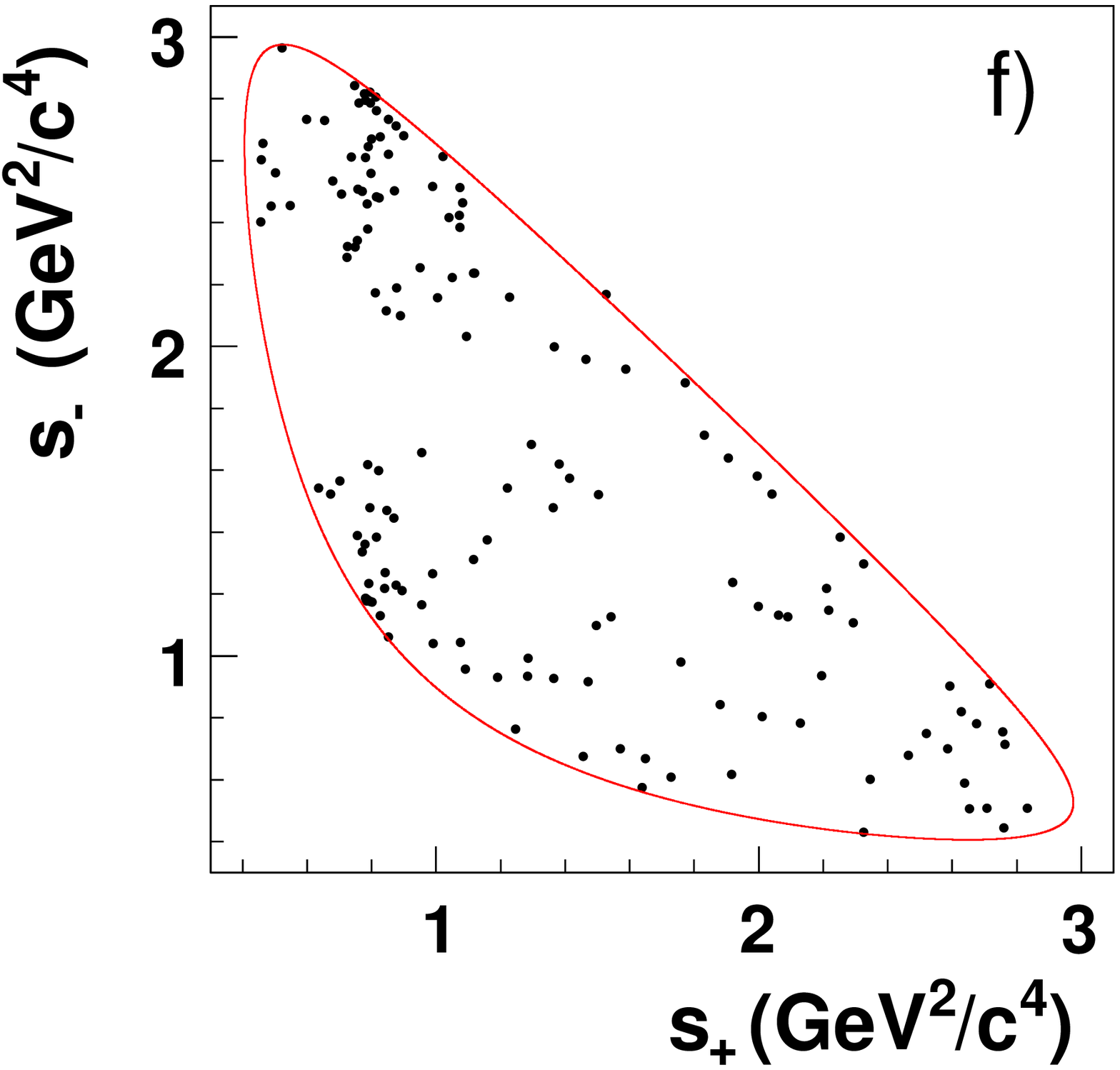}& 
\includegraphics[width=0.24\textwidth]{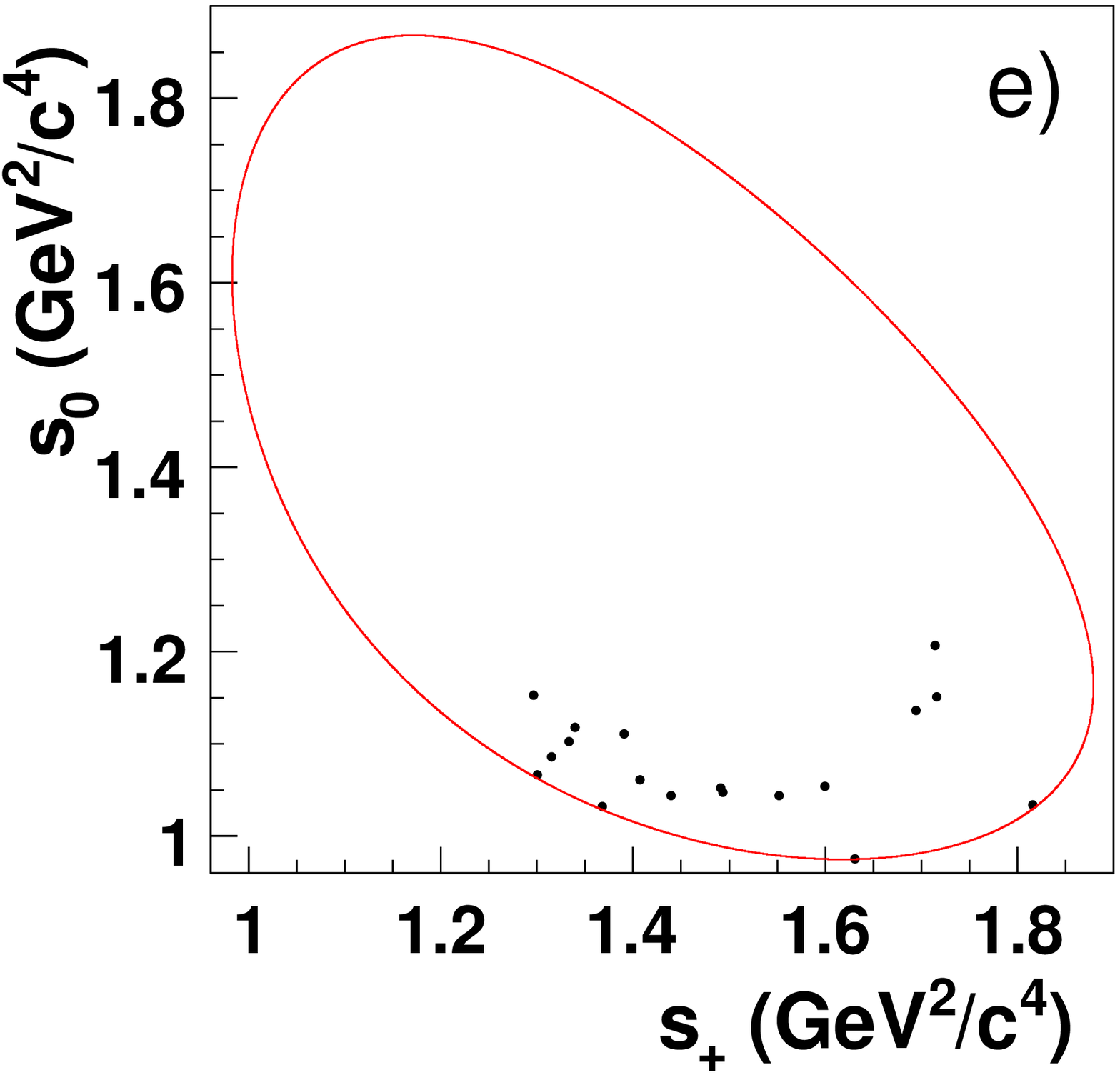}& 
\includegraphics[width=0.24\textwidth]{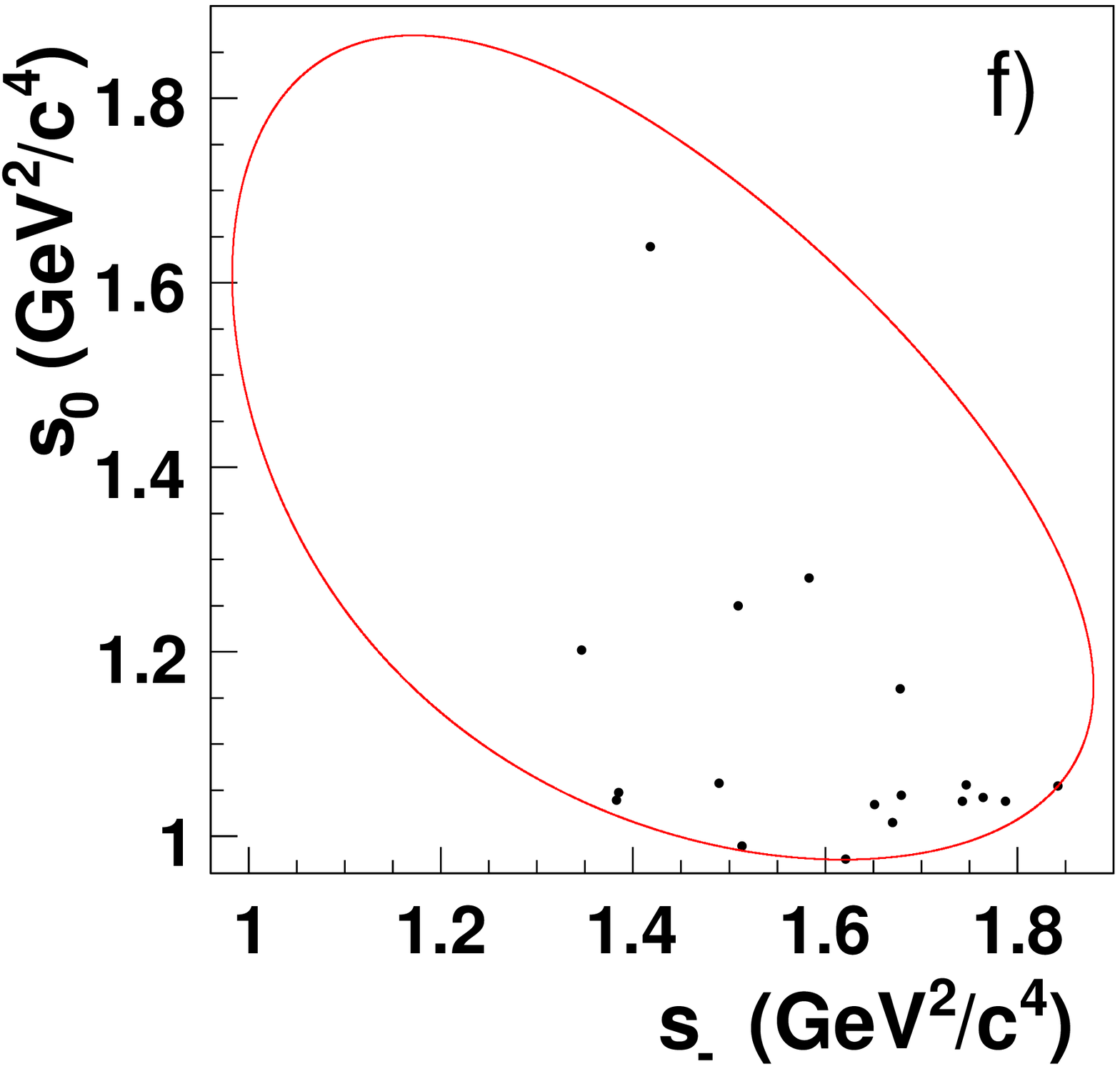} \\ 
\includegraphics[width=0.24\textwidth]{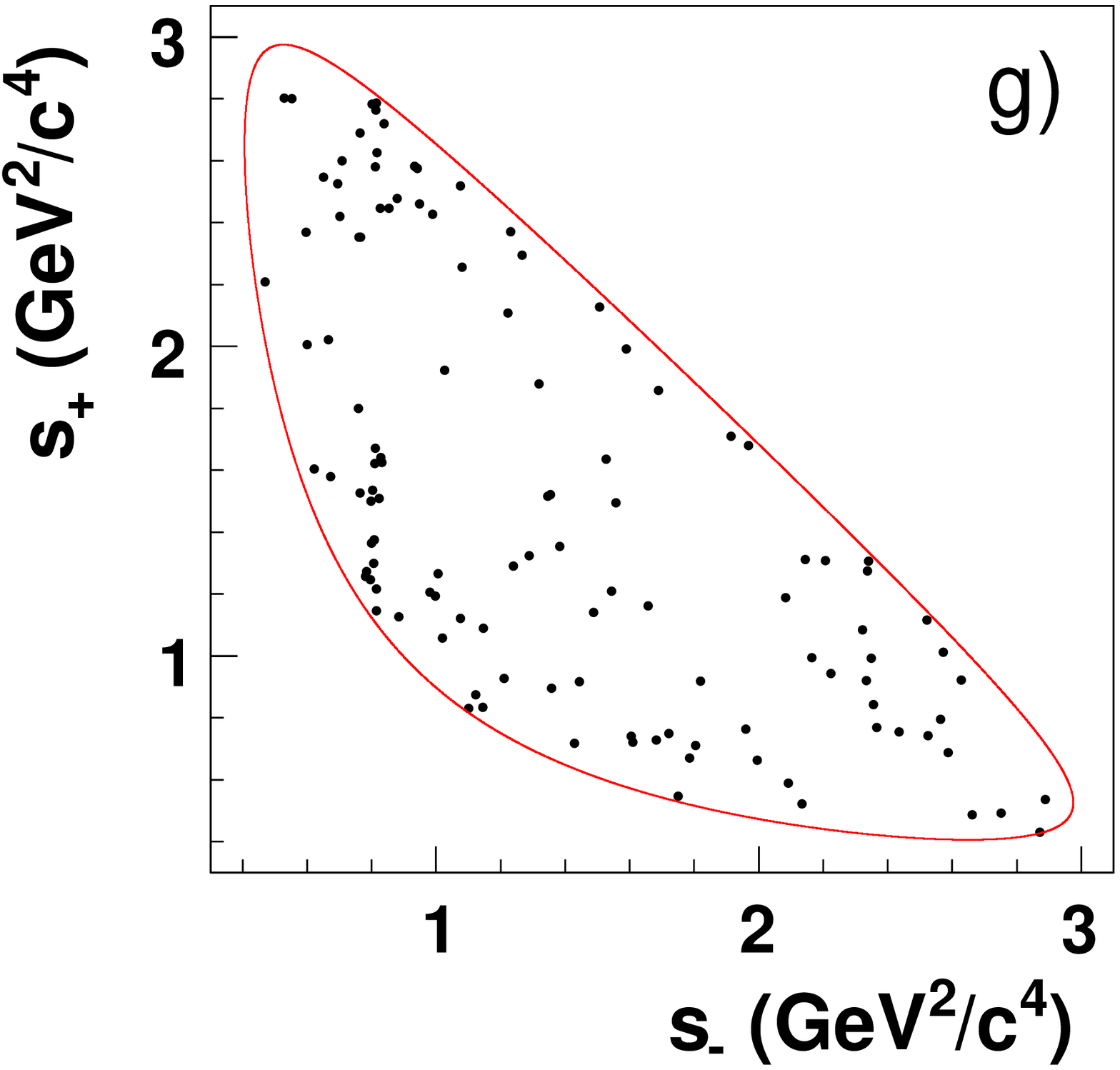}& 
\includegraphics[width=0.24\textwidth]{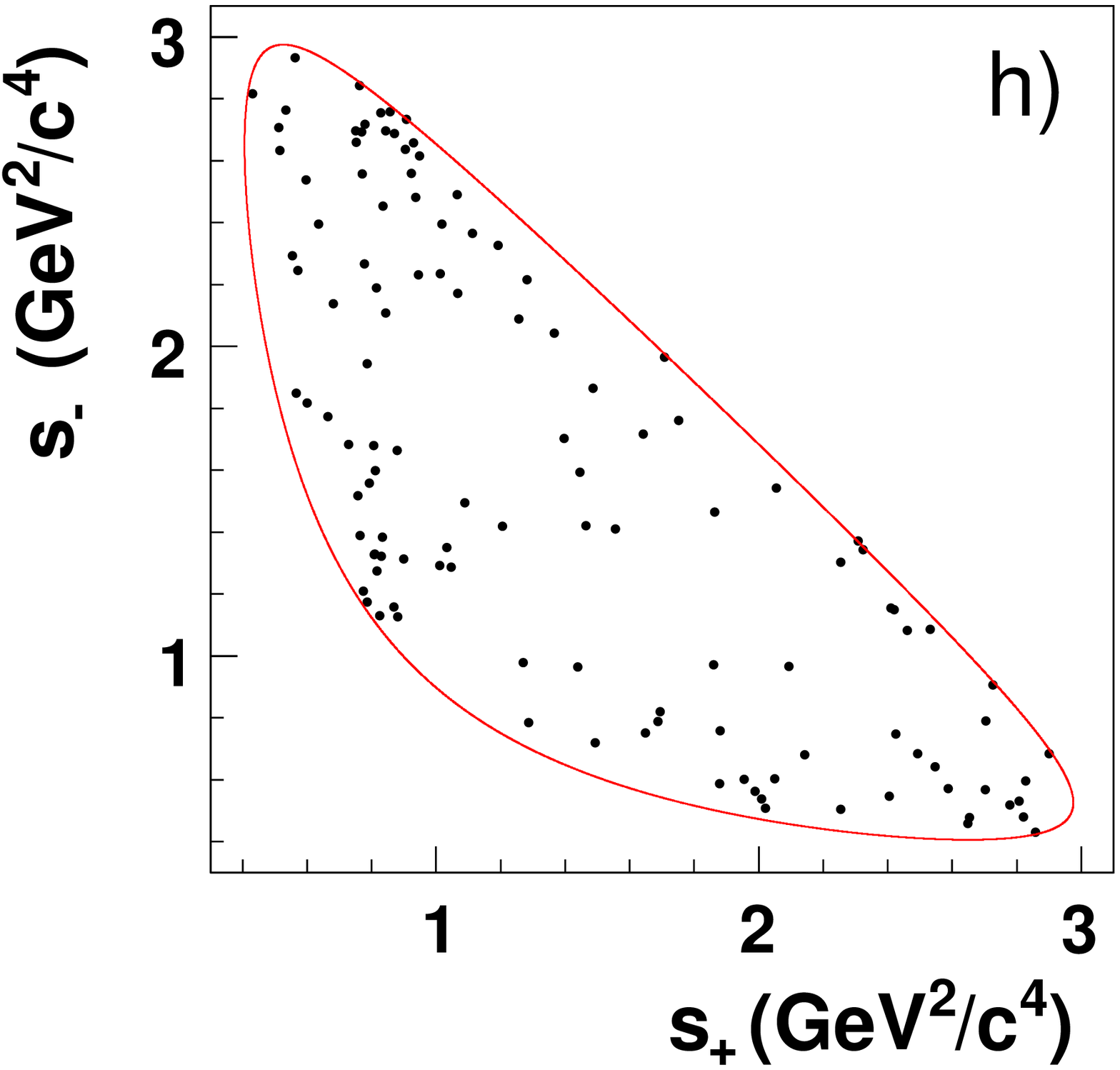}& 
\includegraphics[width=0.24\textwidth]{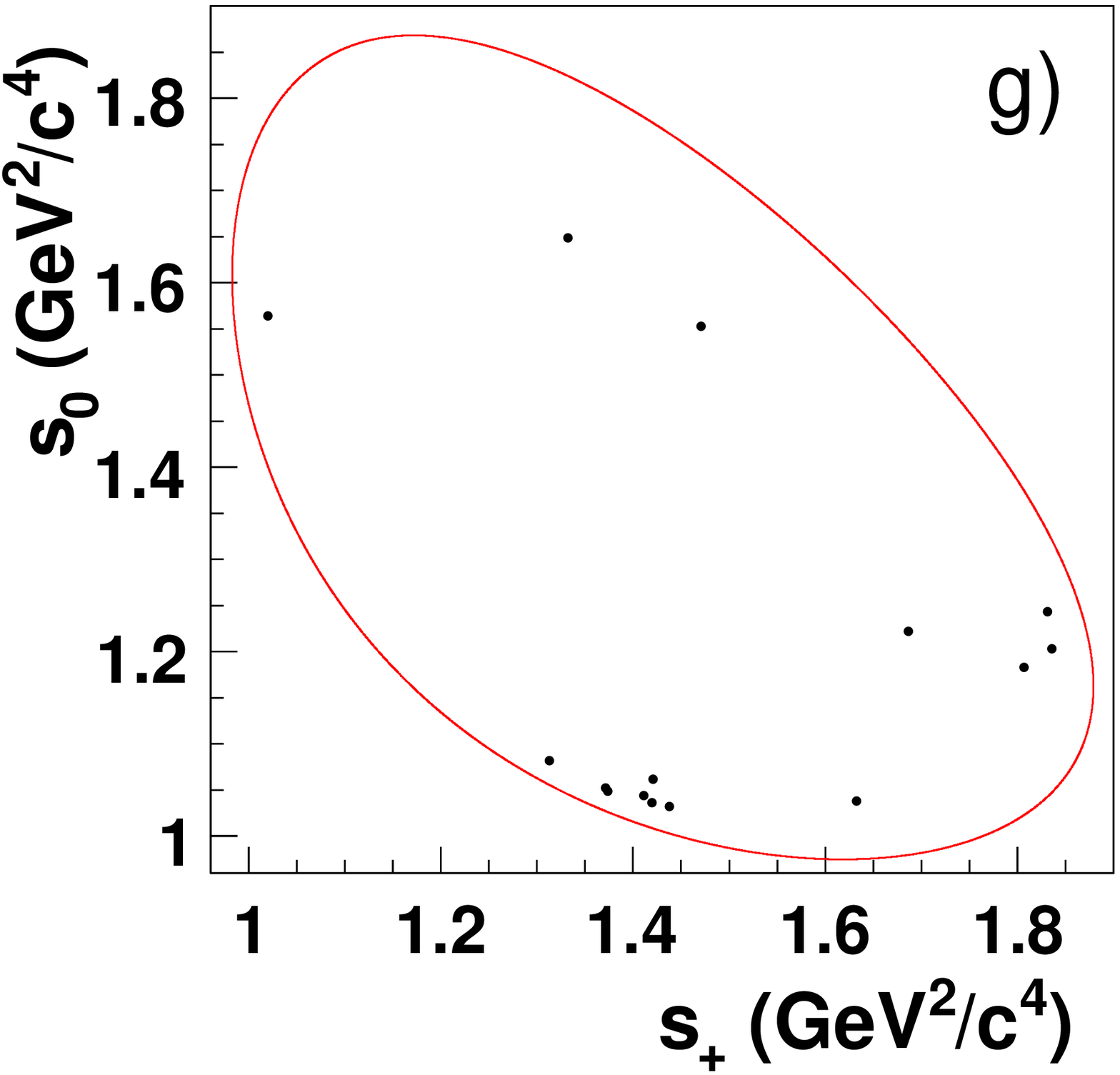}& 
\includegraphics[width=0.24\textwidth]{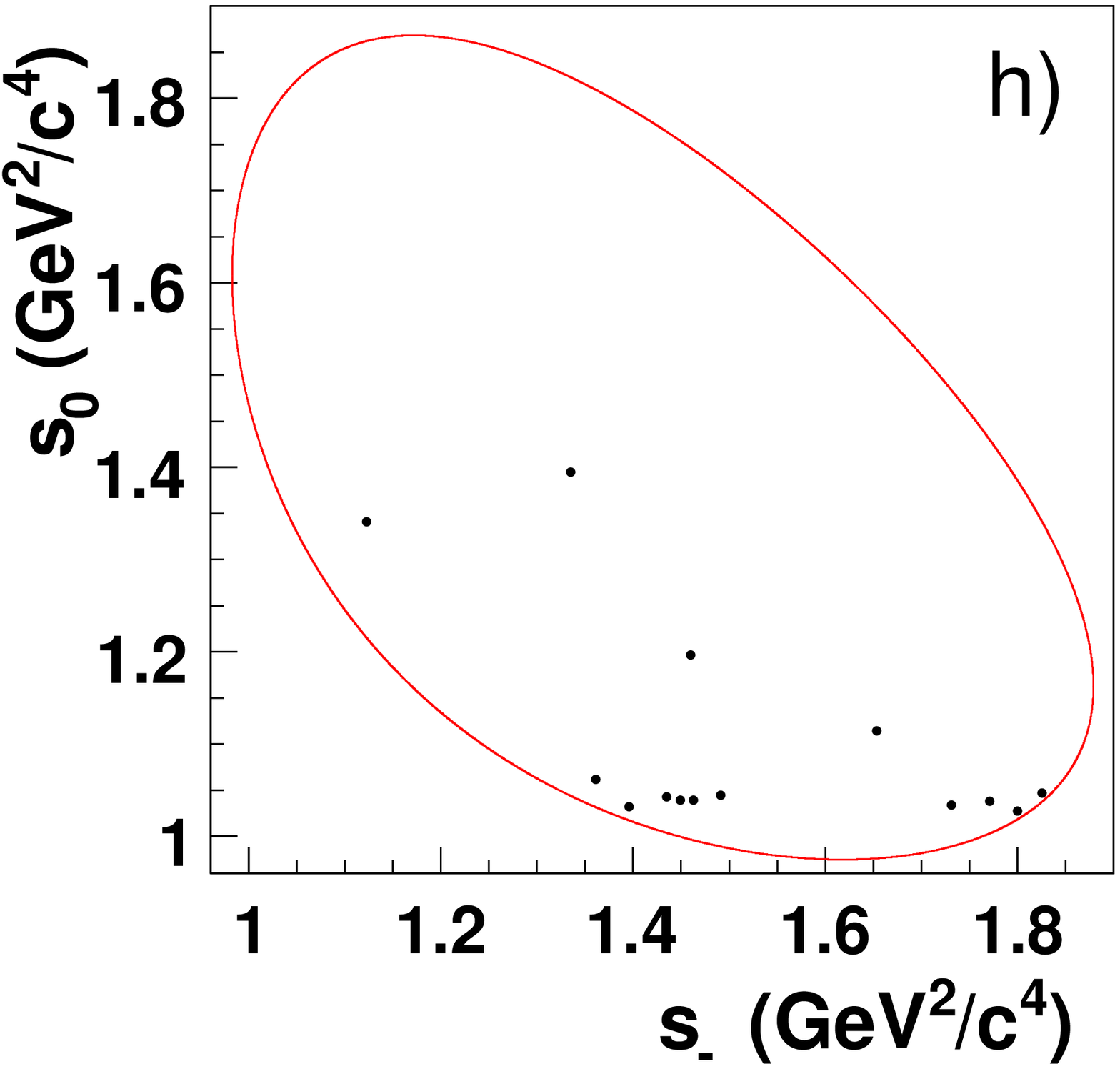} \\ 
\end{tabular} 
\caption{\label{fig:btodk_dp} (color online).  
The DP distributions for  
(a)(b) $\Bmp\to \D \Kmp$, 
(c)(d) $\Bmp\to \Dstar[\D\piz] \Kmp$, 
(e)(f) $\Bmp\to \Dstar[\D\gamma] \Kmp$, and 
(g)(h) $\Bmp\to \D \Kstarmp$ decays, with \Dtokspipi (left panel) and \Dtokskk (right panel). 
The distributions are for events in the signal region defined through the requirements 
$\mes>5.272$~\gevcc, $|\de|<30$~\mev, and $\fisher>-0.1$,  
after all the selection criteria are applied, 
and are shown separately for \Bm (first and third columns) and \Bp (second and last column) decays. 
 For \Bm and \Bp decays the variables \sminus and \splus are interchanged. 
The contours (solid red lines) represent the kinematical limits of the \D decay. 
} 
\end{figure}

\newpage 
\onecolumngrid

\begin{table}[hbt!] 
\caption{\label{tab:cartesian-exp-syst}  
Summary of the main contributions to the experimental systematic error on the \CP parameters. 
All contributions have been evaluated using the same procedure as in our previous analysis~\cite{ref:babar_dalitzpub2008}. 
The statistical contribution to the total error has been decreased, as consequence of the use of larger data and Monte Carlo (with full detector simulation) samples.  
For example, larger simulated continuum samples help to significantly reduce the uncertainty arising from the modeling of the DP distributions for  
background events containing misreconstructed \D mesons. 
} 
\begin{ruledtabular} 
\begin{tabular}{lcccccccccccc} 
\\[-0.15in] 
 Source                               & \xbm  & \ybm  & \xbp  & \ybp   & \xbstm & \ybstm & \xbstp & \ybstp & \xsm  & \ysm  & \xsp  & \ysp \\   [0.01in] \hline 
 \mes, \DeltaE, \fisher shapes        & 0.001 & 0.001 & 0.001 & 0.001 & 0.004 & 0.006 & 0.008 & 0.004 & 0.006 & 0.003 & 0.004 & 0.002 \\ % a+b+c+d+e+f+g+h+l 
 Real \Dz\ fractions                  & 0.002 & 0.001 & 0.001 & 0.001 & 0.003 & 0.003 & 0.002 & 0.002 & 0.004 & 0.001 & 0.001 & 0.001 \\ % i 
 Charge-flavor correlation            & 0.003 & 0.003 & 0.002 & 0.001 & 0.005 & 0.005 & 0.008 & 0.002 & 0.001 & 0.001 & 0.003 & 0.001 \\ % j+k 
 Efficiency in the DP                 & 0.003 & 0.001 & 0.003 & 0.001 & 0.001 & 0.001 & 0.001 & 0.001 & 0.003 & 0.001 & 0.002 & 0.001 \\ % o 
 Background DP distributions          & 0.005 & 0.002 & 0.005 & 0.003 & 0.003 & 0.002 & 0.004 & 0.004 & 0.010 & 0.004 & 0.007 & 0.002 \\ % m+n 
 $B^-\to D^{*0}K^-$ cross-feed        & --    & --    & --    & --    & 0.002 & 0.003 & 0.009 & 0.002 & --    & --    & --    & --    \\ % q 
% Invariant mass resolution           & 0.001 & 0.001 & 0.001 & 0.001 & 0.002 & 0.001 & 0.001 & 0.003 & 0.001 & 0.001 & 0.001 & 0.001 \\ 
 \CP violation in $D\pi$ and \BB      & 0.002 & 0.001 & 0.001 & 0.001 & 0.017 & 0.001 & 0.008 & 0.004 & 0.017 & 0.002 & 0.011 & 0.001 \\ % u+v 
 Non-\Kstar $\Bm\to\D\KS\pim$ decays    & --    & --    & --    & --     & --     & --     & --     & --     & 0.020 & 0.026 & 0.025 & 0.036 \\ % y+z 
 \hline 
 Total experimental                   & 0.007 & 0.004 & 0.006 & 0.004 & 0.019 & 0.009 & 0.017 & 0.008 & 0.029 & 0.027 & 0.029 & 0.036 \\ 
\end{tabular} 
\end{ruledtabular} 
\end{table}

\begin{table}[hbt!] 
\caption{\label{tab:cartesian-model-syst}  
Summary of the main contributions to the \Dz decay amplitude model systematic uncertainty on the \CP parameters. 
We evaluate the different contributions using a similar, but not identical, procedure to that adopted in our previous analysis~\cite{ref:babar_dalitzpub2008}. 
The reference \Dz decay amplitude models and parameters are used to generate 10 data-sized signal samples  
of pseudo-experiments of \Dstarp\to\Dz\pip and \Dstarm\to\Dzb\pim events, and 10 \Bmp\to\DDstar\Kmp and \Bmp\to\D\Kstarmp 
signal samples 100 times larger than each measured signal yield in data, with \Dztokshh. 
The \CP parameters are generated with values in the range found in data.  
We then compare experiment-by-experiment the values of \zbzbstmp and \zsmp obtained from the \CP fits using the reference amplitude models and a set of  
alternative models obtained by repeating the \Dztokshh amplitude analyses on the pseudo-experiments with alternative assumptions~\cite{ref:dmixing-kshh}.  
This technique, although it requires large computing resources, helps to reduce statistical contributions to the amplitude model uncertainties arising from changes in  
sensitivity between alternative models (e.g. alternative K-matrix solutions and P-vector mass dependence in the $\pi\pi$ S-wave parameterization).  
A variety of studies using data have been performed to test the consistency of the results using this procedure 
with those obtained in our previous analysis, where the alternative models were obtained by repeating the \Dztokshh amplitude analyses on data.  
Nevertheless, the largest decrease in the amplitude model uncertainty compared to our previous result is a consequence of the improvements in the experimental 
analysis of tagged \D mesons~\cite{ref:dmixing-kshh}, which is reflected in smaller experimental systematic uncertainties on the \Dz decay amplitudes  
(variations of the reconstruction efficiency across the DP, modeling of the DP distributions for background events containing misreconstructed \D mesons, mistag rates, etc.), 
and thus smaller amplitude model uncertainties on the \CP parameters. 
} 
\begin{ruledtabular} 
\begin{tabular}{lcccccccccccc} 
\\[-0.15in] 
 Source                               & \xbm  & \ybm  & \xbp  & \ybp   & \xbstm & \ybstm & \xbstp & \ybstp & \xsm  & \ysm  & \xsp  & \ysp \\   [0.01in] \hline 
 Mass and width of Breit-Wigner's     & 0.001 & 0.001 & 0.001 & 0.002 & 0.001 & 0.002 & 0.001 & 0.002 & 0.001 & 0.002 & 0.001 & 0.002 \\  
 $\pi\pi$ S-wave parameterization     & 0.001 & 0.001 & 0.001 & 0.001 & 0.001 & 0.001 & 0.001 & 0.002 & 0.001 & 0.001 & 0.001 & 0.002 \\ 
 $\K\pi$ S-wave parameterization      & 0.001 & 0.004 & 0.003 & 0.008 & 0.001 & 0.006 & 0.002 & 0.004 & 0.003 & 0.002 & 0.003 & 0.007 \\ 
 Angular dependence                   & 0.001 & 0.001 & 0.002 & 0.001 & 0.001 & 0.001 & 0.001 & 0.002 & 0.002 & 0.001 & 0.002 & 0.001 \\ 
 Blatt-Weisskopf radius               & 0.001 & 0.001 & 0.001 & 0.001 & 0.001 & 0.001 & 0.001 & 0.001 & 0.001 & 0.001 & 0.002 & 0.001 \\ 
 Add/remove resonances                & 0.001 & 0.001 & 0.001 & 0.001 & 0.001 & 0.002 & 0.001 & 0.001 & 0.001 & 0.001 & 0.001 & 0.002 \\ 
 DP efficiency                        & 0.003 & 0.002 & 0.003 & 0.001 & 0.001 & 0.001 & 0.001 & 0.001 & 0.004 & 0.002 & 0.003 & 0.001 \\ 
 Background DP shape                  & 0.001 & 0.001 & 0.001 & 0.001 & 0.001 & 0.001 & 0.001 & 0.001 & 0.001 & 0.001 & 0.001 & 0.001 \\ 
 Mistag rate                          & 0.003 & 0.003 & 0.002 & 0.001 & 0.001 & 0.001 & 0.001 & 0.001 & 0.003 & 0.003 & 0.001 & 0.001 \\ 
 Effect of mixing                     & 0.003 & 0.001 & 0.003 & 0.001 & 0.001 & 0.001 & 0.001 & 0.001 & 0.003 & 0.001 & 0.003 & 0.001 \\ 
 DP complex amplitudes                & 0.001 & 0.001 & 0.001 & 0.002 & 0.001 & 0.001 & 0.001 & 0.002 & 0.002 & 0.001 & 0.001 & 0.002 \\  
\hline 
 Total \Dz decay amplitude model      & 0.006 & 0.006 & 0.007 & 0.009 & 0.002 & 0.007 & 0.003 & 0.006 & 0.007 & 0.006 & 0.006 & 0.008 \\ 
\end{tabular} 
\end{ruledtabular} 
\end{table}

\begin{table}[!htb] 
\caption{\label{tab:xyresults-kspipi-kskk-only} 
\CP-violating complex parameters  
$\zbzbstmp = \xbxbstmp + i \ybybstmp$ 
and  
$\zsmp = \xsmp + i \ysmp$ 
as obtained from the \CP fit to \kspipi and \kskk final states separately. 
The first error is statistical, the second is the experimental systematic uncertainty and the third is  
the systematic uncertainty associated with the \Dz decay amplitude models. 
These results yield for the weak phase  
$\g = \left( 61^{+19}_{-17} \right)^\circ \ \{3,3\}^\circ$  
and  
$\g = \left( 87^{+43}_{-37} \right)^\circ \ \{8,3\}^\circ$,  
respectively. 
} 
\begin{center} 
\begin{ruledtabular} 
\begin{tabular}{lrrrr} 
           & \multicolumn{2}{c}{\kspipi} & \multicolumn{2}{c}{\kskk} \\ %\hline 
           & Real part (\%)\phzz\phzz\phz &  Imaginary part (\%)\phzz & 
             Real part (\%)\phzz\phzz\phz &  Imaginary part (\%)\phzz \\ [0.025in] \hline  
 $\zbm$    & $\phz\phm3.6\pm\phz4.6\pm0.9\pm0.6\phzz$    & $\phz\phm6.7\pm\phz4.9\pm0.4\pm0.6\phz$ & 
             $\phz\phm12.6\pm\phz7.6\pm1.5\pm0.5\phzz$   & $\phz\phm4.4\pm11.7\pm\phz2.3\pm1.2\phz$ \\ 
 $\zbp$    & $ -8.3\pm\phz4.1\pm0.7\pm0.8\phzz$          & $\phz-0.8\pm\phz4.9\pm0.4\pm1.0\phz$ & 
             $-19.0\pm\phz8.7\pm2.2\pm0.5\phzz$          & $\phz-2.0\pm18.8\pm\phz6.0\pm1.5\phz$ \\ 
 $\zbstm$  & $ -8.9\pm\phz5.8\pm1.7\pm0.2\phzz$          & $\phz-7.1\pm\phz6.8\pm0.9\pm0.7\phz$ &  
             $-17.0\pm11.0\pm2.0\pm0.4\phzz$             & $\phz8.8\pm17.1\pm\phz3.6\pm1.2\phz$ \\ 
 $\zbstp$  & $\phm15.4\pm\phz5.9\pm1.4\pm0.4\phzz$       & $\phz-3.6\pm\phz8.7\pm0.9\pm0.7\phz$ & 
             $\phm11.7\pm11.7\pm4.2\pm0.4\phzz$          & $\phz-2.5\pm16.4\pm\phz1.9\pm0.5\phz$ \\ 
 $\zsm$    & $\phm12.8\pm10.5\pm3.4\pm0.8\phzz$          & $\phm12.0\pm10.5\pm2.5\pm0.6\phz$ & 
             $\phz-11.7\pm20.8\pm8.2\pm0.8\phzz$         & $\phm14.3\pm22.4\pm\phz9.5\pm1.4\phz$ \\ 
 $\zsp$    & $-9.6\pm\phz9.2\pm3.2\pm0.8\phzz$           & $\phz\phm3.8\pm10.9\pm3.7\pm0.9\phz$ & 
             $-36.6\pm20.1\pm5.8\pm0.6\phzz$             & $-17.1\pm39.9\pm13.5\pm1.7\phz$ \\ 
\end{tabular} 
\end{ruledtabular} 
\end{center} 
\end{table}

\begin{table}[!htb] 
\caption{\label{tab:rhostat}  
Statistical correlation coefficients for the vector \zvec of measurements, 
(in order) \xbm, \ybm, \xbp, \ybp, \xbstm, \ybstm, \xbstp, \ybstp, \xsm, \ysm, \xsp, \ysp,  
as obtained from the \CP fit to \kspipi and \kskk final states (upper panel), 
and to \kspipi (bottom left panel) and \kskk (bottom right panel) separately. 
Only lower off-diagonal terms are written, in \%. 
} 
\begin{center} 
% Kspipi+KsKK 
$\left( 
\begin{tabular}{rrrrrrrrrrrrr} 
100 \\ 
 -2 & 100 \\ 
 0 & 0 & 100 \\ 
 0 & 0 & 6 & 100 \\ 
 0 & 0 & 0 & 0 & 100 \\ 
 0 & 0 & 0 & 0 & -1 & 100 \\ 
 0 & 0 & 0 & 0 & 0 & 0 & 100 \\ 
 0 & 0 & 0 & 0 & 0 & 0 & -3 & 100 \\ 
 0 & 0 & 0 & 0 & 0 & 0 & 0 & 0 & 100 \\ 
 0 & 0 & 0 & 0 & 0 & 0 & 0 & 0 & -20 & 100 \\ 
 0 & 0 & 0 & 0 & 0 & 0 & 0 & 0 & 0 & 0 & 100 \\ 
 0 & 0 & 0 & 0 & 0 & 0 & 0 & 0 & 0 & 0 & 10 & 100 \\ 
\end{tabular} 
\right)$ 
\\ 
% Kspipi 
$\left( 
\begin{tabular}{rrrrrrrrrrrrr} 
100 \\ 
 -4 & 100 \\ 
 0 & 0 & 100 \\ 
 0 & 0 & 4 & 100 \\ 
 0 & 0 & 0 & 0 & 100 \\ 
 0 & 0 & 0 & 0 & 4 & 100 \\ 
 0 & 0 & 0 & 0 & 0 & 0 & 100 \\ 
 0 & 0 & 0 & 0 & 0 & 0 & -7 & 100 \\ 
 0 & 0 & 0 & 0 & 0 & 0 & 0 & 0 & 100 \\ 
 0 & 0 & 0 & 0 & 0 & 0 & 0 & 0 & -18 & 100 \\ 
 0 & 0 & 0 & 0 & 0 & 0 & 0 & 0 & 0 & 0 & 100 \\ 
 0 & 0 & 0 & 0 & 0 & 0 & 0 & 0 & 0 & 0 & 10 & 100 \\ 
\end{tabular} 
\right)$ 
% KsKK 
$\left( 
\begin{tabular}{rrrrrrrrrrrrr} 
100 \\ 
 5 & 100 \\ 
 0 & 0 & 100 \\ 
 0 & 0 & 13 & 100 \\ 
 0 & 0 & 0 & 0 & 100 \\ 
 0 & 0 & 0 & 0 & -10 & 100 \\ 
 0 & 0 & 0 & 0 & 2 & 0 & 100 \\ 
 0 & 0 & 0 & 0 & 0 & 0 & 15 & 100 \\ 
 0 & 0 & 0 & 0 & 0 & 0 & 0 & 0 & 100 \\ 
 0 & 0 & 0 & 0 & 0 & 0 & 0 & 0 & -19 & 100 \\ 
 0 & 0 & 0 & 0 & 0 & 0 & 0 & 0 & 0 & 0 & 100 \\ 
 0 & 0 & 0 & 0 & 0 & 0 & 0 & 0 & 0 & 0 & -7 & 100 \\ 
\end{tabular} 
\right)$ 
\end{center} 
\end{table}

\begin{table}[!htb] 
\caption{\label{tab:rhoexp}  
Experimental systematic correlation coefficients for the vector \zvec of measurements, 
(in order) \xbm, \ybm, \xbp, \ybp, \xbstm, \ybstm, \xbstp, \ybstp, \xsm, \ysm, \xsp, \ysp,  
defined as 
$\rho_{ij} = {\cal C}_{ij}/\sqrt{{\cal C}_{ii}{\cal C}_{jj}}$, where 
${\cal C}_{ij} = \overline{(\zvec-\zvecbest)_i (\zvec-\zvecbest)_j}$, with  
\zvecbest the vector of best measurements, 
for \kspipi and \kskk final states (upper panel), 
and \kspipi (bottom left panel) and \kskk (bottom right panel) separately. 
Only lower off-diagonal terms are written, in \%. 
} 
\begin{center} 
% Kspipi+KsKK 
$\left( 
\begin{tabular}{rrrrrrrrrrrrr} 
100 \\ 
-23 & 100 \\ 
-2 & 18 & 100 \\ 
-5 & -5 & 0 & 100 \\ 
-32 & -10 & -13 & 5 & 100 \\ 
-15 &   13 &  -2  & -5 & -2 & 100 \\ 
-12 & -9 & -9 & 5 & 44 & -60 & 100 \\ 
12 & 17 & 16 & -9 & -50 & 19 & -44 & 100 \\ 
-19 & -8 & -15 & -7 & 49 & 0 & 25 & -25 & 100 \\ 
3 & -6 & -3 & -3 & -8 & -5 & -2 & 5 & 63 & 100 \\ 
-20 & 0 & -7 & 0 & 33 & 0 & 16 & -14 & 73 & 65 & 100 \\ 
-1 & -6 & -3 & 0 & -3 & -4 & -3 & 2 & 66 & 97 & 74 & 100 \\ 
\end{tabular} 
\right)$ 
\\ 
% Kspipi 
$\left( 
\begin{tabular}{rrrrrrrrrrrrr} 
100 \\ 
13 & 100 \\ 
-3 & -8 & 100 \\ 
-9 & -6 & 2 & 100 \\ 
-28 & -27 & 16 & 10 & 100 \\ 
 3 &  2 &  -5  & -10 & 3 & 100 \\ 
-4 & -1 & 10 & 5 & 16 & -72 & 100 \\ 
16 & 7 & -12 & -16 & -48 & 39 & -57 & 100 \\ 
-11 & -9 & 5 & -6 & 40 & -14 & 22 & -29 & 100 \\ 
2 & -1 & -4 & -6 & -16 & -4 & -5 & 2 & 64 & 100 \\ 
-11 & -12 & 4 & 0 & 28 & -13 & 10 & -21 & 74 & 70 & 100 \\ 
-2 & -6 & -2 & -3 & -3 & -6 & -3 & -3 & 71 & 96 & 82 & 100 \\ 
\end{tabular} 
\right)$ 
% KsKK 
$\left( 
\begin{tabular}{rrrrrrrrrrrrr} 
100 \\ 
-72 & 100 \\ 
-78 & 59 & 100 \\ 
-91 & 78 & 88 & 100 \\ 
-41 & 30 & 17 & 34 & 100 \\ 
 10 & -11 &  -12  & -5 & -52 & 100 \\ 
-2 & -11 & -7 & -1 & 60 & -52 & 100 \\ 
-20 & 47 & 1 & 19 & 44 & -42 & 46 & 100 \\ 
-13 & 5 & -8 & 3 & -1 & -2 & 3 & 9 & 100 \\ 
12 & -6 & 6 & -4 & 7 & 2 & 0 & -6 & -98 & 100 \\ 
-25 & 32 & 40 & 36 & 7 & 15 & -12 & 3 & -80 & 81 & 100 \\ 
16 & -21 & -31 & -26 & -16 & 0 & 2 & -5 & 85 & -85 & -95 & 100 \\ 
\end{tabular} 
\right)$ 
\end{center} 
\end{table}

\begin{table}[!htb] 
\caption{\label{tab:rhomodel}  
Amplitude model systematic correlation coefficients for the vector \zvec of measurements,  
(in order) \xbm, \ybm, \xbp, \ybp, \xbstm, \ybstm, \xbstp, \ybstp, \xsm, \ysm, \xsp, \ysp,  
defined as previously, 
for \kspipi and \kskk final states (upper panel), 
and \kspipi (bottom left panel) and \kskk (bottom right panel) separately. 
Only lower off-diagonal terms are written, in \%. 
} 
\begin{center} 
% Kspipi+KsKK 
$\left( 
\begin{tabular}{rrrrrrrrrrrrr} 
100 \\ 
9  & 100 \\ 
83 & -31   & 100   \\ 
25 & -56 &   57    & 100  \\ 
28 & -64 &   71 &  61  & 100   \\ 
6   & -74 &   44 &  93   & 64  & 100 \\ 
24 &   80 & -15 &  -74 &-50 & -86 & 100 \\ 
0 & 51 & -15 &-74 & -16 & -76 & 74  & 100 \\ 
86 & -23 & 91 & 54 & 52 & 39 & -12 & -22  & 100 \\ 
11 & 85 & -20 & -18 & -54 & -41 & 52 & 16 & -11 & 100 \\ 
77 & -38 & 96 & 61 & 73 & 50 & -23 & -18 & 91 & -26 & 100 \\ 
30 & -51 & 56 & 95 & 56 & 89 & -70 & -77 & 53 & -14 & 58 & 100 \\ 
\end{tabular} 
\right)$ 
\\ 
% Kspipi 
$\left( 
\begin{tabular}{rrrrrrrrrrrrr} 
100 \\ 
7  & 100 \\ 
83 & -33   & 100   \\ 
30 & -60 &   62    & 100  \\ 
31 & -61 &   71 &  61  & 100   \\ 
9   & -77 &   47 &  94   & 60  & 100 \\ 
22 &   84 & -19 &  -74 &-51 & -87 & 100 \\ 
-2 & 63 & -23 &-80 & -25 & -84 & 81  & 100 \\ 
86 & -31 & 94 & 64 & 59 & 49 & -18 & -32  & 100 \\ 
4 & 91 & -31 & -40 & -60 & -59 & 69 & 42 & -28 & 100 \\ 
75 & -47 & 96 & 72 & 76 & 61 & -35 & -36 & 93 & -43 & 100 \\ 
35 & -61 & 61 & 96 & 55 & 91 & -74 & -85 & 65 & -41 & 71 & 100 \\ 
\end{tabular} 
\right)$ 
% KsKK 
$\left( 
\begin{tabular}{rrrrrrrrrrrrr} 
100 \\ 
48  & 100 \\ 
22 & -36   & 100   \\ 
-33 & -48 &   23    & 100  \\ 
-36 & -82 &   53 &  56  & 100   \\ 
-52   & -84 &   38 &  57   & 88  & 100 \\ 
43 &   40 & -5 &  -31 &-41 & -47 & 100 \\ 
29 & 36 & -16 &-31 & -42 & -45 & 30  & 100 \\ 
59 & 49 & 6 & -33 & -42 & -53 & 37 & 26  & 100 \\ 
42 & 86 & -35 & -43 & -77 & -80 & 33 & 29 & 46 & 100 \\ 
75 & 66 & 4 & -44 & -57 & -70 & 49 & 39 & 62 & 59 & 100 \\ 
-49 & -69 & 33 & 56 & 81 & 83 & -48 & -49 & -48 & -60 & -66 & 100 \\ 
\end{tabular} 
\right)$ 
\end{center} 
\end{table}

\begin{figure}[htb!] 
\begin{tabular}{cc} 
\includegraphics[width=0.44\textwidth]{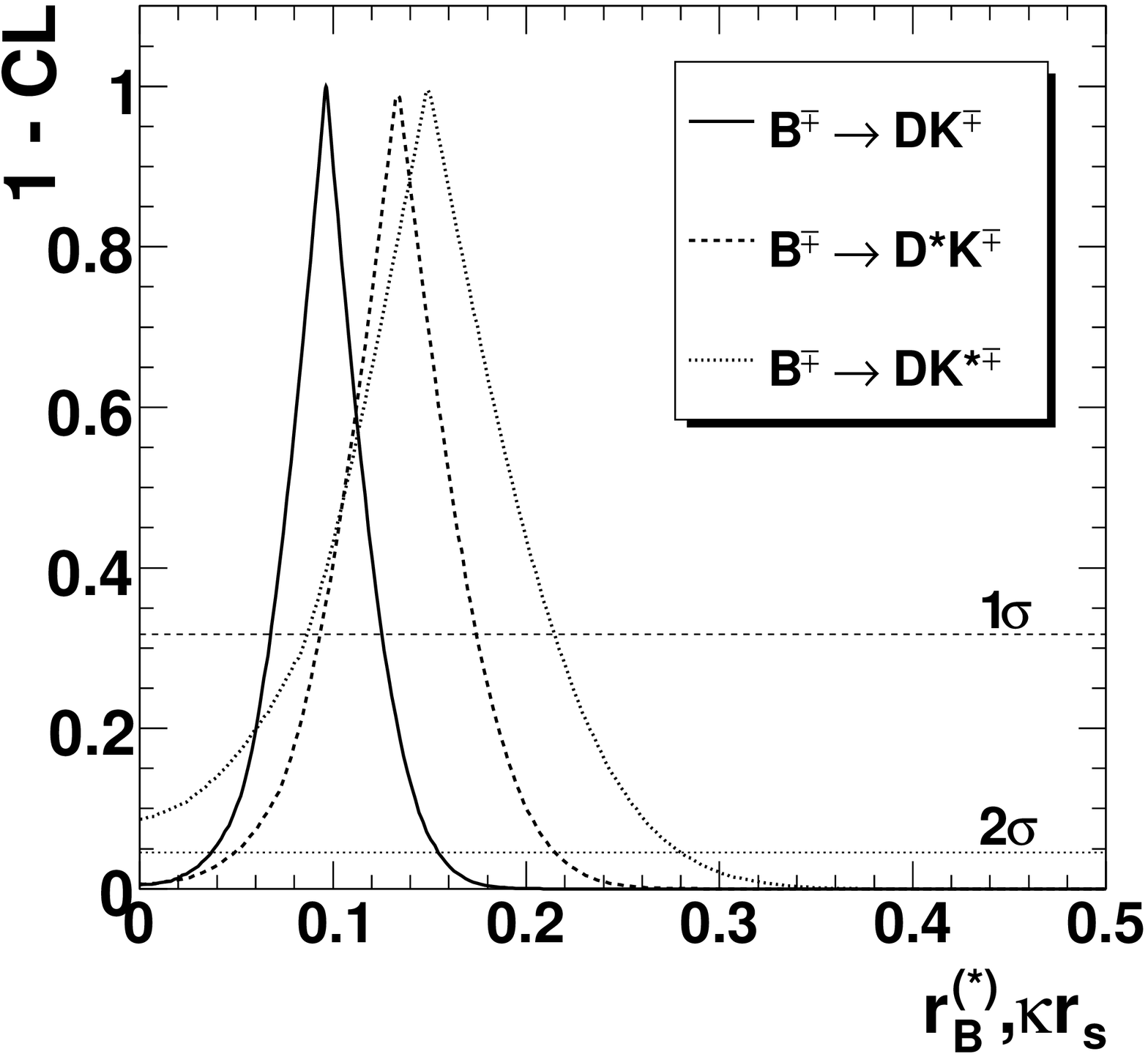} & 
\includegraphics[width=0.44\textwidth]{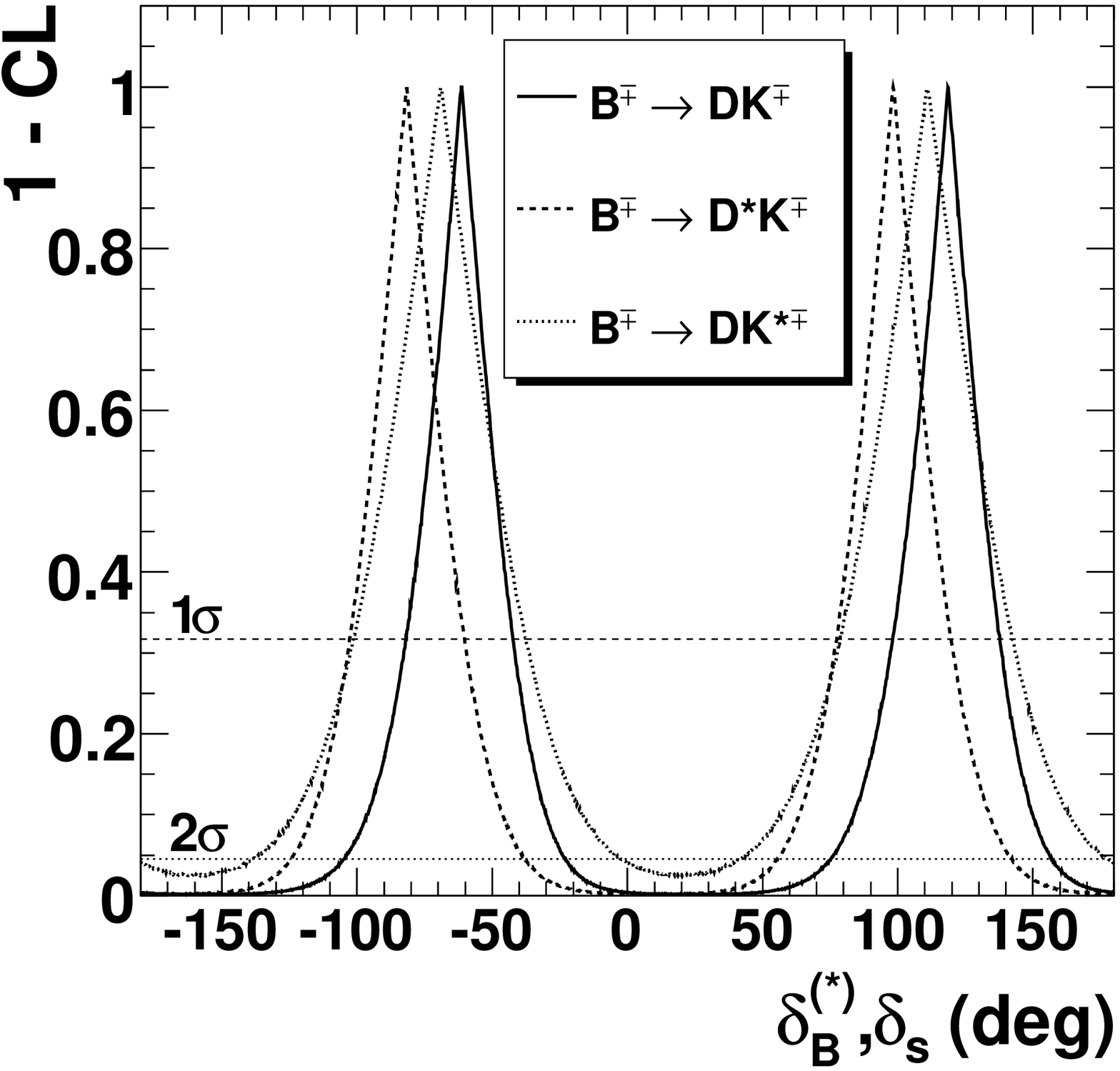} 
\end{tabular} 
\caption{\label{fig:scans-toymc-rb+delta} $1 - {\rm CL}$ as a function of (left panel) \rb, \rbst, and $\krs$,  
and (right panel) \deltab, \deltabst, and \deltas, for 
$\Bmp \to \D\Kmp$, $\Bmp \to \Dstar\Kmp$, and $\Bmp \to \D\Kstarmp$ decays, including statistical and 
systematic uncertainties. The dashed (upper) and dotted (lower) horizontal lines  
correspond to the one- and two-standard deviation intervals, respectively.} 
\end{figure}

% DK xy scans for Kspipi and KsKK only 
 
\begin{figure}[hbt!] 
\begin{tabular}{ccc} 
\includegraphics[width=0.24\textwidth]{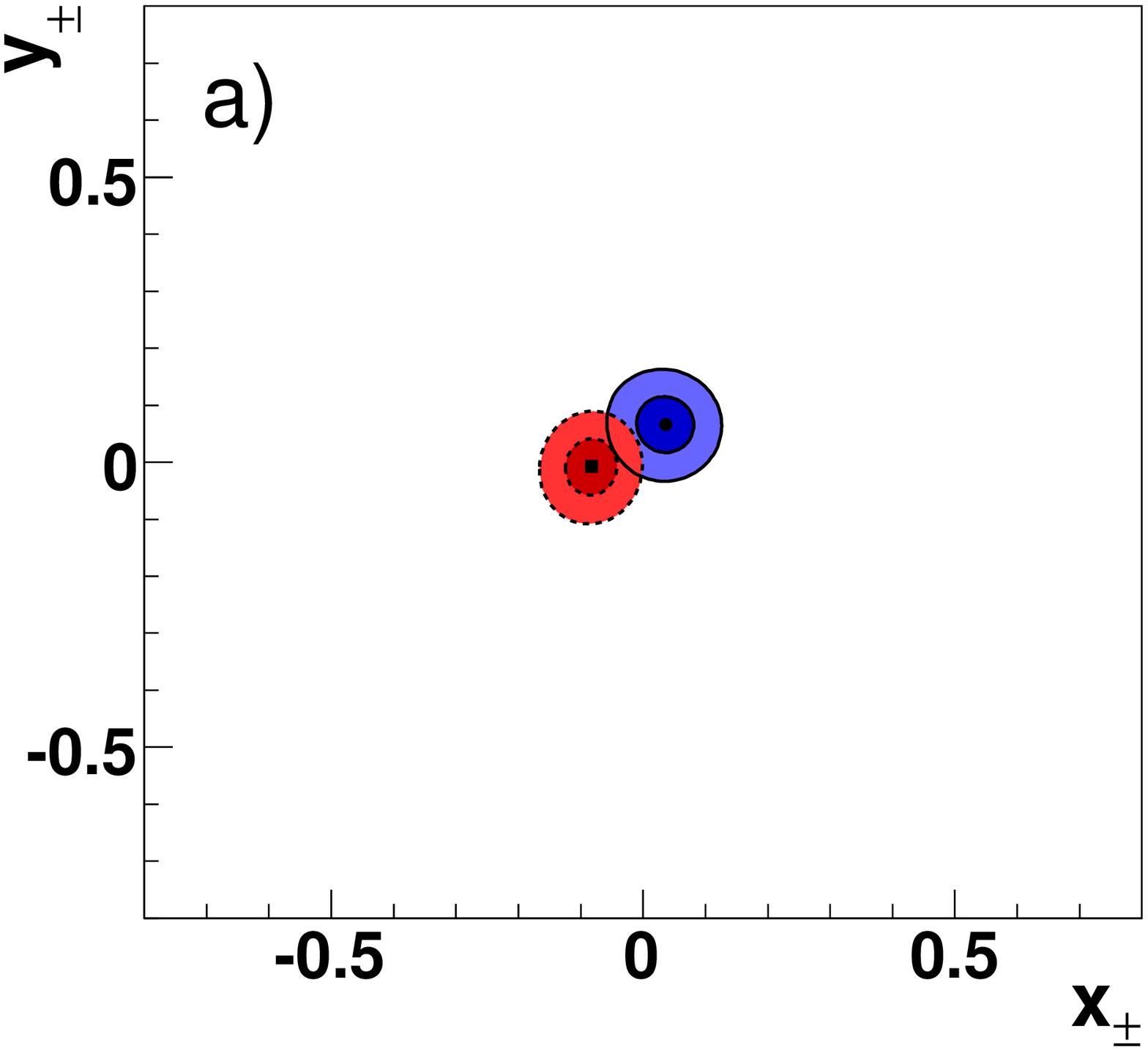}& 
\includegraphics[width=0.24\textwidth]{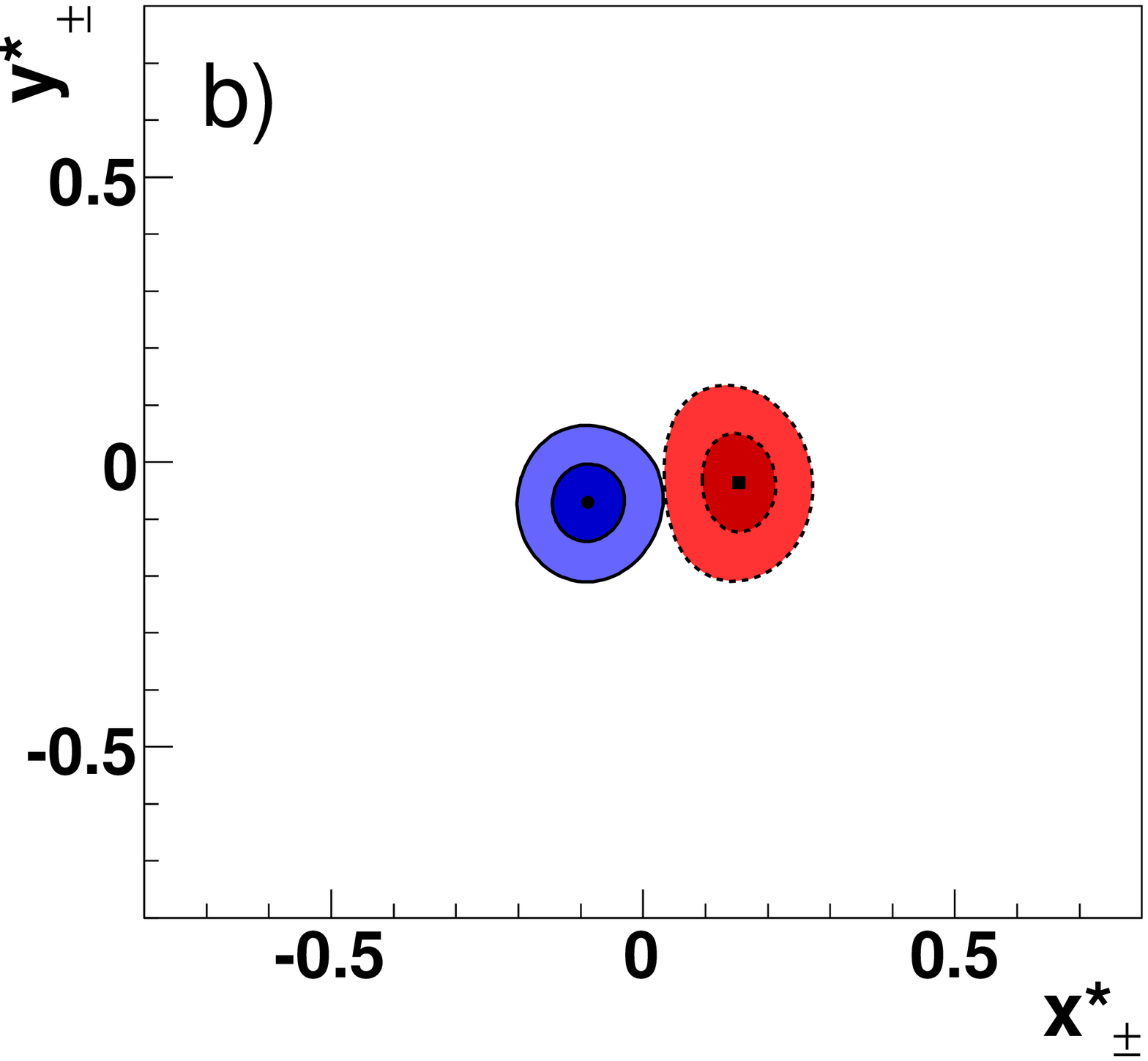} & 
\includegraphics[width=0.24\textwidth]{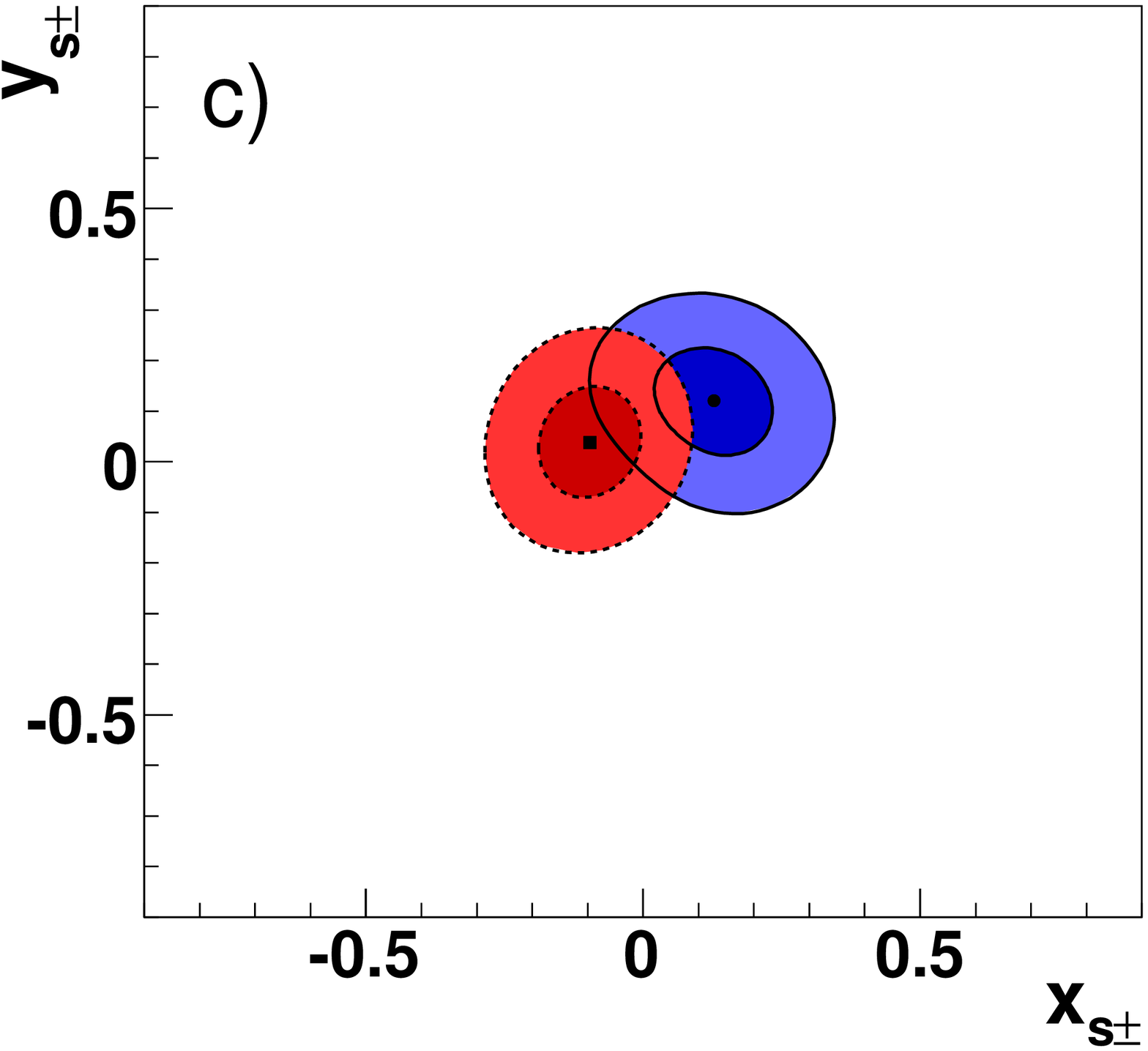} \\ 
\includegraphics[width=0.24\textwidth]{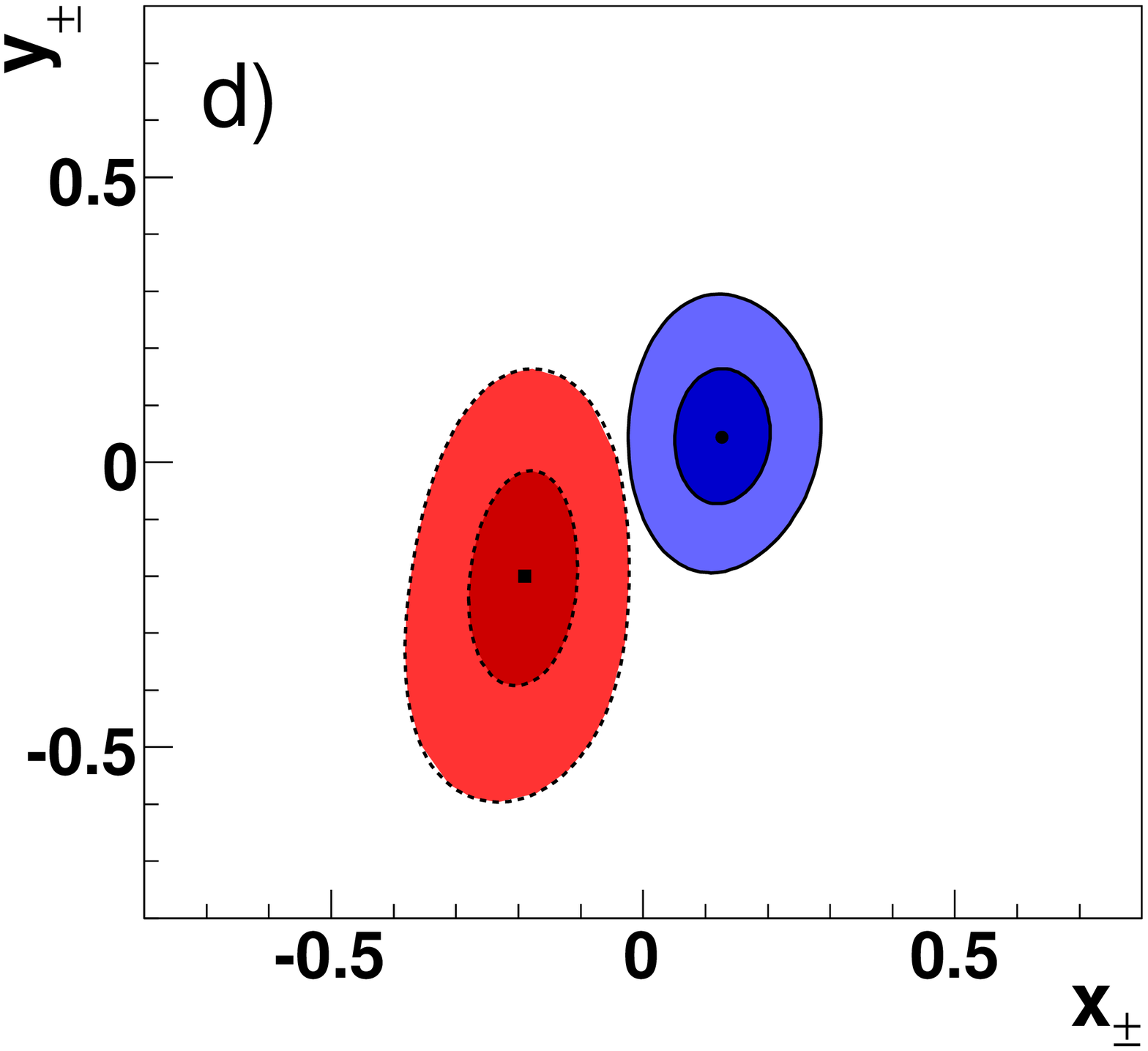}& 
\includegraphics[width=0.24\textwidth]{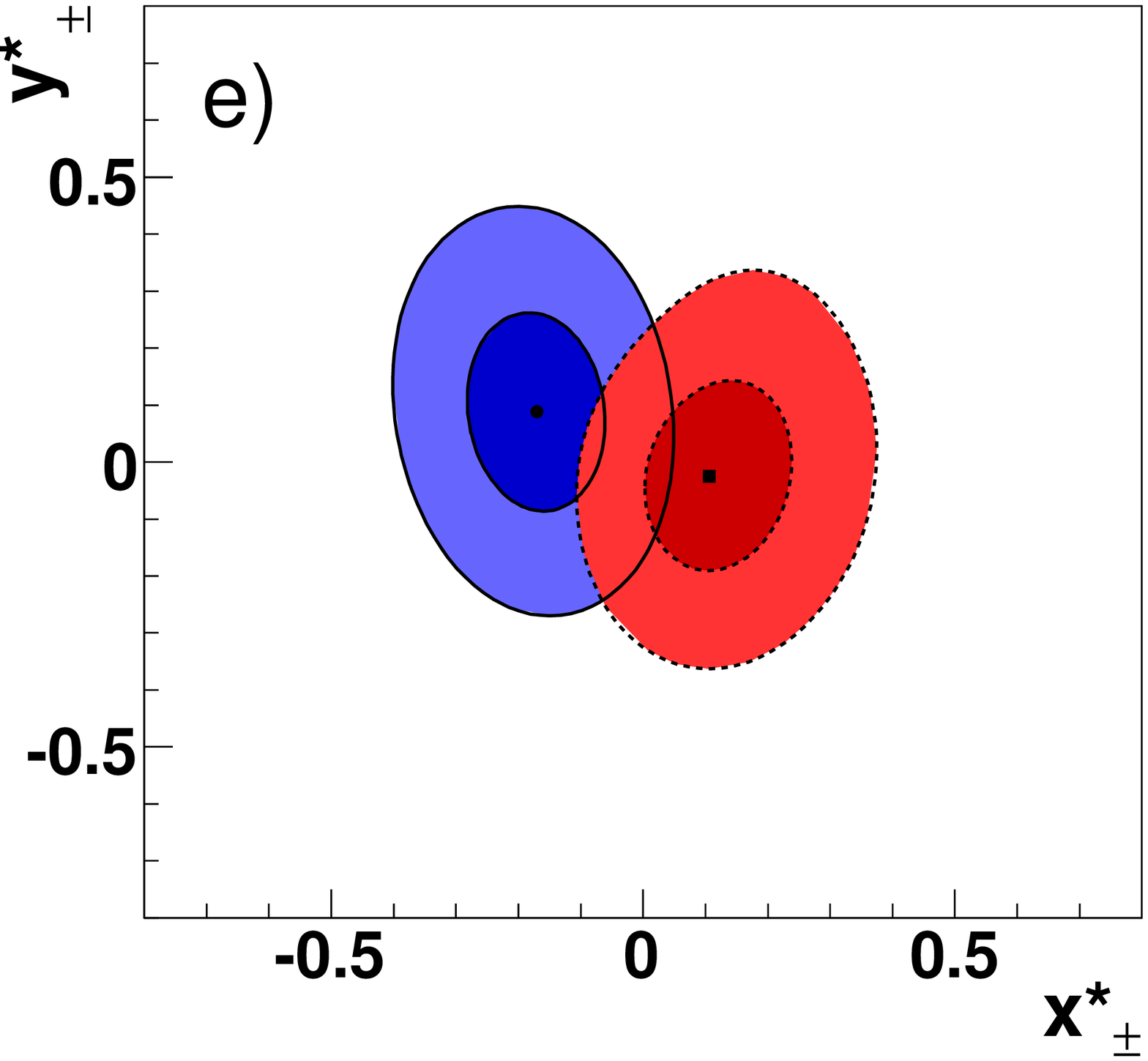} & 
\includegraphics[width=0.24\textwidth]{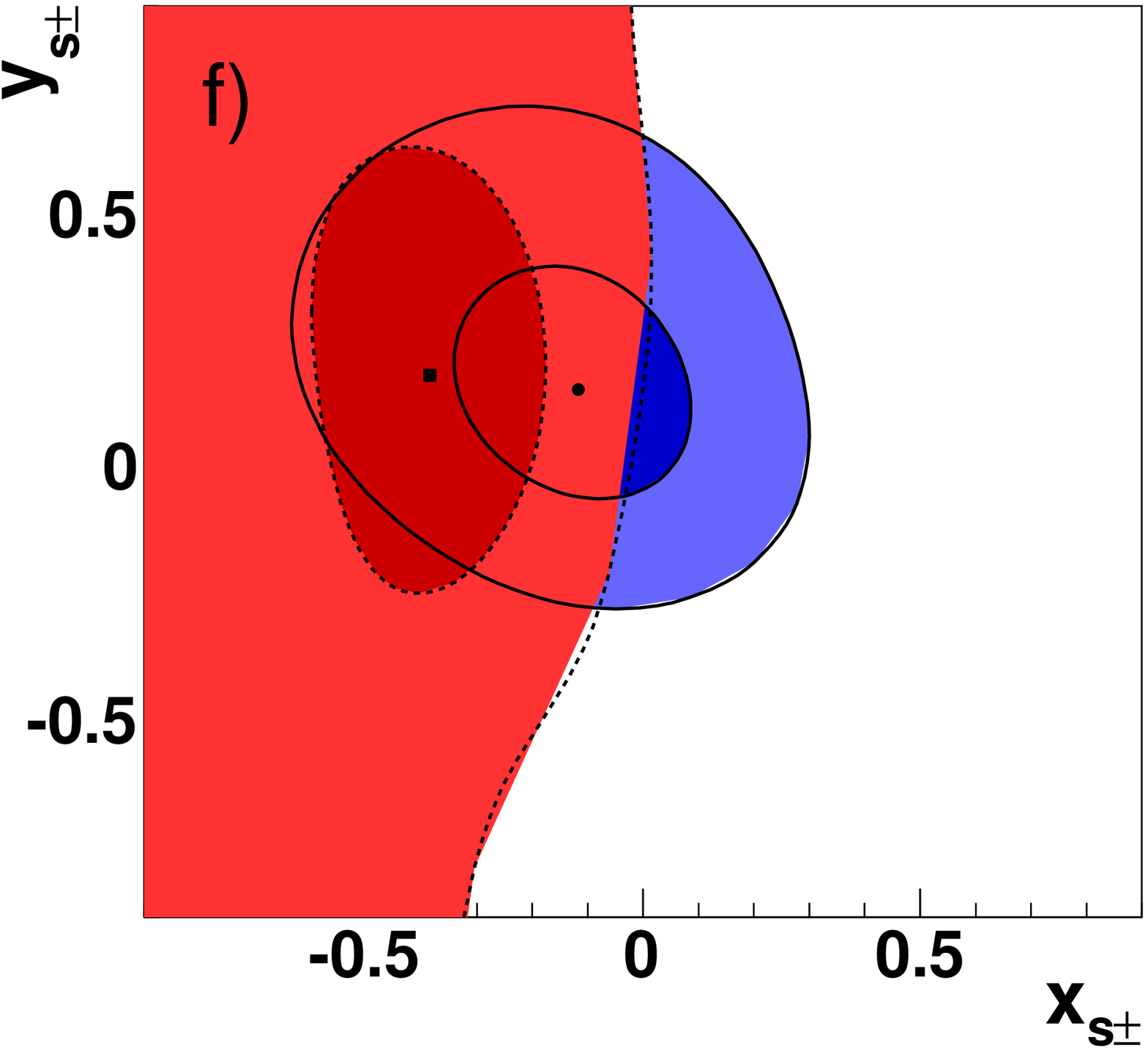} \\ 
\end{tabular} 
\caption{\label{fig:cart_CL_kspipi_kskk} (color online).  
Contours at $39.3\%$ (dark) and $86.5\%$ (light) 2-dimensional CL in the  
(a)(d) \zbmp, (b)(e) \zbstmp, and (c)(f) \zsmp planes, corresponding to one- and two-standard deviation regions (statistical only),  
for \Bm (solid lines) and \Bp (dotted lines) decays, from the \CP fit to the signal samples performed separately for  
(a)-(c) \Dtokspipi and (d)-(f) \Dtokskk decays. The results from the two subsets are statistically consistent. 
} 
\end{figure}

% Dpi xy scans for Kspipi and KsKK only 
 
\begin{figure}[hbt!] 
\begin{tabular}{cc} 
\includegraphics[width=0.24\textwidth]{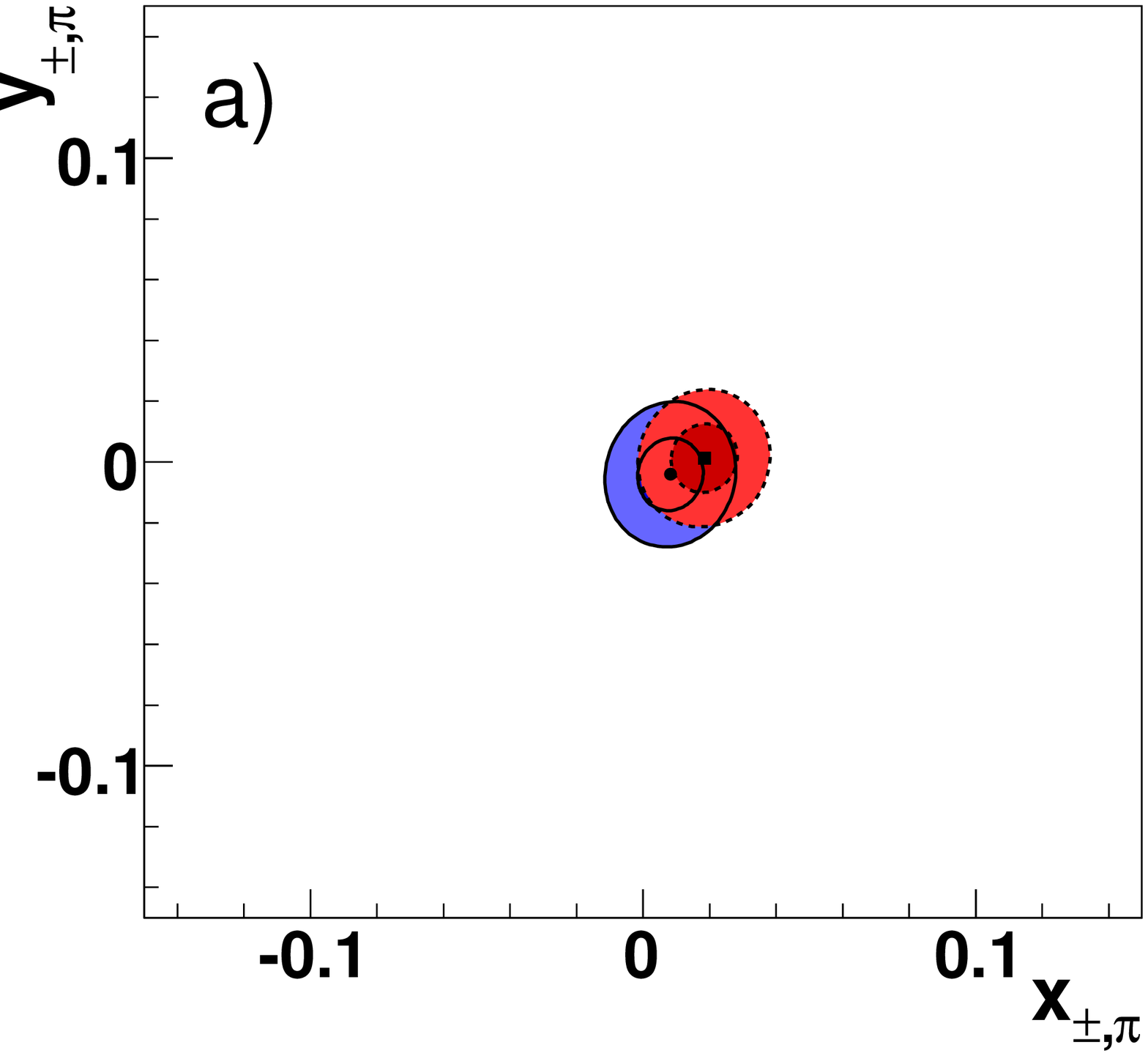}& 
\includegraphics[width=0.24\textwidth]{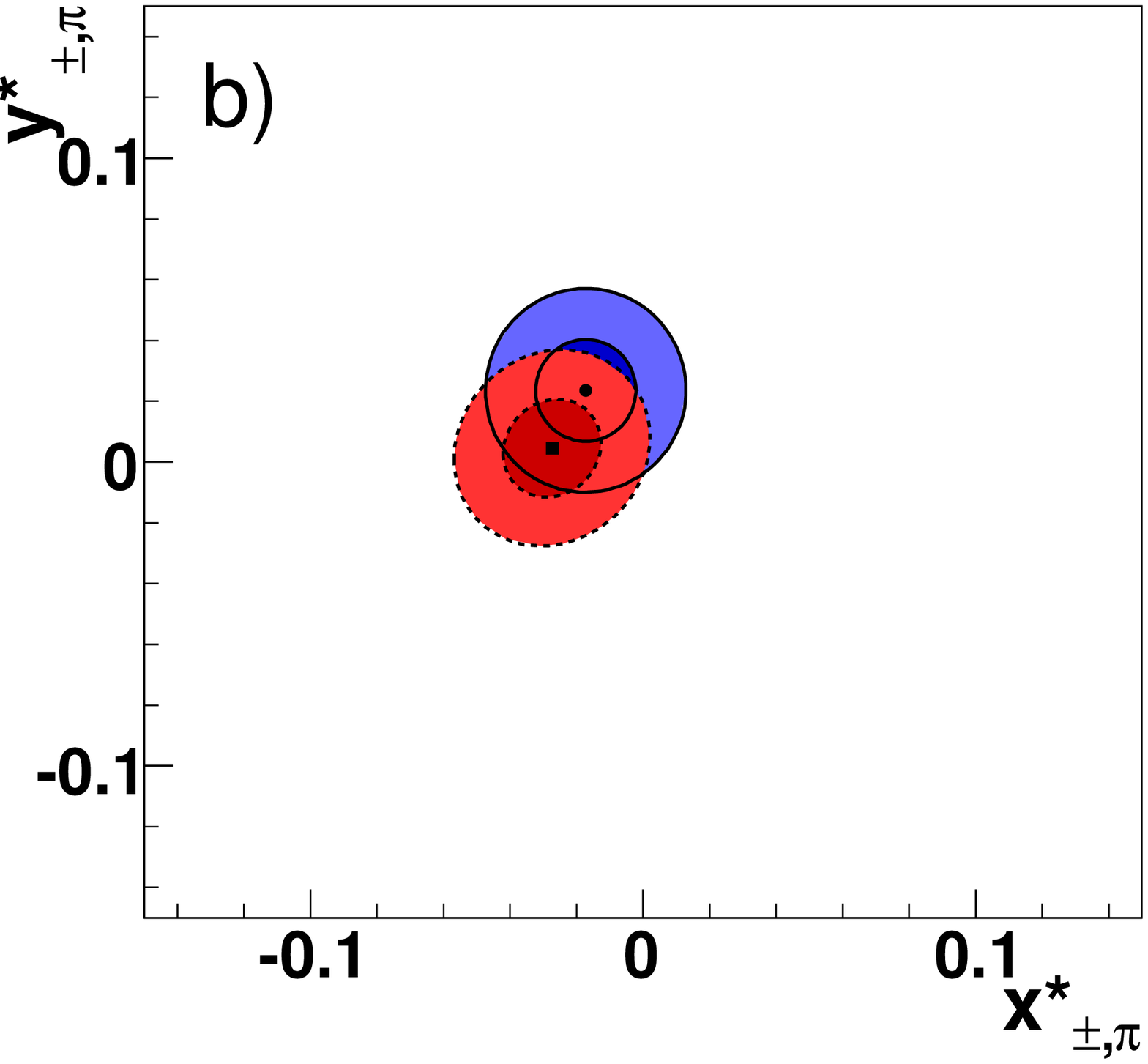} \\ 
\includegraphics[width=0.24\textwidth]{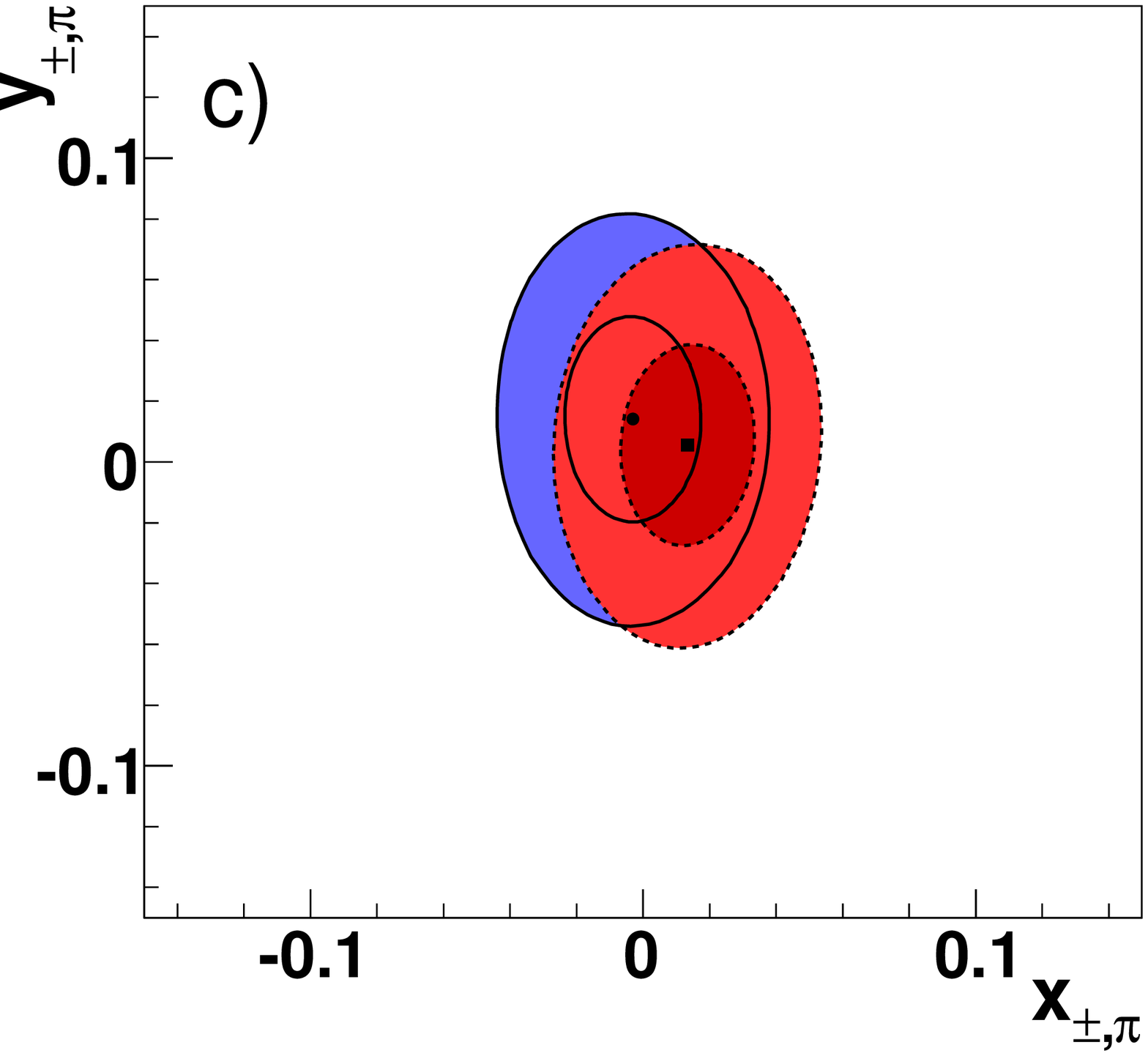}& 
\includegraphics[width=0.24\textwidth]{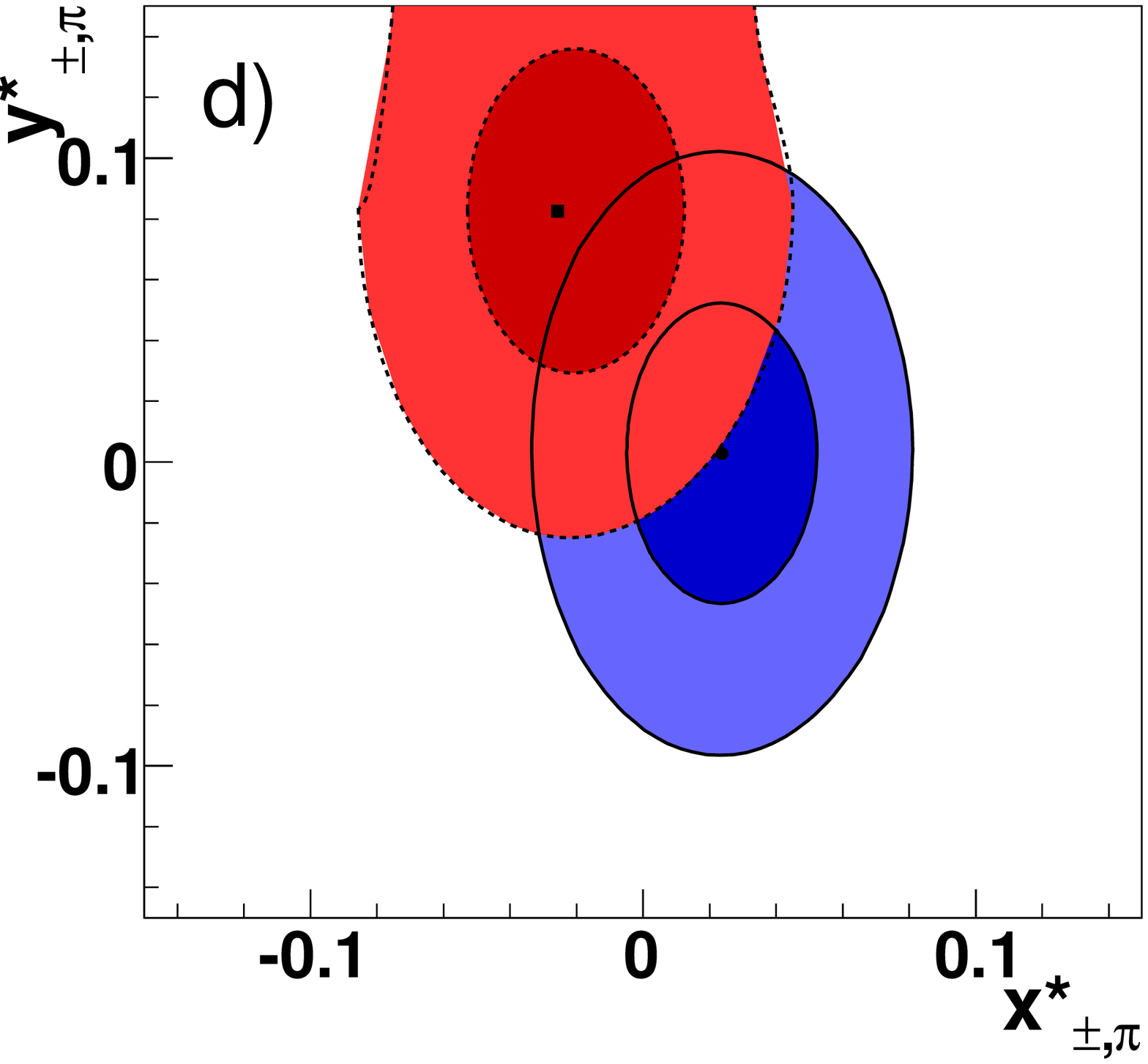} \\ 
\end{tabular} 
\caption{\label{fig:cart_CL_Dpi_kspipi_kskk} (color online).  
Contours at $39.3\%$ (dark) and $86.5\%$ (light) 2-dimensional CL in the  
(a)(c) \zpibmp and (b)(d) \zpibstmp planes, corresponding to one- and two-standard deviation regions (statistical only),  
for \Bm (solid lines) and \Bp (dotted lines) decays, from the \CP fit to the $\Bmp \to \DDstar \pimp$ control samples 
performed separately for (a)(b) \Dtokspipi and (c)(d) \Dtokskk decays.  
In this case we expect the \zpibmp and \zpibstmp contours close to the origin up to $\sim 0.01$, since $r_{\B,\pi}^{(*)} \approx 0.01$ and 
the experimental resolutions are of the same order.  
Deviations from this pattern could be an indication that the DP distributions are not well described by the amplitude models~\cite{ref:babar_dalitzpub2008}. 
The results from all the subsets are consistent with the expectations. 
Note the differences in scale when comparing to Fig.~\ref{fig:cart_CL_kspipi_kskk}. 
} 
\end{figure}


\begin{thebibliography}{99}
\bibitem{ref:chargeconj} Reference to the charge-conjugate state is implied here and throughout the text unless otherwise specified.
\end{thebibliography}
\end{document}